Radial distribution and its spherical harmonics expansion in liquid $CO_2$ at 250 K and 50 bar based on pressure-constant Monte Carlo simulation using Kihara Potential Model

(Second revision)

Koji Kobashi

Former Research Assistant, Physics Department, Colorado State University, Fort Collins, CO, USA, and Former Senior Researcher, Kobe Steel, Ltd., Japan

**Abstract**

The purpose of this article is to compute the radial distribution function of liquid $CO_2$ at 250 K and 50 bar and its expansion coefficients by spherical harmonics as a function of intermolecular distance up to 20 Å for all possible combinations of angular indexes $l$, $l'$, and $|m| \leq 8$, in contrast to past works of limited combinations. The positions and orientations of $CO_2$ molecules were determined by pressure-constant Monte Carlo simulation using a Kihara potential model. It was found that the radial distribution function had an intense peak at 4.05 Å and a broad band at around 7.85 Å. The expansion coefficients had prominent structures below approximately 7.5 Å, showing positive and/or negative peaks, which indicated that the orientational correlations of the central molecule were intense with the molecules present in that region. It was also found that the peaks of some expansion coefficients were not small even when $l$ = 8 on the contrary to a normal assumption that expansion coefficients would be smaller as $l$ increases. Furthermore, molecular orientations had a tendency to orient along one direction presumably because of the linear form of $CO_2$ molecule.

## 1. Introduction

Research of linear molecules such as $CO_2$ is of interest as the molecule has the orientational degrees of freedom, expressed by $\omega = (\theta, \varphi)$, in addition to the positional degree of freedom as expressed by the molecular center coordinate $(x, y, z)$. The theory of molecular fluids has been well established and excellent textbooks have been published [1-4], some of which present even computer programs for Monte Carlo (MC) simulation [1, 3, 4]. Thermodynamic data of liquid $CO_2$ such as molar volume, internal energy, and heat capacity at different pressures $P$ and temperatures $T$ are available [5, 6]. The boundaries of solid-liquid, solid-gas, and gas-liquid have been determined both analytically [5, 7] and experimentally. A phase diagram of $CO_2$ is shown in Fig. 1 [4, 5, 8, 9]. The red dot indicates the data point (250 K, 50 bar, 1 bar = $10^{-4}$ GPa) used in the present study.

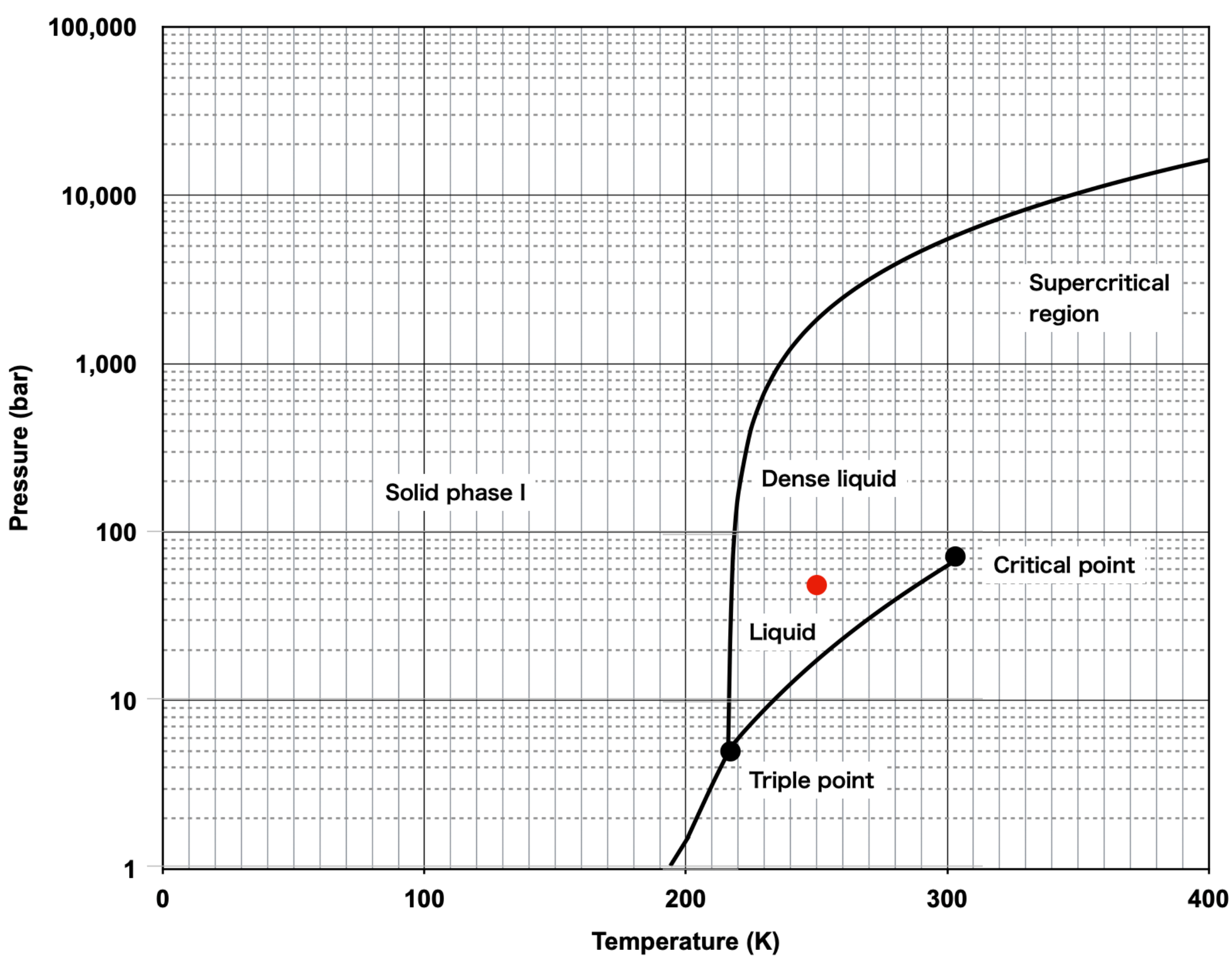

Fig. 1. Phase diagram of $CO_2$: the red point (250 K, 50 bar) indicates the conditions used in the present study. Solid phase I has a crystal structure of space group *Pa3*, and exists beyond $10^5$ bar (10 GPa).

Liquids consisting of simple linear molecules such as $N_2$ and $CO_2$ have been studied in many articles [10-12] but most of them used 12-6 Lennard-Jones (LJ) potentials that included the energy ε and the core diameter σ, and actual energy, length, and volume were normalized by ε or σ to simplify and generalize the computed results. Since the present article used a 9-6 LJ potential for the Kihara core potential (see Appendix I), it was difficult to compare the results of the present study with those of past theoretical articles, and therefore no comparison with past works was made. Rather, the present study focussed on computing the radial distribution function $g(r)$ and its expansion by spherical harmonics $Y_{lm}(\omega)$ of liquid $CO_2$ at 250K and 50 bar, in which pressure-constant MC (*NPT*-MC) simulation was done using the Kihara potential. The Kihara potential has been used previously to study solid $CO_2$ phase I under ambient and high pressures ≤ 10 GPa [13-17], and is described in Appendix I. In addition, the computational procedure used in the present study is explained in Appendix II. Before showing the computed results, $g$(r) and its expansion by spherical harmonics are explained in Sec. 2, the results and discussion are described in Sec. 3, and finally the conclusion is given in Sec. 4.

## 2. Radial distribution function and its expansion by spherical harmonics

In liquid $CO_2$, molecules surrounding a central molecule under consideration are pushed away from the central molecule by the repulsive intermolecular potentials. These molecules tend to accumulate around the minimum of the intermolecular potentials although the molecular distribution is broadened by thermal motion of the molecules. The radial distribution function $g(r)$ expresses the degree of the molecular density at a distance $r$ from the central molecule in comparison with the average number density of molecules $N/V$, where $N$ is the number of molecules in volume $V$. According to the *NPT*-MC simulation in the present study with $N$ = 2048, the edge length of the computed cubic box $A_L$ = 53.2 Å, hence $V = A_L^3$, and thus the molar volume was 44.3 $cm^3$/mol.

As shown in Fig. 2, the radial distribution function $g(r)$ is defined by the following equation [1-4, 10, 11]:

$$g(r) = \frac{< N(r) >_{shell}}{Nn\frac{4\pi}{3}\{(r + \Delta r)^3 - (r - \Delta r)^3\}} \quad , \tag{1}$$

where $<N(r)>_{shell}$ denotes the number of molecules with their centers lying within a spherical shell defined by ($r$ - $\Delta r$/2, $r$ + $\Delta r$/2) like molecule 2. Here, $r$ represents the radius of the spherical shell measured from the molecular center of molecule 1. The shell thickness was $\Delta r$ = 0.1 Å in the present study. As a result, $g(r)$ and its expansion coefficients by spherical harmonics are depicted by step-type graphs, as will be seen later. In the denominator of Eq. 1, $n$ is the average number density of molecules in the cubic box, $n = N/V$. The denominator also includes the volume of the spherical shell, $(4\pi/3)\{(r + \Delta r/2)^3 - (r - \Delta r/2)^3\}$. Above Eq. 1 states that $g(r)$ is the ratio of the number of molecules in the spherical shell at a distance $r$ against the average number density of molecules $n$ in the cubic box. After repeated *NPT*-MC computations, (i) the volume of the cubic box was determined from the average of last 100 data in the final round of computation, subsequently (ii) 100 sets of *NVT*-MC computations were performed, each generating $g(r)$, and lastly (iii) the final $g(r)$ was obtained by averaging the 100 sets of $g(r)$. See Appendix II for details.

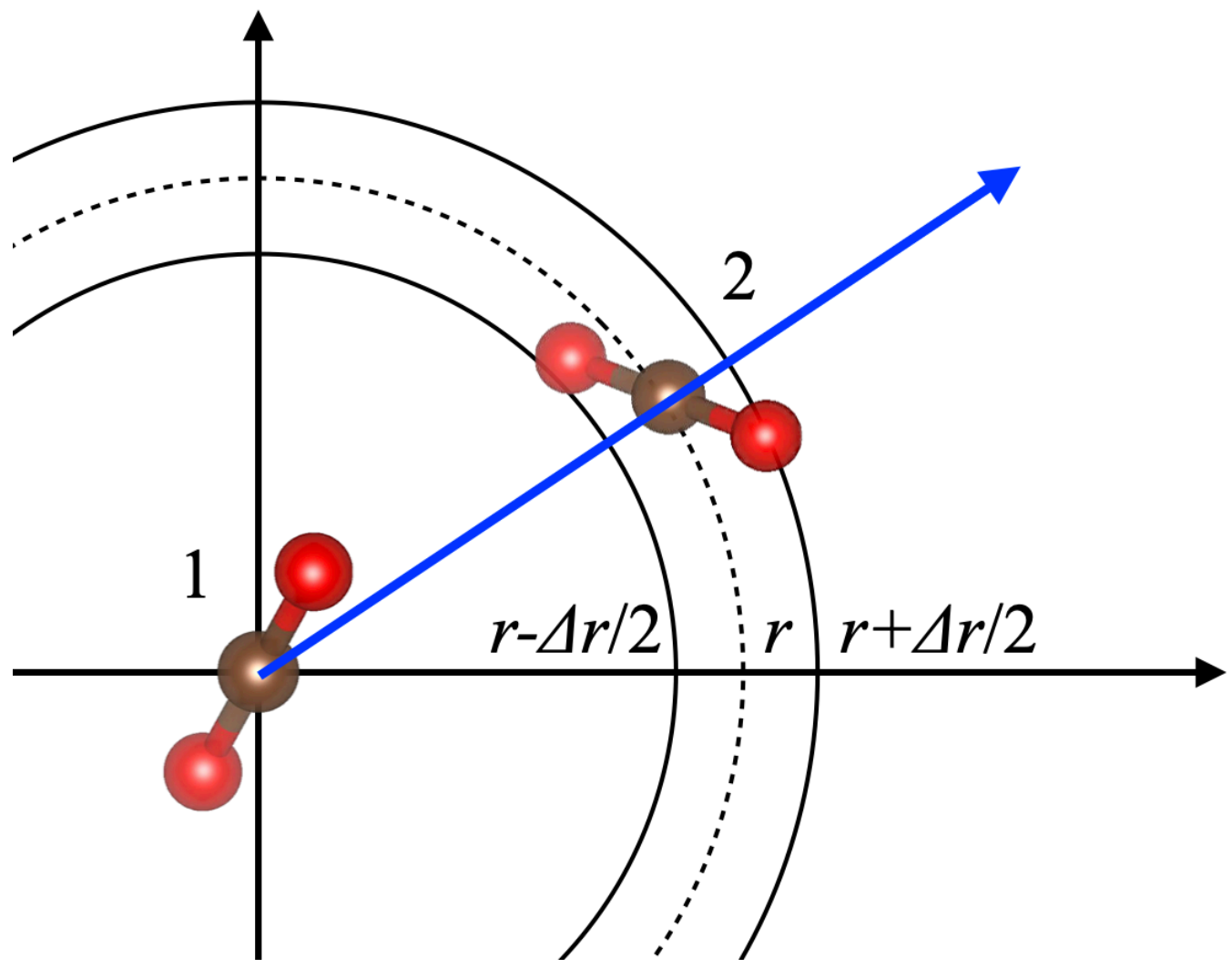

Fig. 2. Concept of $g(r)$: molecule 2 lies in a spherical shell at a distance $r$ from the molecular center of molecule 1 at the origin. The shell thickness is $\Delta r$.

Define $\omega_i = (\theta_i, \varphi_i)$ as the orientation of the $i$-th molecule, where $i$ runs from 1 through 2048, and also define $\Omega$ as a representation of angular variables $\Omega = \{\omega_i\}$ of all molecules. The radial distribution function $g(r)$ can be expanded with spherical harmonics $Y_{lm}(\omega_i)$ as follows [1-4, 10, 11]:

$$g(r, \omega_1, \omega_2) = 4\pi \sum_{l,l',m} g_{ll'm}(r)\, Y_{lm}(\omega_1)\, Y_{l'-m}(\omega_2) \quad , \tag{2}$$

and hence,

$$g_{ll'm}(r) = 4\pi\, g(r) < Y^*_{lm}(\omega_1)\, Y^*_{l'-m}(\omega_2) >_{shell} \quad . \tag{3}$$

It should be noted that $g(r)$ in Eq. 1 is equivalent to $g_{000}(r)$ in Eq. 3:

$$g_{000}(r) = g(r) \quad . \tag{4}$$

Note also that the suffixes 1 and 2 to $\omega$ in Eqs. 2 and 3 do not indicate specific molecules but represent different molecules. Furthermore, in Eqs. 2 and 3, all of $l$ and $l'$ are even numbers as $g_{ll'm}(r)$ vanishes if $l$ and/or $l'$ are odd numbers because $CO_2$ molecule possesses point symmetry. It should also be noted that $\omega_i$, and hence $\Omega$, refers to a coordinate system in which the blue line in Fig. 2 is the main axis. This causes a complication to analytically determine $\varphi_i$, and some techniques have been used [18, 19]. In the present study, a different method was used and presented in Appendix III. Note also that $m$ in Eqs. 2 and 3 should be $|m| \leq \min(l, l')$, and $<Y_{lm}^*(\omega_i)\, Y_{l'-m}^*(\omega_j)>_{shell}$ is the ensemble average of all molecules with their molecular center-to-center distances within a spherical shell. From the relation below:

$$Y_{lm}^*(\omega) = (-1)^m\, Y_{l-m}(\omega), \tag{5}$$

it follows that:

$$< Y_{lm}^*(\omega_1)\, Y_{l'-m}^*(\omega_2) >_{shell} = < Y_{l-m}(\omega_1)\, Y_{l'm}(\omega_2) >_{shell} \quad , \tag{6}$$

and hence the right-hand side of Eq. 6 was used in computation. The explicit forms of $Y_{lm}(\theta, \varphi)$ are listed in Ref. 20. Since $g_{ll'm}(r)$ expresses orientational correlations of molecules in liquid $CO_2$, it is preferable that the value is real in order to depict it. Define $\Lambda$ as:

$$\Lambda = < Y_{l-m}(\omega_1)\ Y_{l'm}(\omega_2) >_{\text{shell}}\ , \tag{7}$$

then there are following cases (i) to (iii) that $\Lambda$ is real:

(i) if $m = 0$, $Y_{l0}(\omega)$ is real [20] and hence $\Lambda$ is real.

(ii) if $l = l'$, $\Lambda$ is real because if there is a case that molecule 1 is the central molecule and molecule 2 lies in the spherical shell at a distance $r$, as seen in Fig. 2, then there exists a reversed case that molecule 2 is the central molecule and molecule 1 lies in the spherical shell at the same distance $r$. This means that in $\Lambda$, both $Y_{l-m}(\omega_1)\ Y_{l'm}(\omega_2)$ and $Y_{l-m}(\omega_2)\ Y_{l'm}(\omega_1)$ are included as a pair, and they are added in computing $\Lambda$. From Eq. 5 and the complex conjugate of the following function equals to its real function:

$$\begin{aligned}
&[Y_{l-m}(\omega_1)\ Y_{lm}(\omega_2) + Y_{l-m}(\omega_2)\ Y_{lm}(\omega_1)]^* \\
&= [Y_{l-m}^*(\omega_1)\ Y_{lm}^*(\omega_2) + Y_{l-m}^*(\omega_2)\ Y_{lm}^*(\omega_1)] \\
&= [Y_{lm}(\omega_1)\ Y_{l-m}(\omega_2) + Y_{lm}(\omega_2)\ Y_{l-m}(\omega_1)] \\
&= [Y_{l-m}(\omega_1) Y_{lm}(\omega_2) + Y_{l-m}(\omega_2)\ Y_{lm}(\omega_1)],
\end{aligned}$$

$\Lambda$ is proved to be real if $l = l'$. This also leads to a relation that:

$$g_{ll+m}(r) = g_{ll-m}(r) \tag{8}$$

Hence, $g_{ll+m}(r)$ and $g_{ll-m}(r)$ will be written as $g_{ll+/-m}(r)$, hereafter.

(iii) In other cases, $\Lambda$ is complex in general. To circumvent this problem, define a new function $(1/2)[\Lambda + \Lambda^*]$ for $l \neq l'$:

$$(1/2)[\Lambda + \Lambda^*] = (1/2) < Y_{l-m}(\omega_1)\ Y_{l'm}(\omega_2) + Y_{l-m}^*(\omega_1)\ Y_{l'm}^*(\omega_2) >_{\text{shell}}, \tag{9}$$

which is obviously real. Above Eq. 9 will also be denoted as $g_{ll'+/-m}(r)$, or further abbreviated as ($ll'm$), hereafter. The reason for defining Eq. 9 is because the imaginary part of a spherical harmonics $Y_{lm}(\theta, \varphi)$ arises only from the complex term, $\exp(i\ m\varphi)$, here $i$ being the imaginary unit, included in the spherical harmonics (see Ref. 20). The function $(1/2)[\Lambda + \Lambda^*]$ of Eq. 9 is considered to be the same as that presented by Sweet [18] in his thesis in 1966, and the results of Ref. 18 have been used in later articles by some authors [2, 10]. In practice, however, Eq. 9 is more straightforward, and can be computed without further modification with ordinary desktop computers. More importantly, Eq. 9 can be used for all cases (i) to (iii) even though there are redundant computations.

## 3. Results and discussion

At 250 K and 50 bar, the *NPT*-MC simulation showed that the average edge length $A_L$ of the cubic box of volume $V$ (= $A_L^3$) was $A_L$ = 53.2 Å in equilibrium, hereby the molar volume was 44.3 $cm^3$/mol that was 6.6% greater than the NIST datum, 41.56 $cm^3$/mol. In terms of $A_L$, the difference was 2.1%. The radial distribution function $g(r)$ was computed according to Eq. 3. As seen in Fig. 3 shown below, $g(r)$ started to rise at $r$ = 2.65 Å, had an intense peak at $r$ = 4.05 Å, a broad band at around $r$ = 7.85 Å, and an extremely weak band at around $r$ = 11.85 Å. It should be noted that in the present simulation, the orientation-dependent potentials were the Kihara core potential between molecules that works up to $r \leq 10$ Å, and the electrostatic quadrupole-quadrupole (EQQ) potential that works up to $r \leq 15$ Å. Therefore, the orientational interactions reached up to $r \leq 15$ Å, and thus $g(r)$ in the region of $r > 15$ Å of Fig. 3 simply expresses random molecular correlations.

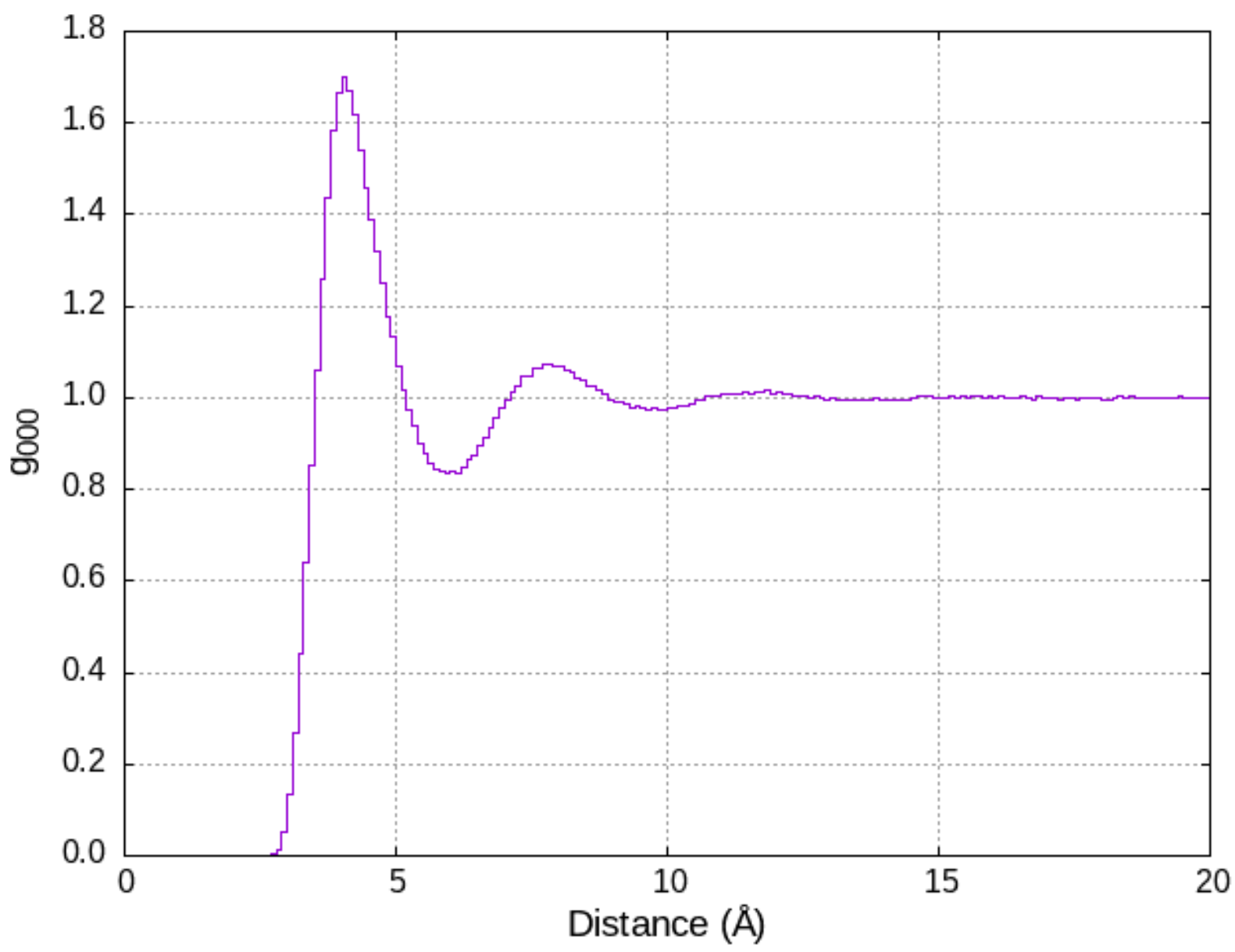

Fig. 3. The radial distribution function $g(r)$ of the molecular centers at 250 K and 50 bar computed in the present study. Note that $g_{000}(r)$ is equivalent to $g(r)$. The graph is drawn using gnuplot [21] in a step form with the step width of $\Delta r$ = 0.1 Å.

As a reference, the same computation as Eq. 1 was carried out for solid $CO_2$ phase I at $T$ = 0 K and $A_L$ = 44.44 Å. The result is shown in Fig. 4. The prominent three lines are located at $r$ = 4.65, 8.15, and 12.35 Å. Apart from the fact that the first peak is most intense, no relation was identified with Fig. 3 as $A_L$ of Fig. 4 was only 83% of Fig. 3.

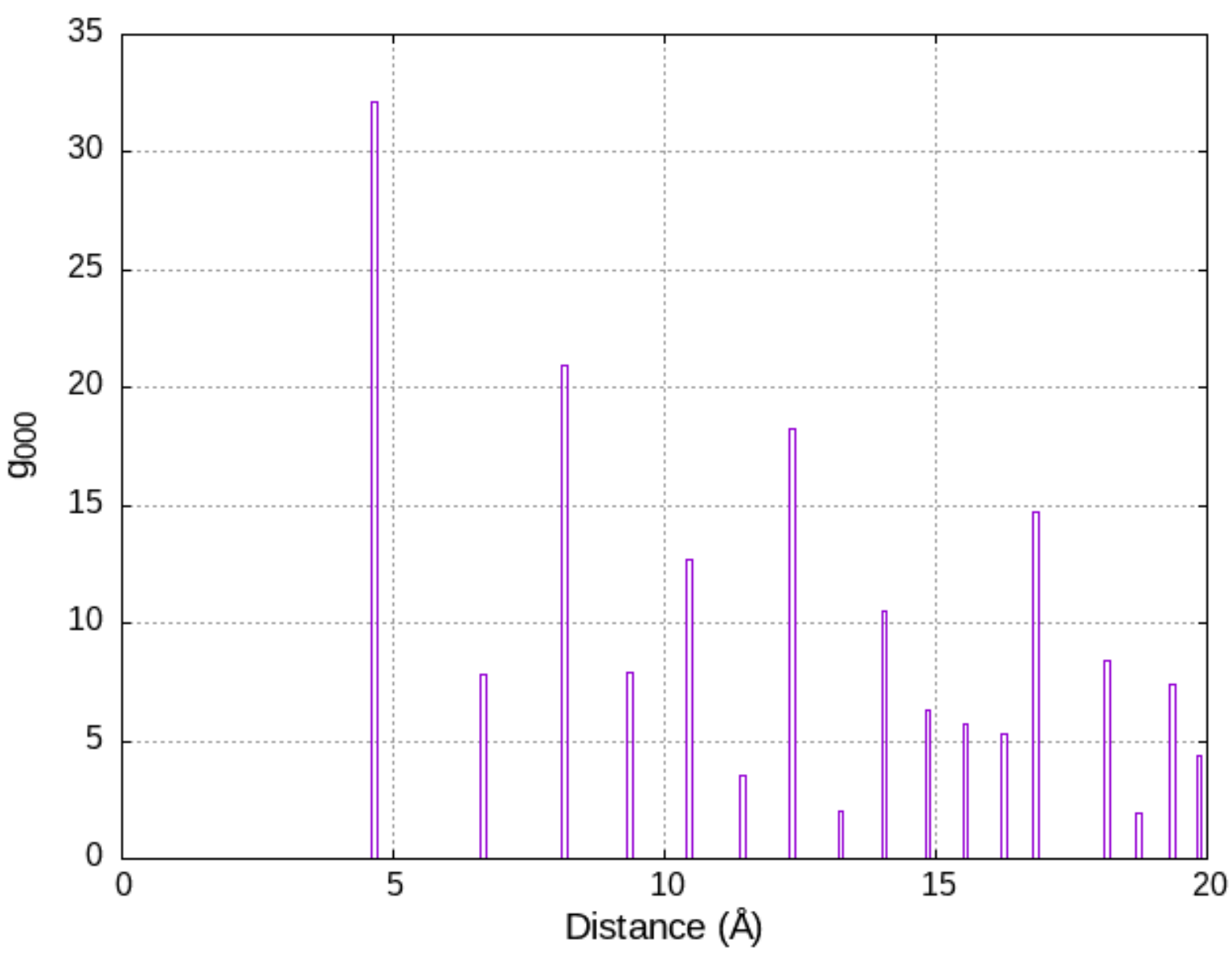

Fig. 4. The radial distribution function of the molecular centers in solid $CO_2$ phase I at $T$ = 0 K and $A_L$ = 44.44 Å computed according to Eq. 1.

Figure 3 can be compared with the results of both diffraction experiments and theoretical study of liquid $CO_2$ by Neuefeind *et al.* [22]: for simulation, Molecular Dynamics (MD) using both 12-6 Lennard-Jones (LJ) inter-atomic potentials and electrostatic potential of point charges, representing the electrostatic quadrupole moment, was used at $T$ = 298.1 K and $P$ = 66 bar. In the MD simulation, the number of $CO_2$ molecules was 1000. Their radial distribution function had a first peak at 3.95 Å and a second peak at around 7.9 Å. The position of the first peak was significantly smaller than the present result, 4.05 Å, presumably because of the difference in $T$ and $P$ used in the present study. It was not clear to what degree the difference of both the potential models and MD *vs*. MC influenced $g(r)$ between Ref. 22 and the present study.

In the following Fig. 5, the computed $g_{ll'm}(r)$ [or $g_{ll'+/-m}(r)$] are shown altogether. Each figure was not labelled but is identified by $g_{ll'm}$ used as the title of the vertical axis: $g_{ll'm}$ will be simply denoted as ($ll'm$), hereafter. General features of the figures were as follows:

(i) Each figure had the first prominent peak at around the main peak position of $g(r)$, $r$ = 4.05 Å (see Fig. 2), or mostly $r \leq 5$ Å.

(ii) The peaks were either positive and/or negative.

(iii) Some figures did not converge to zero at large $r$ and had a finite height even at $r \geq 10$ Å.

These features will be further discussed below.

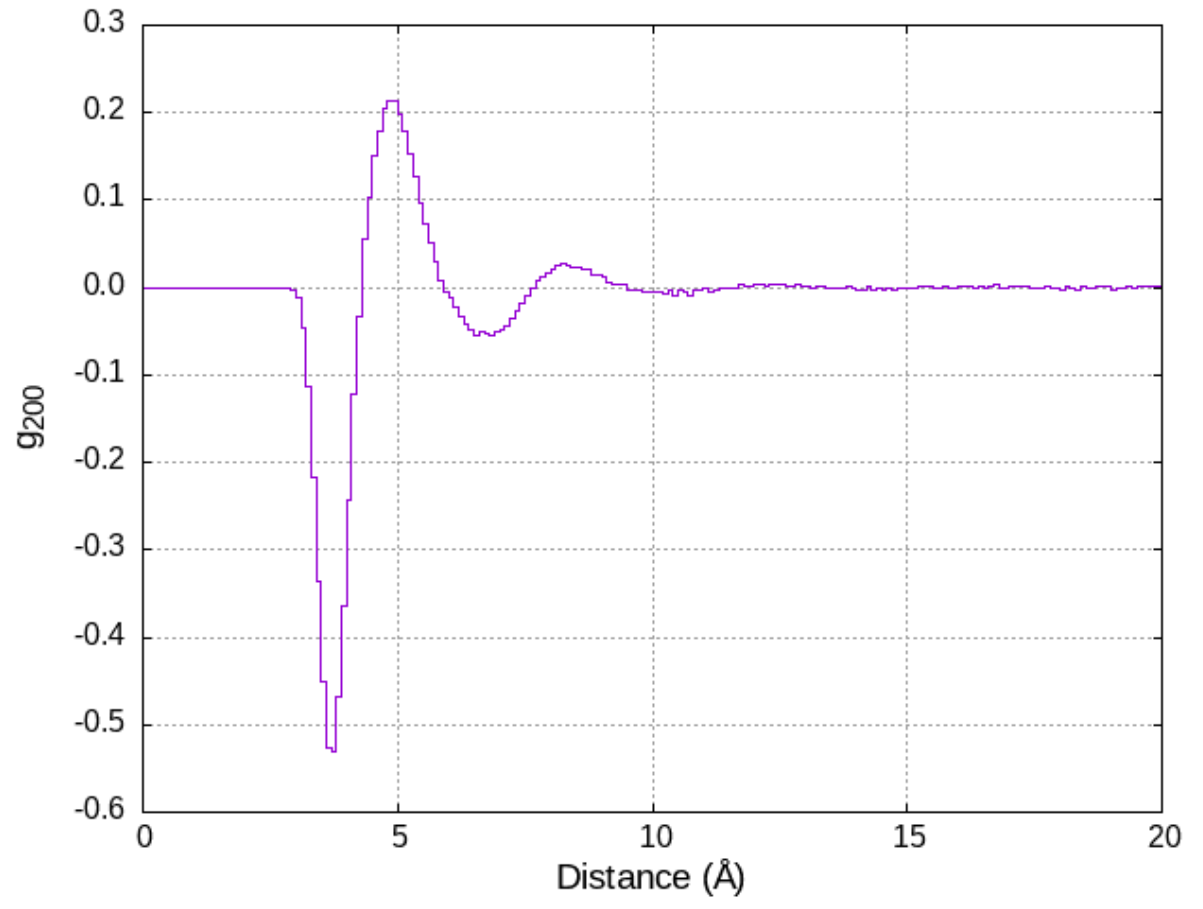

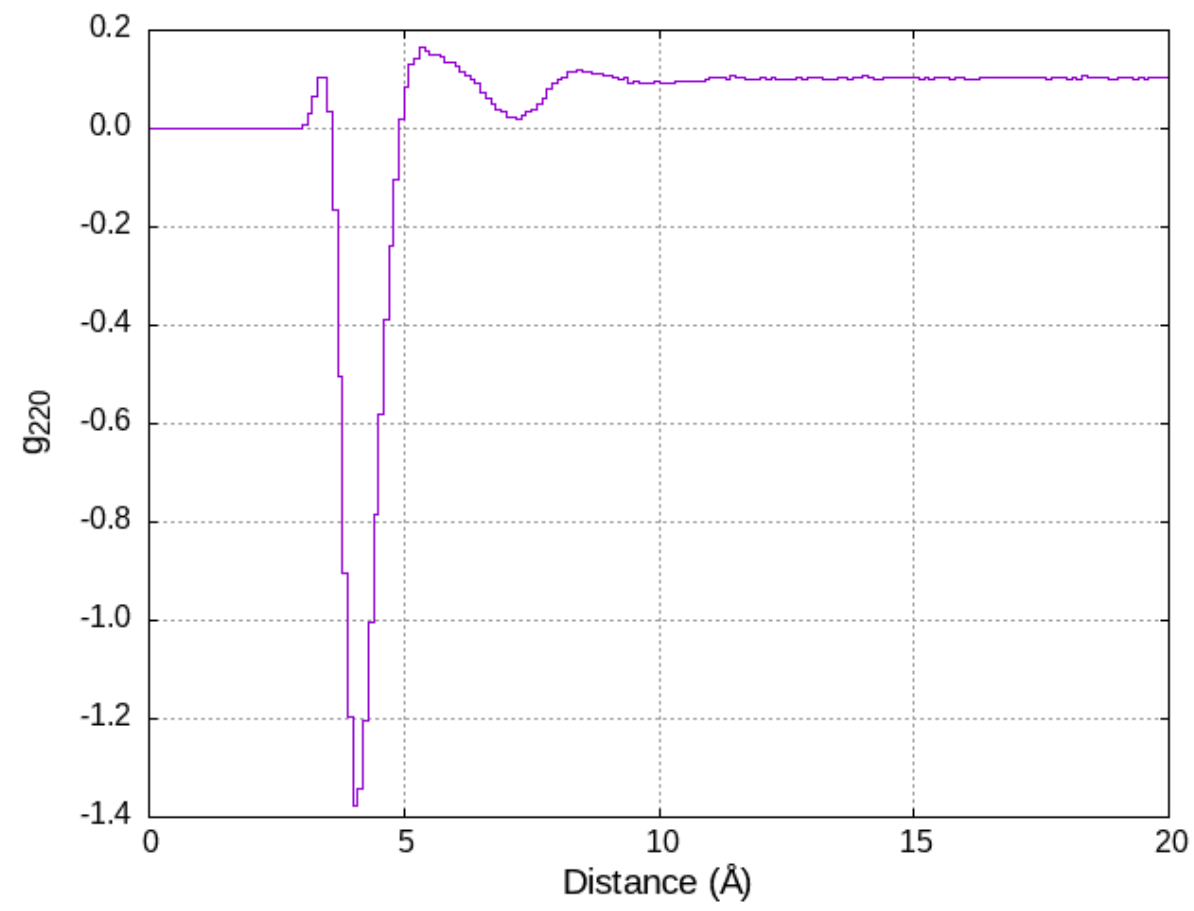

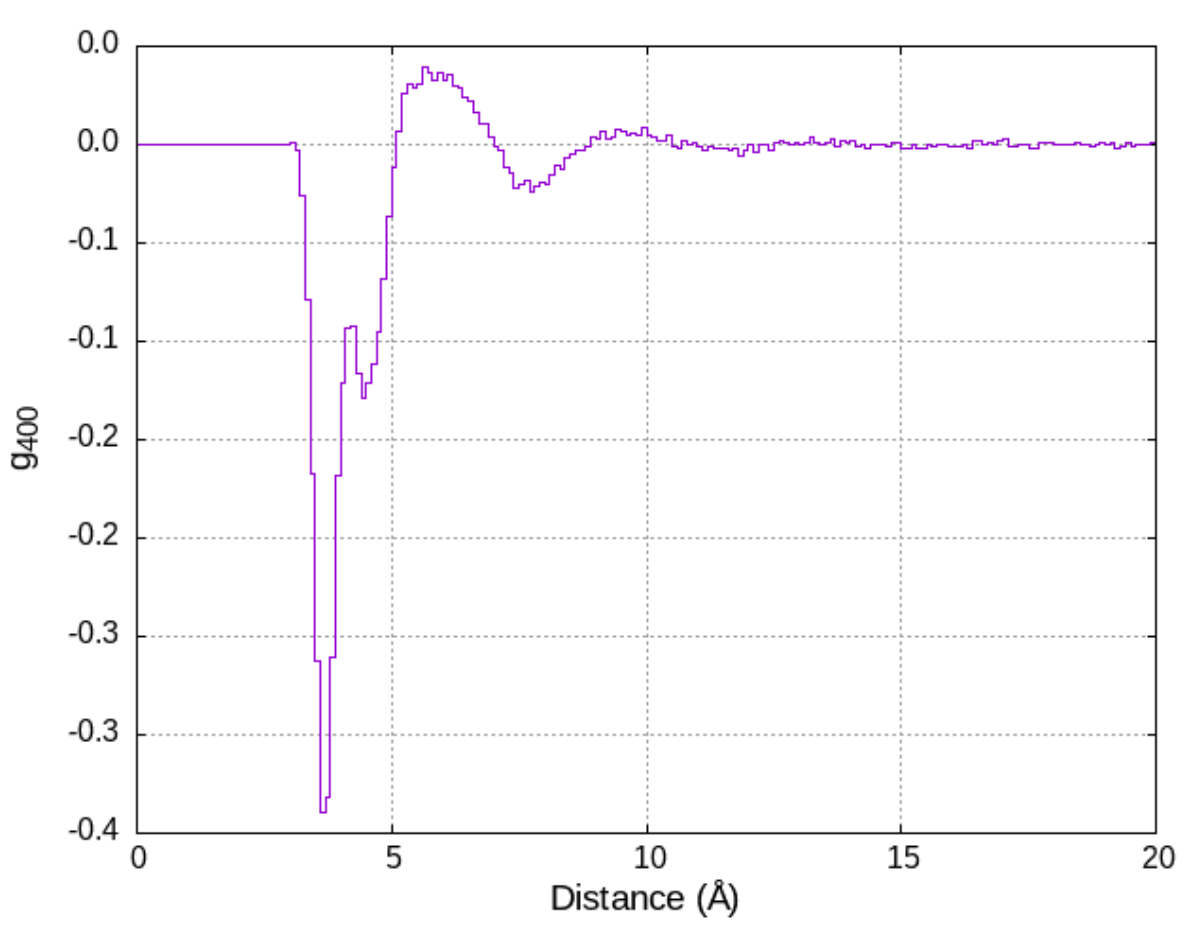

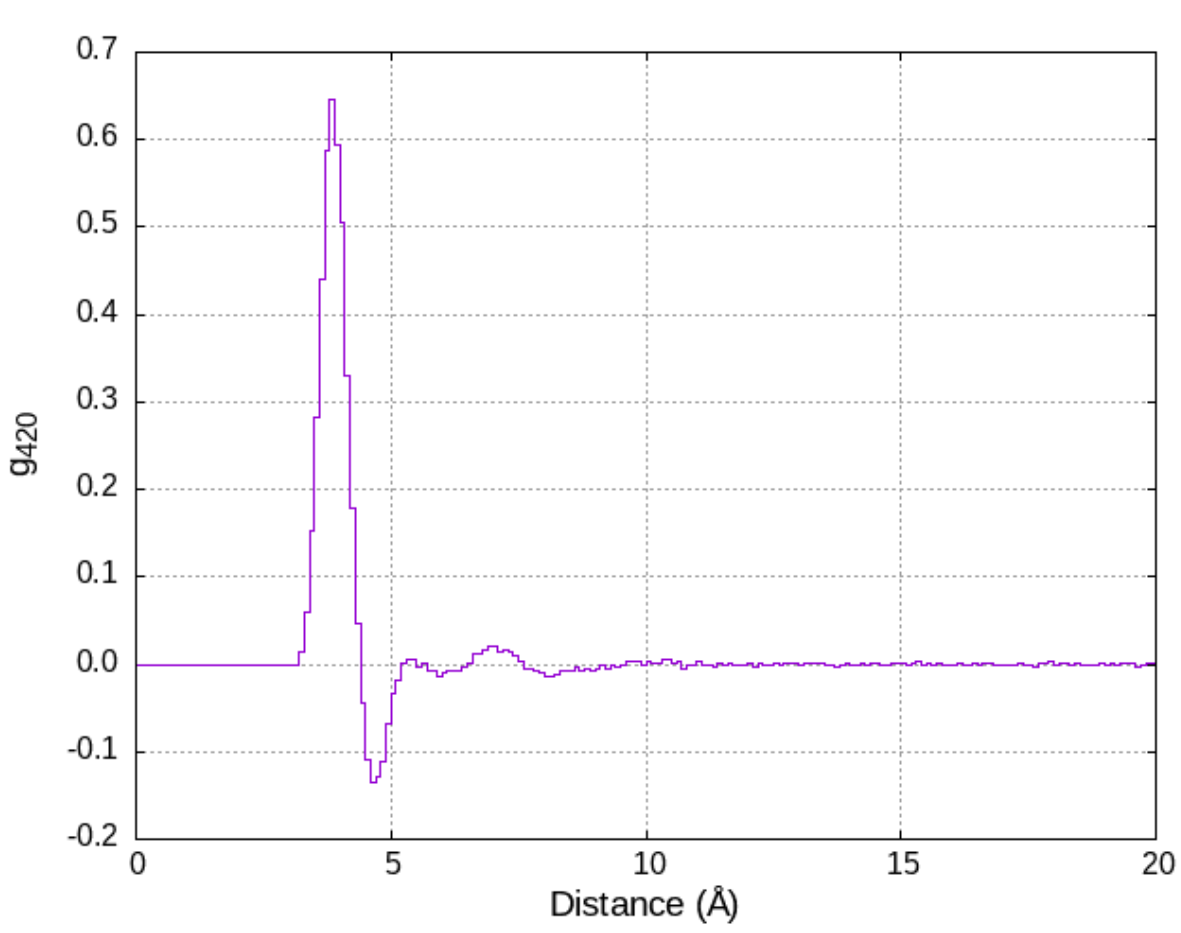

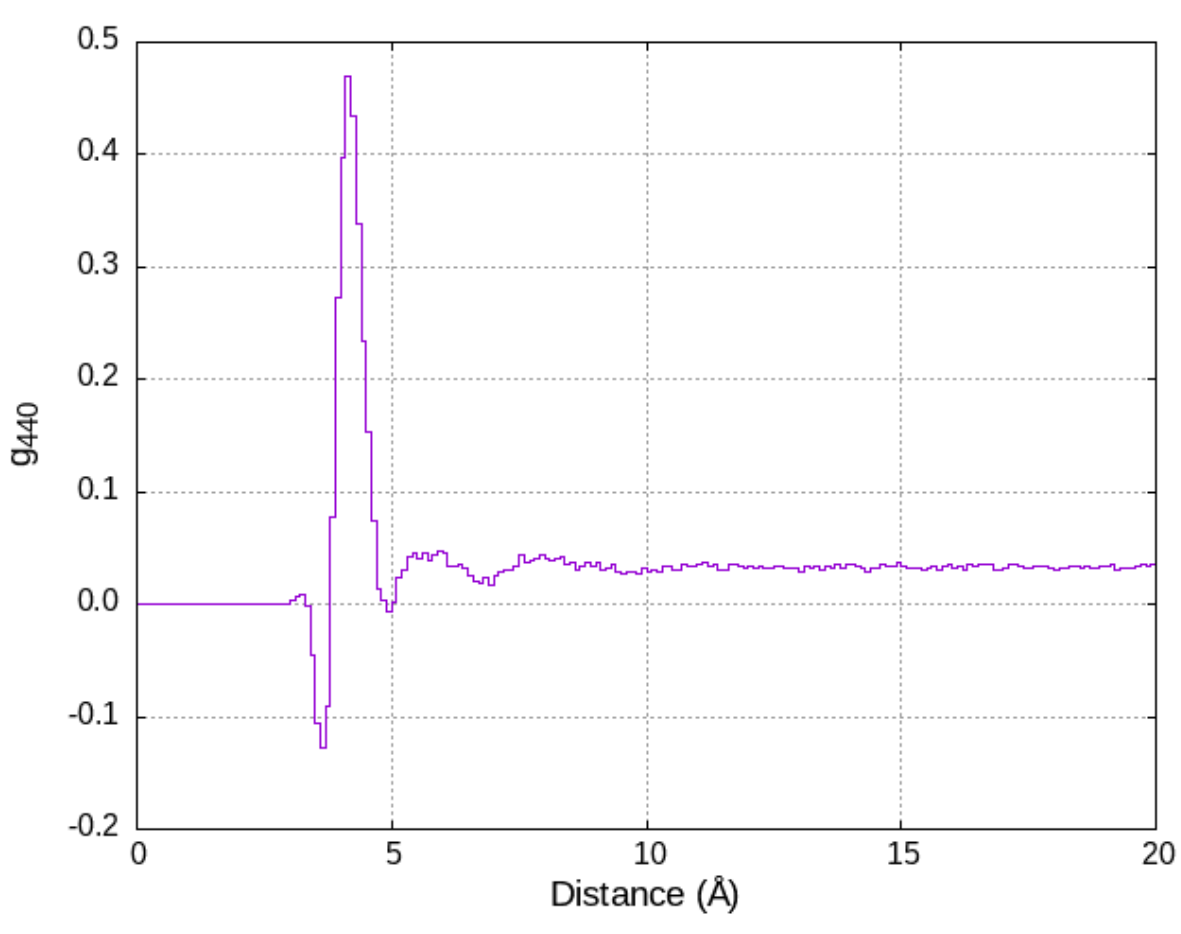

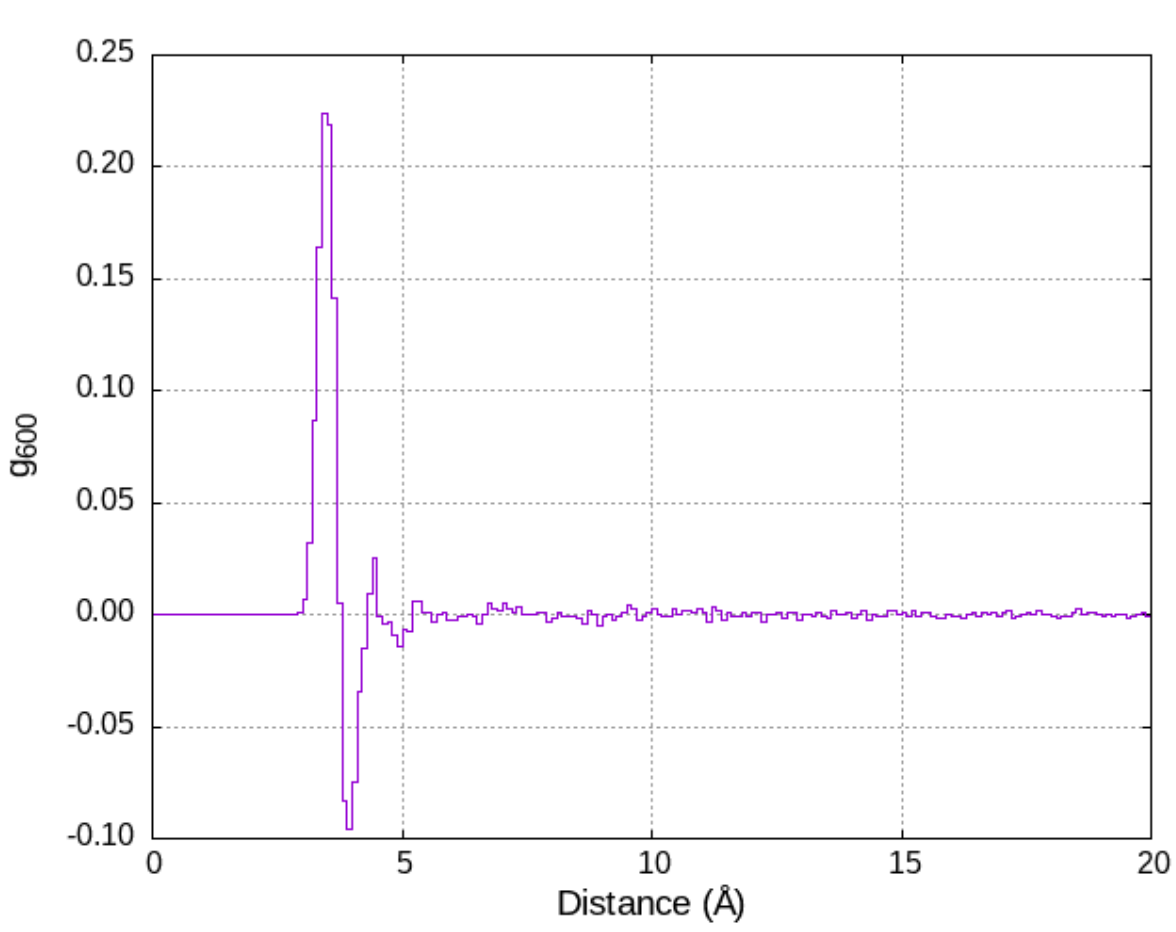

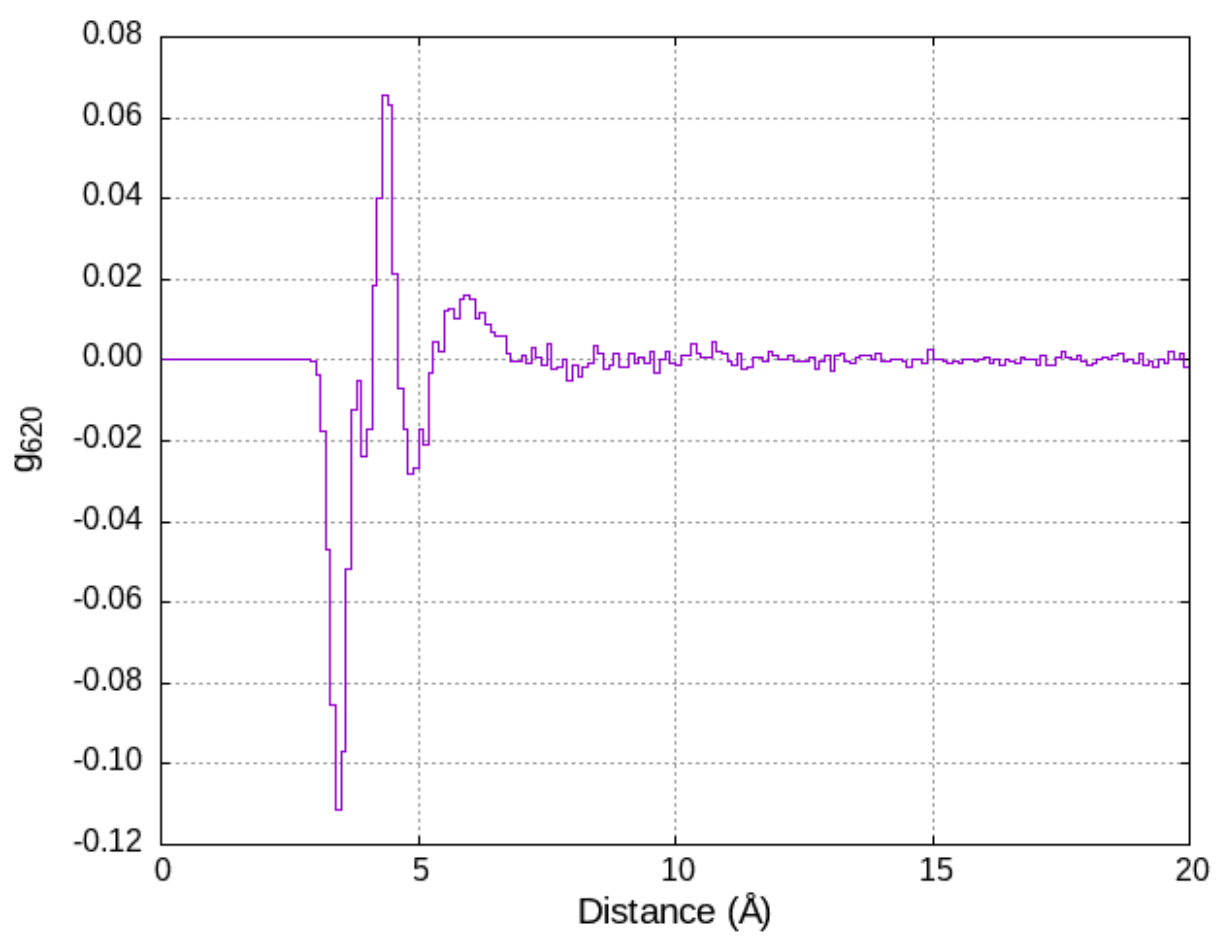

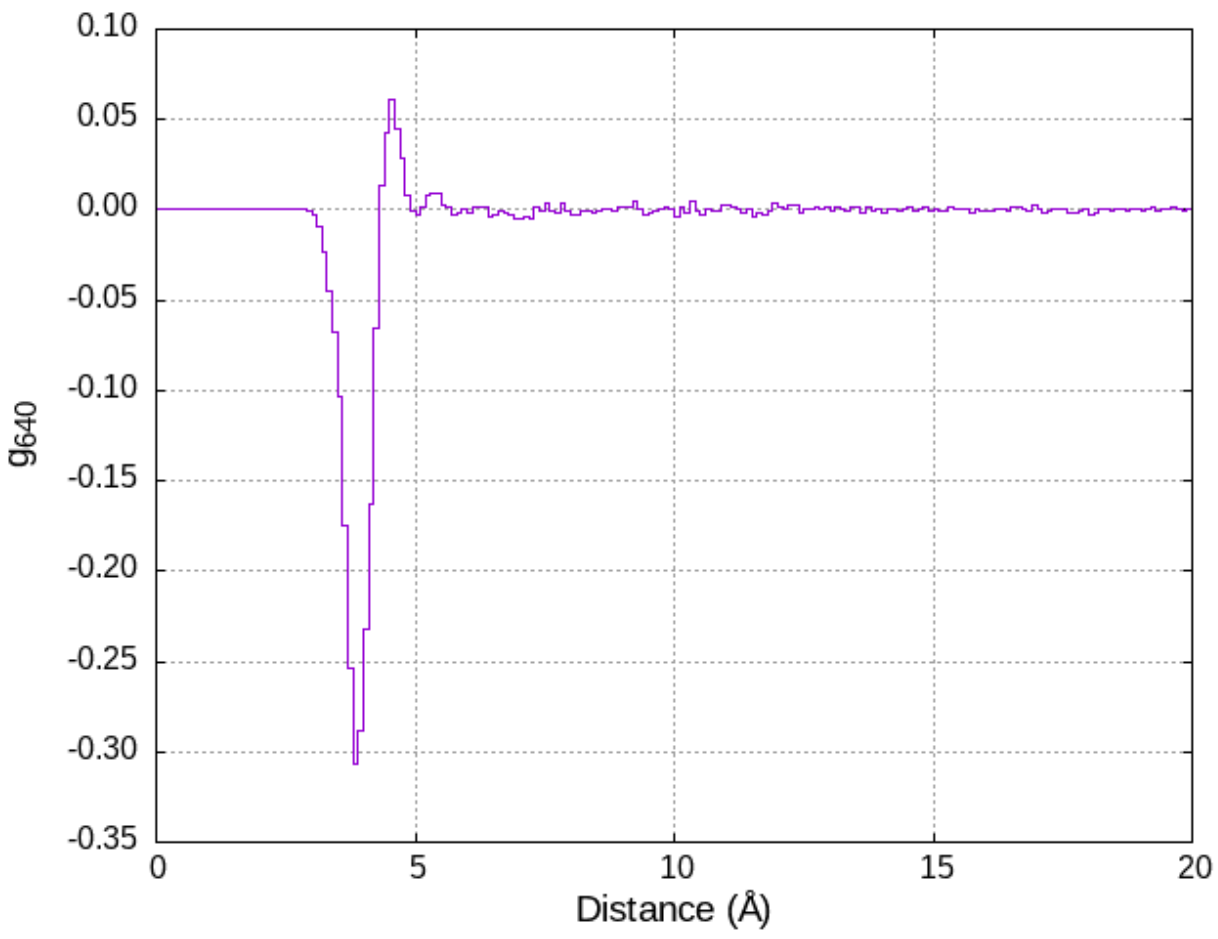

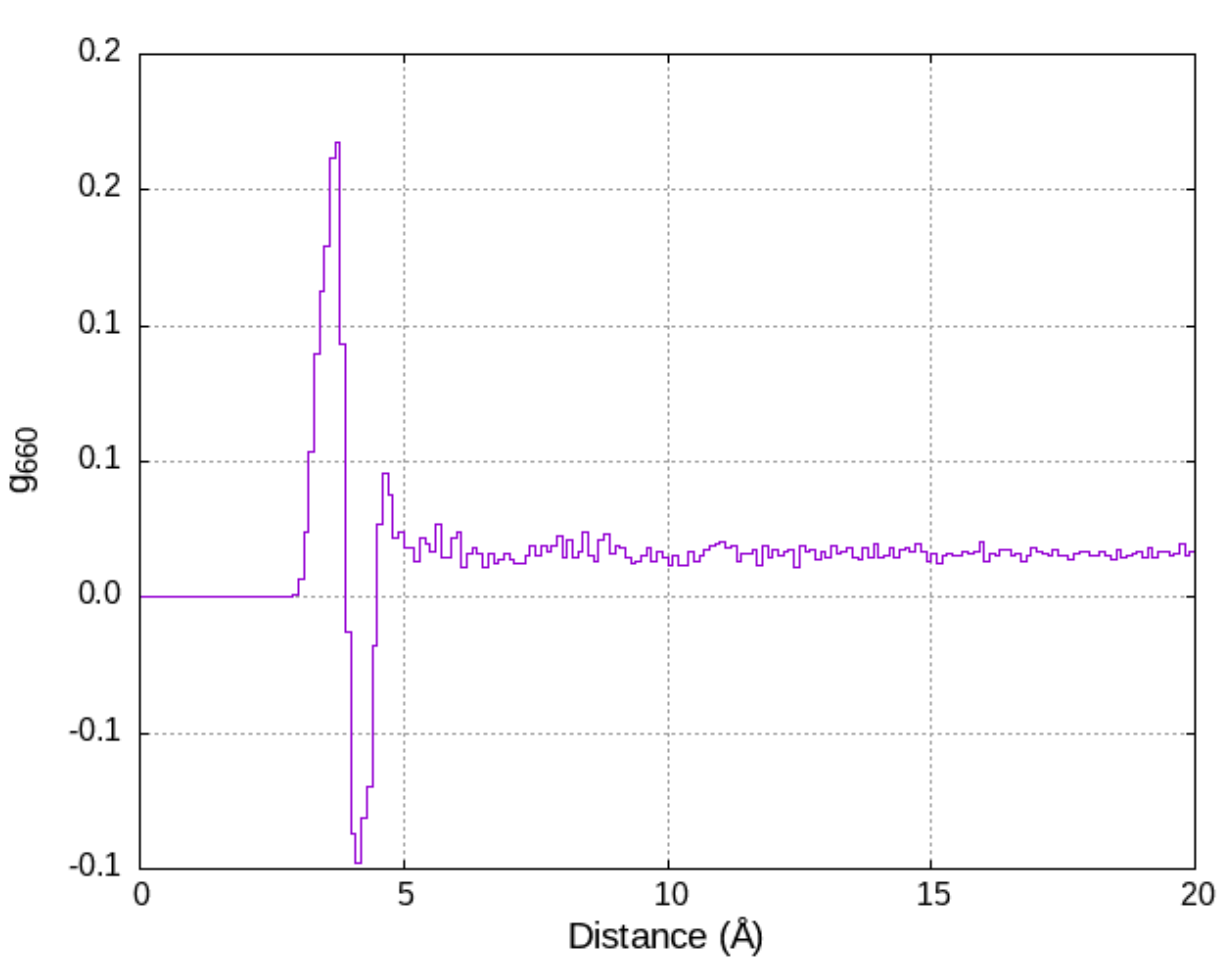

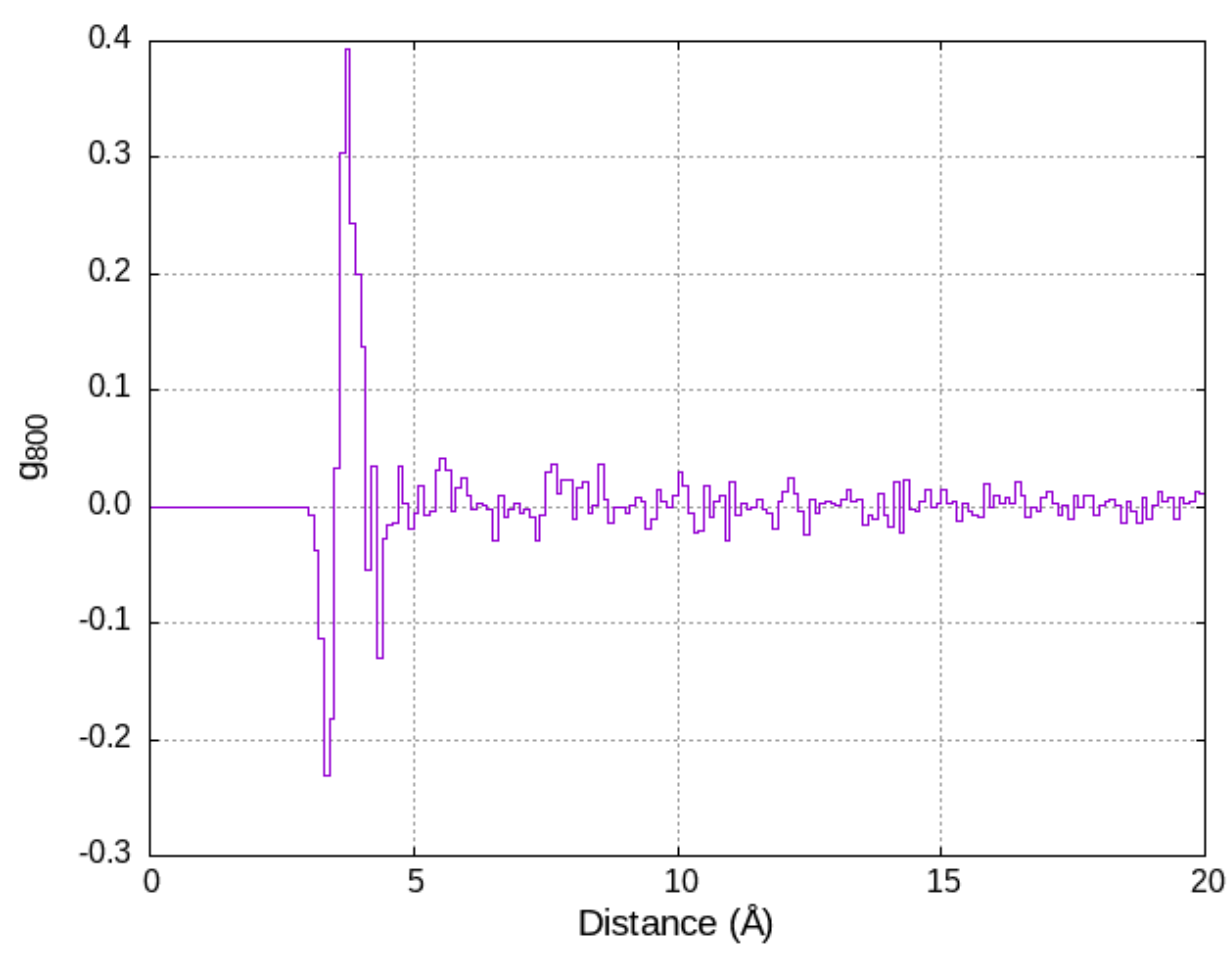

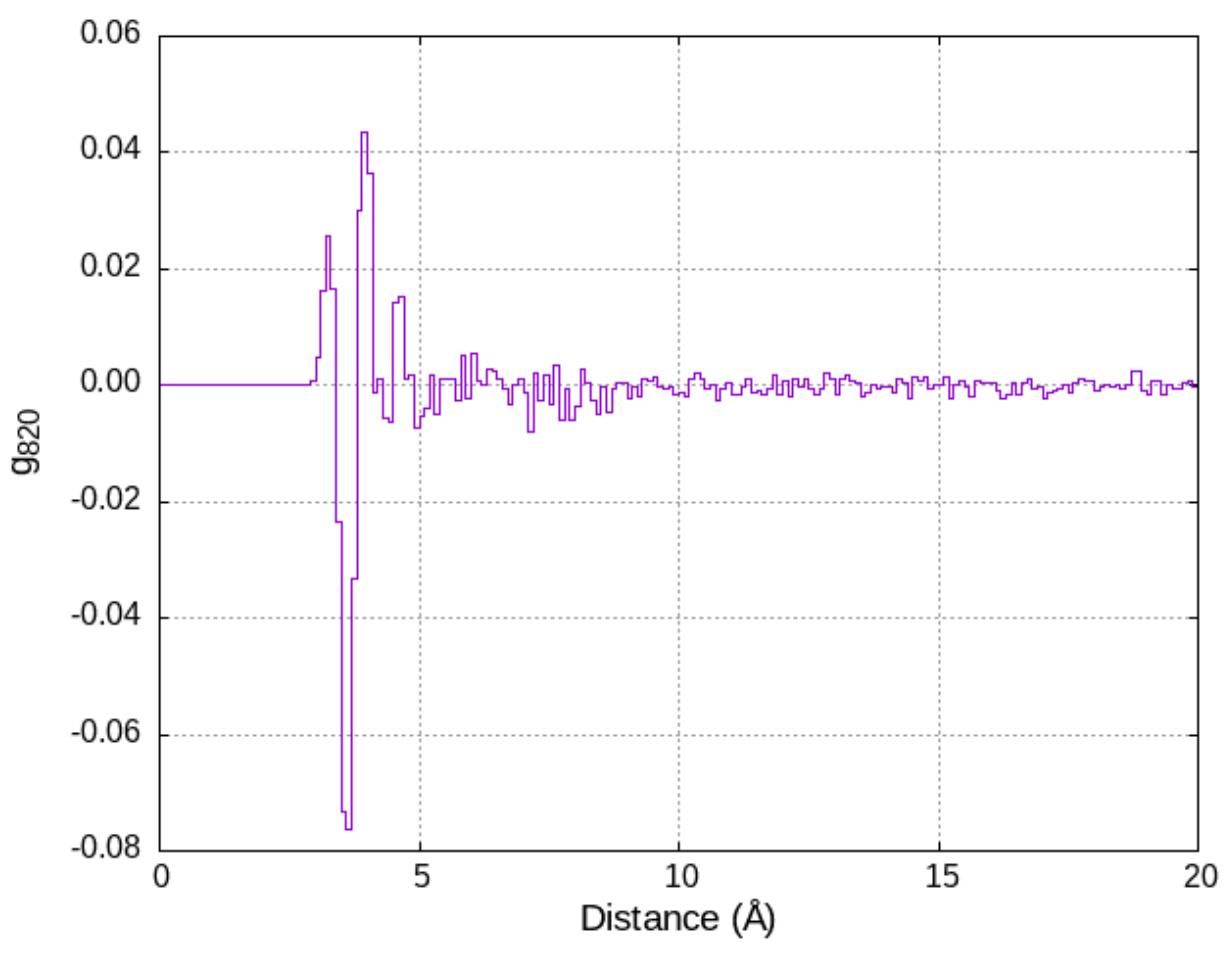

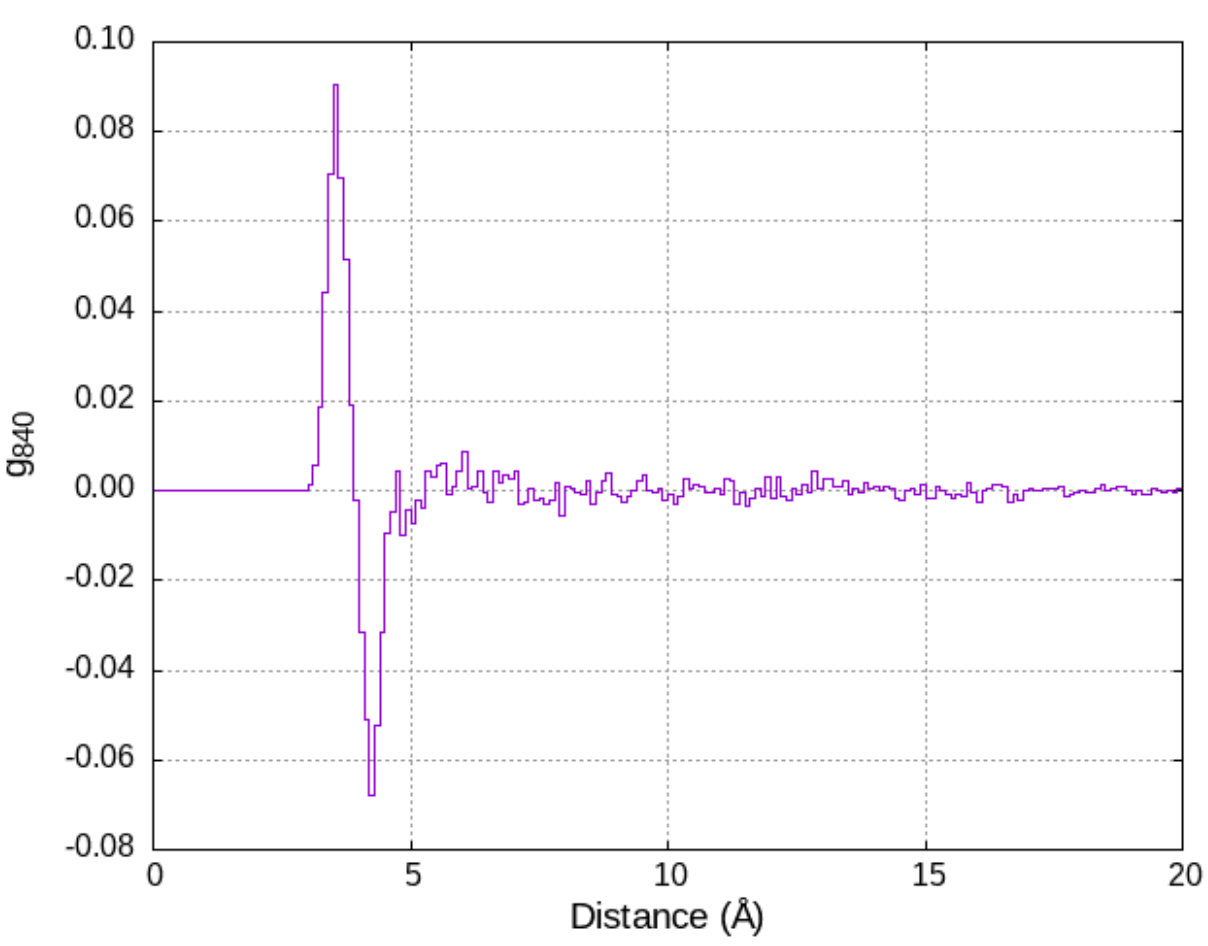

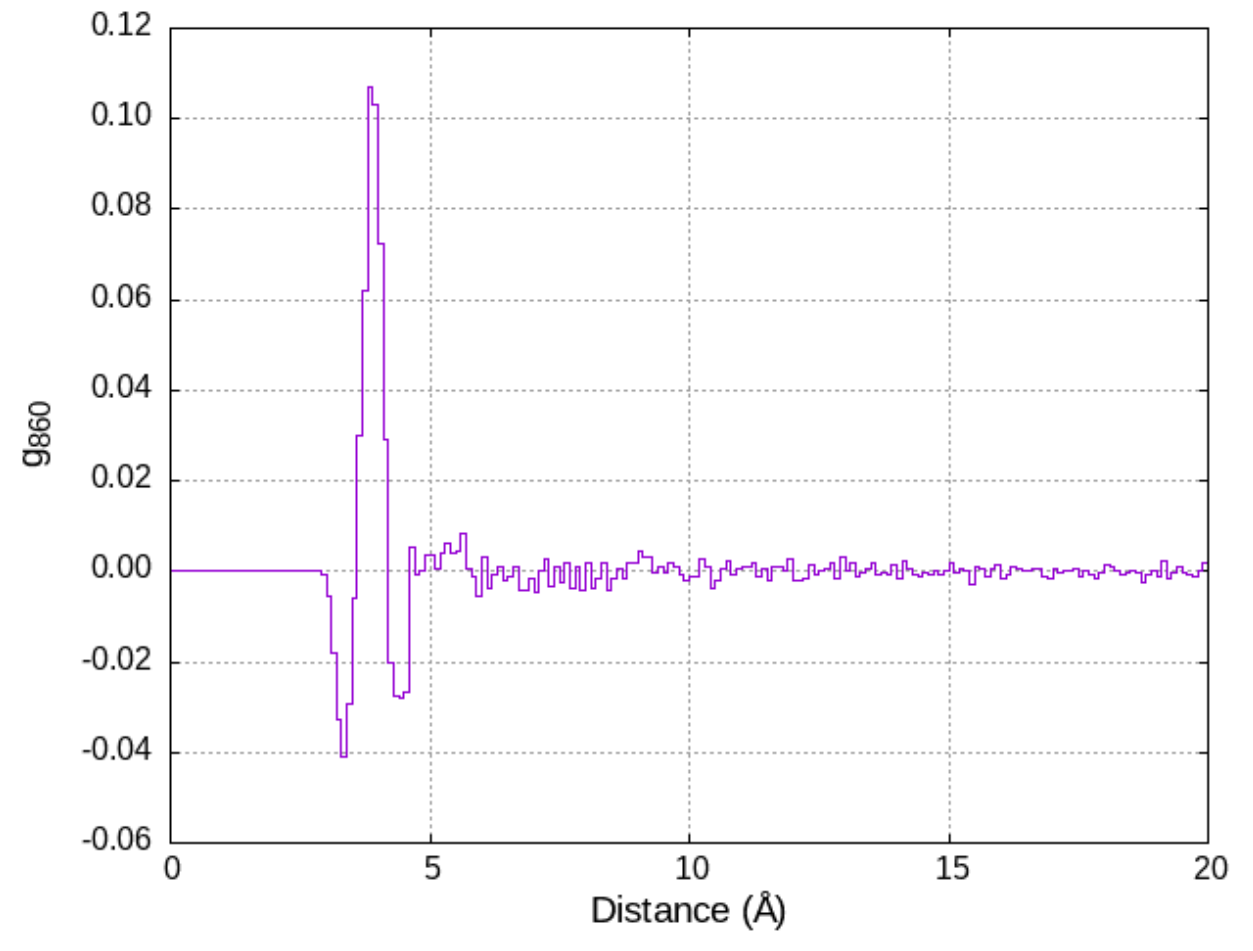

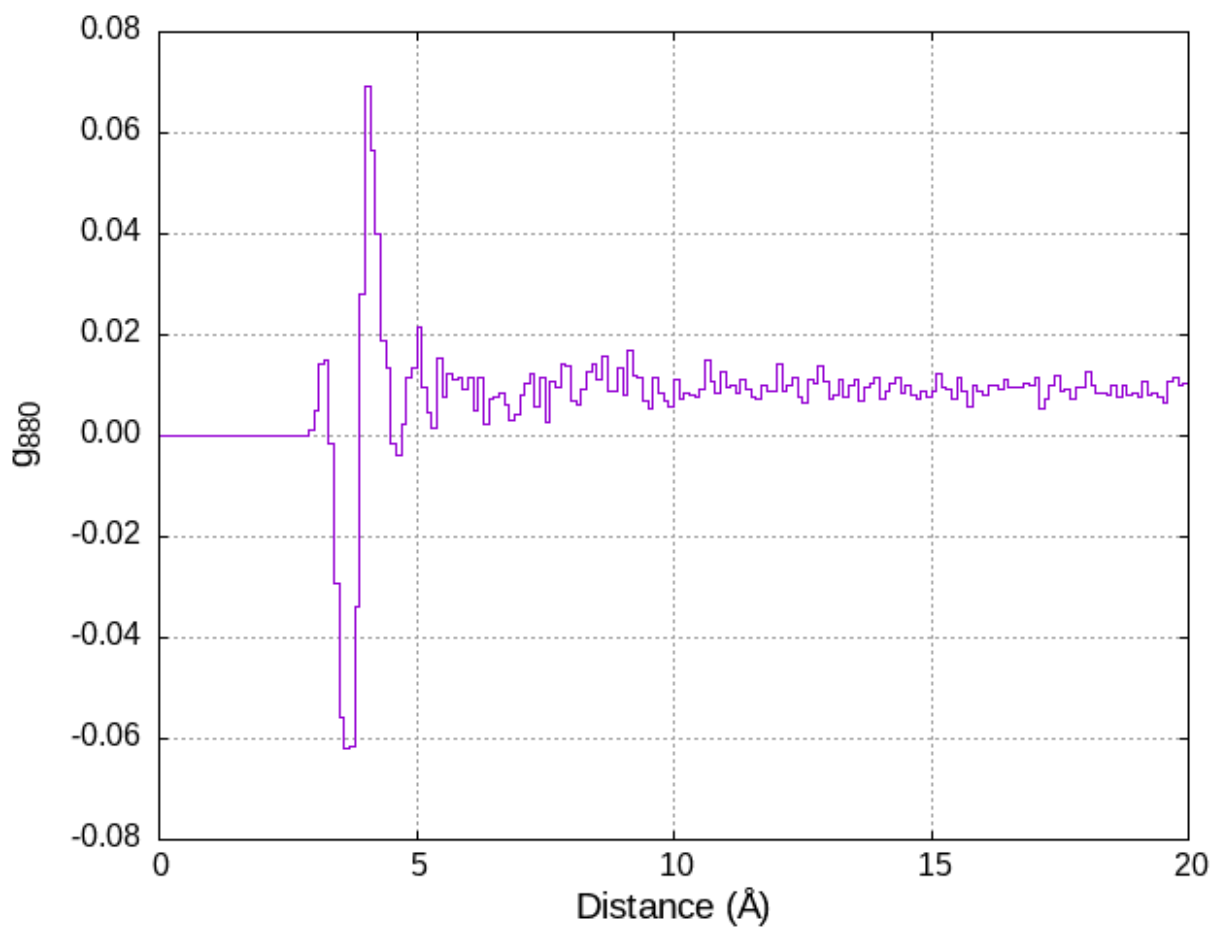

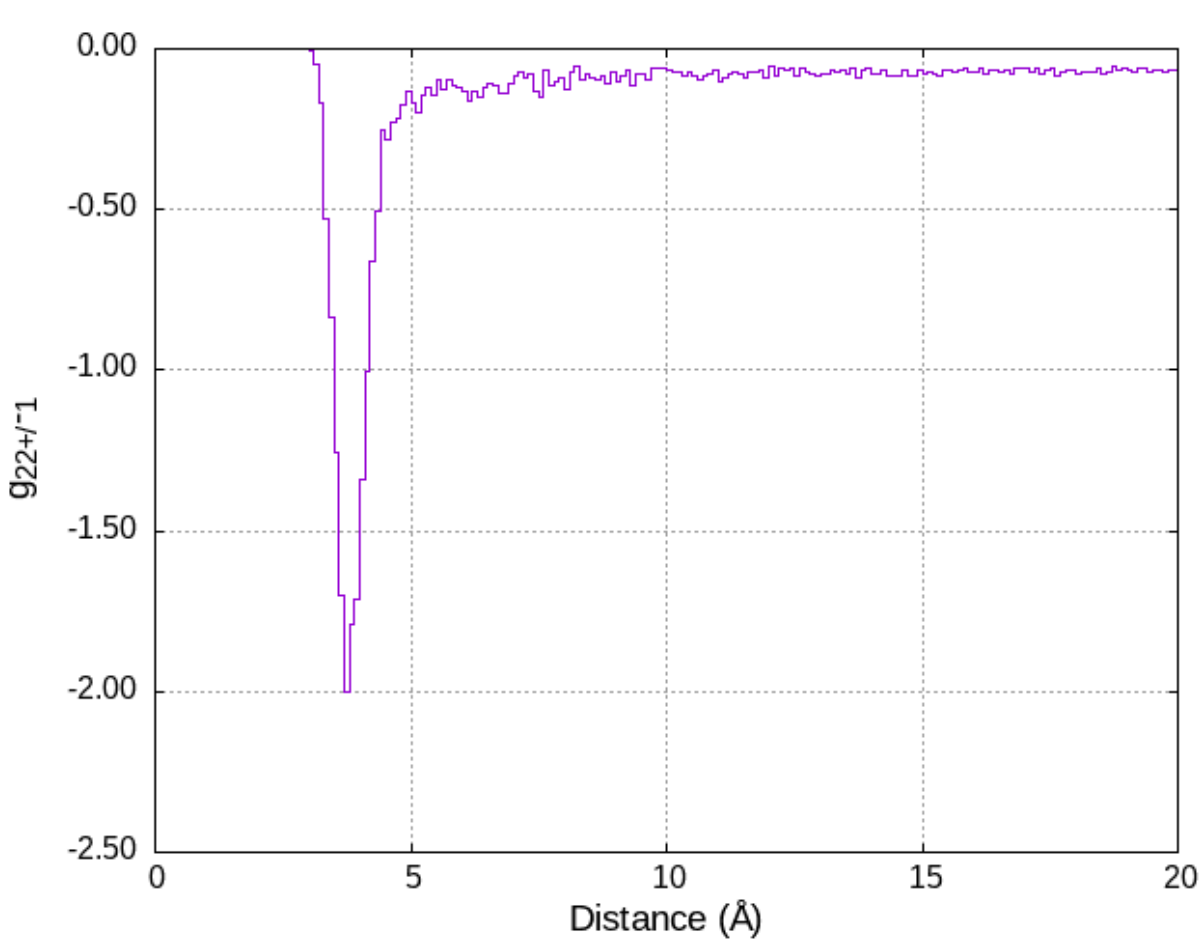

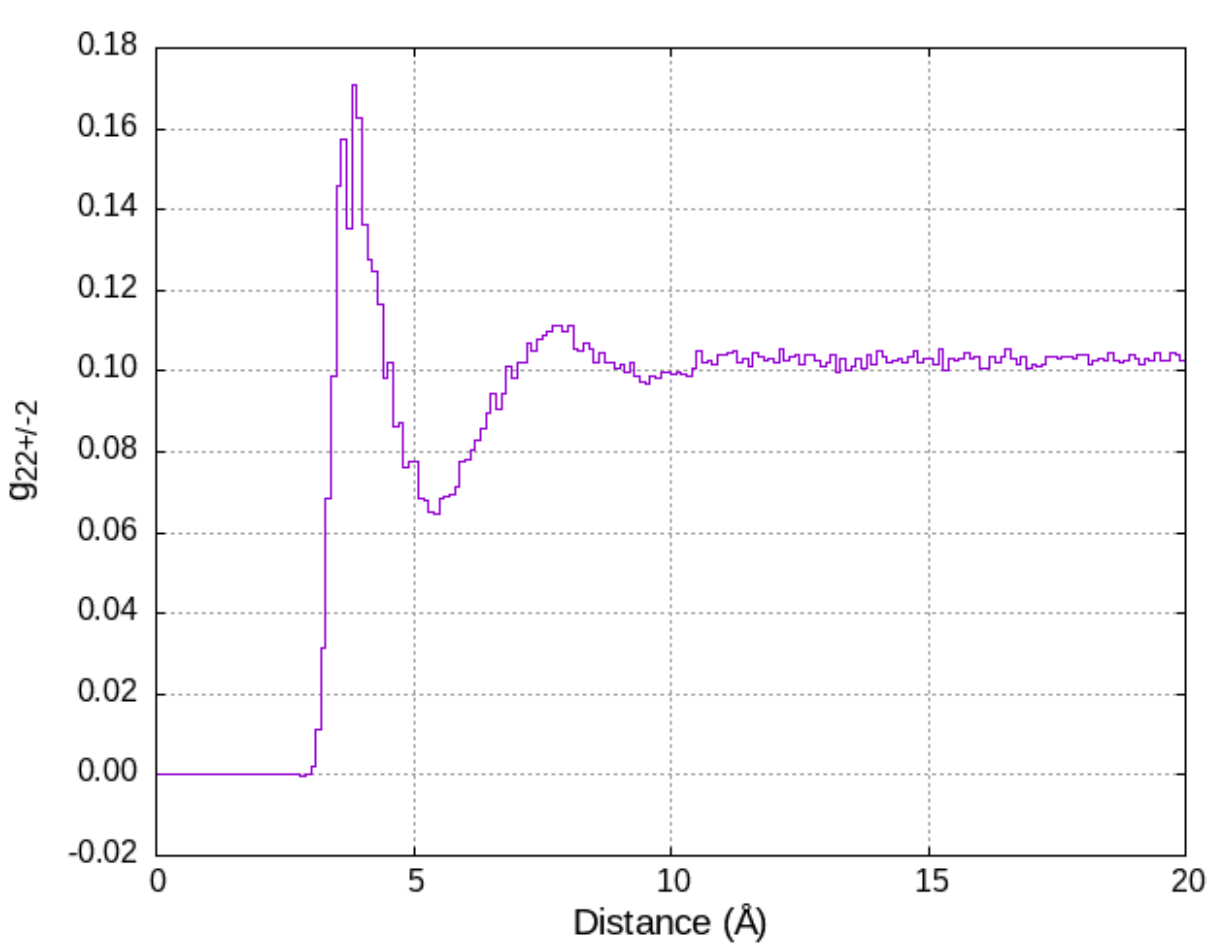

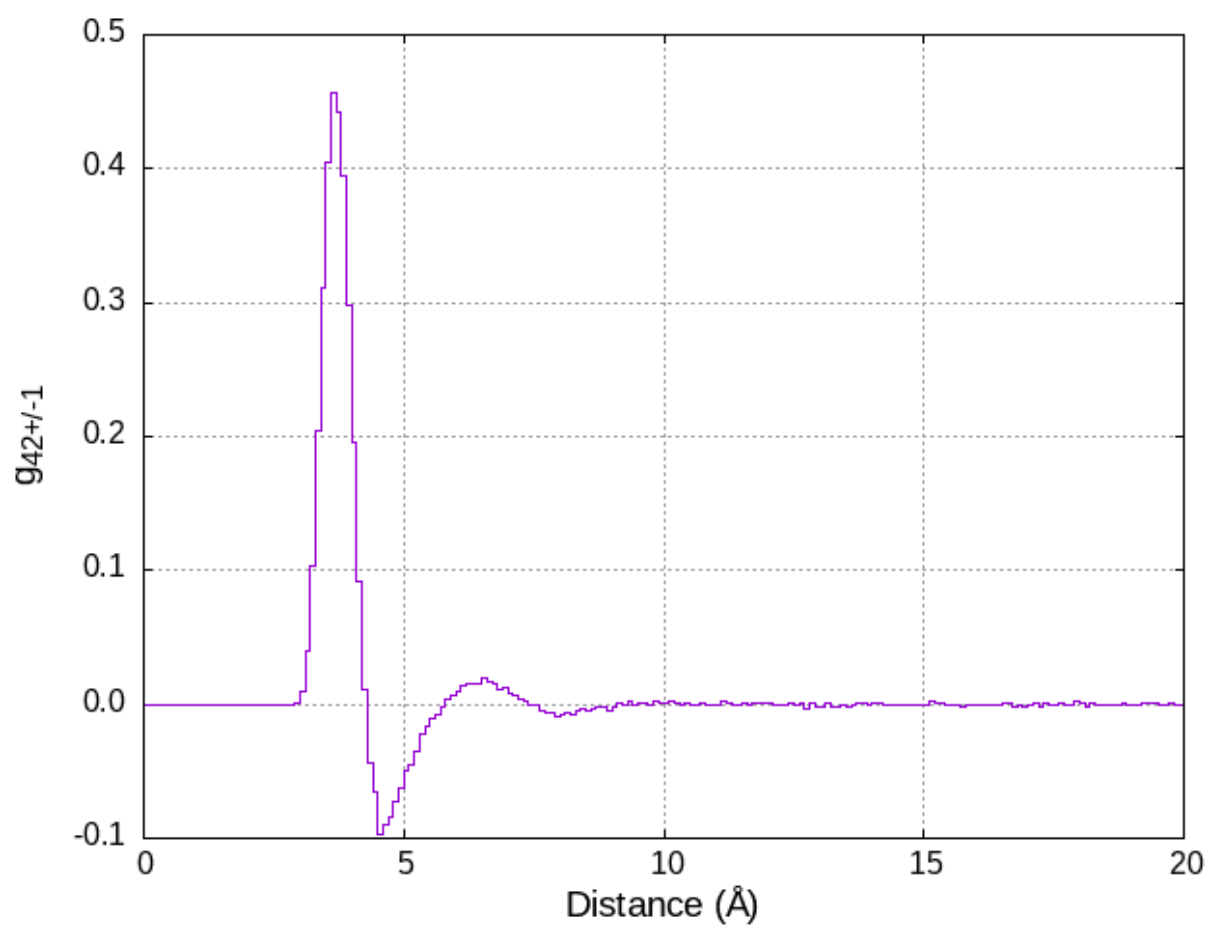

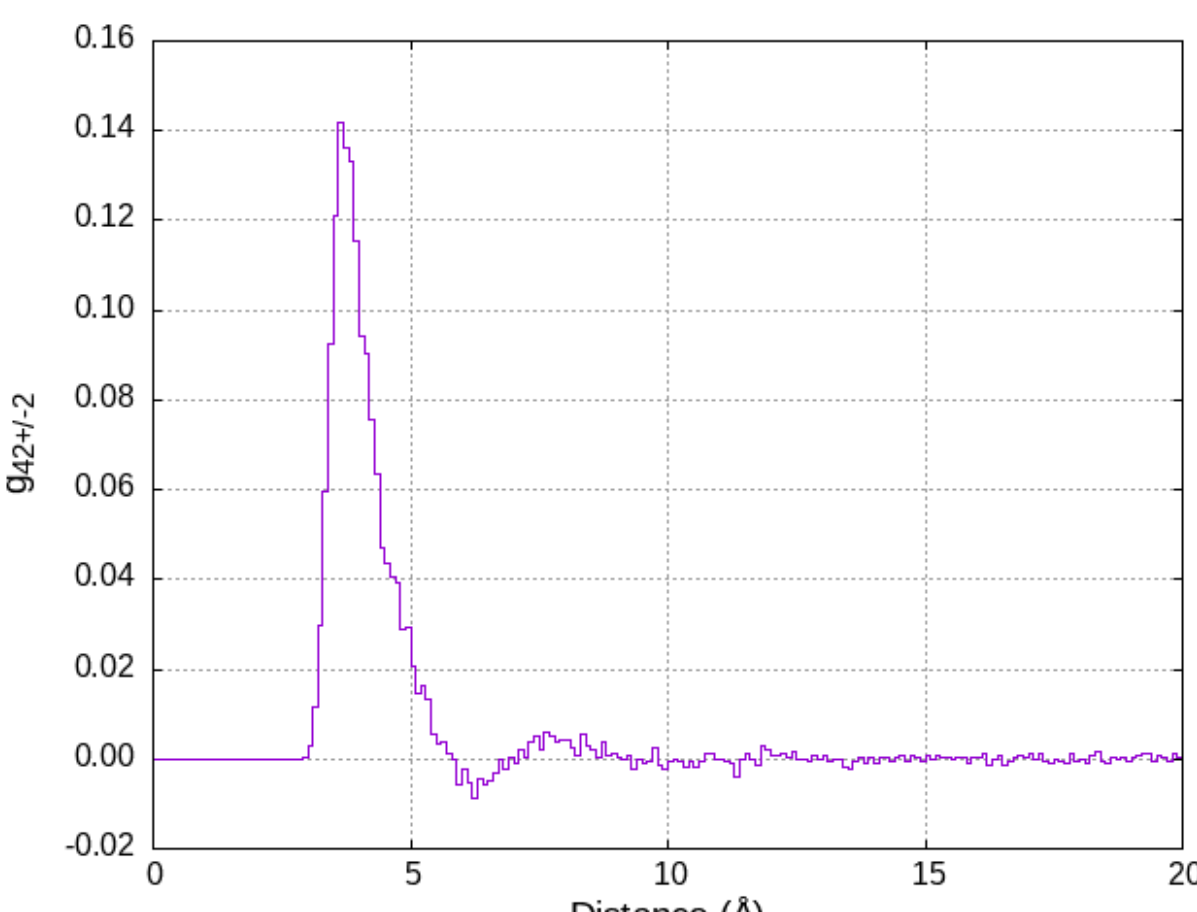

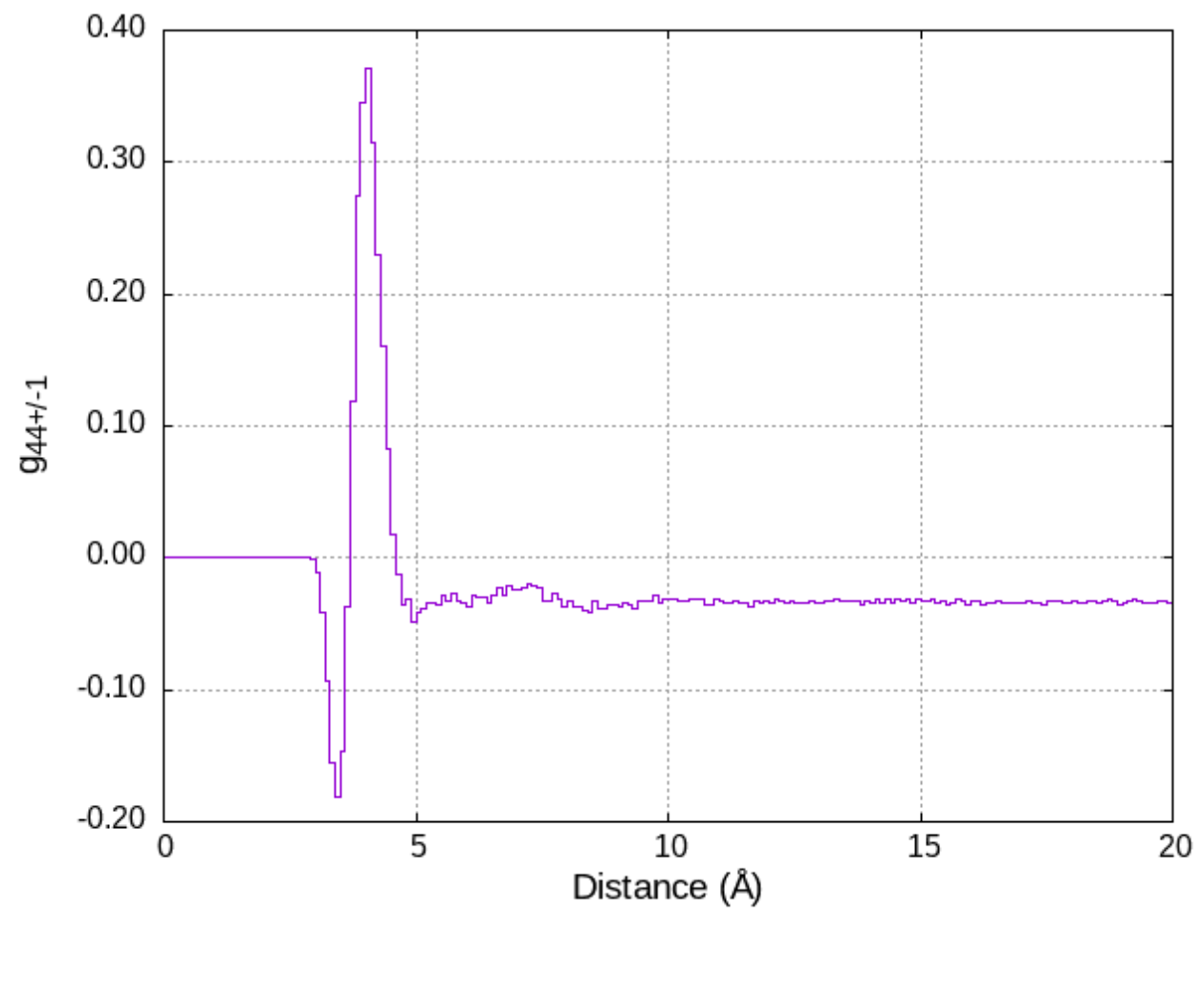

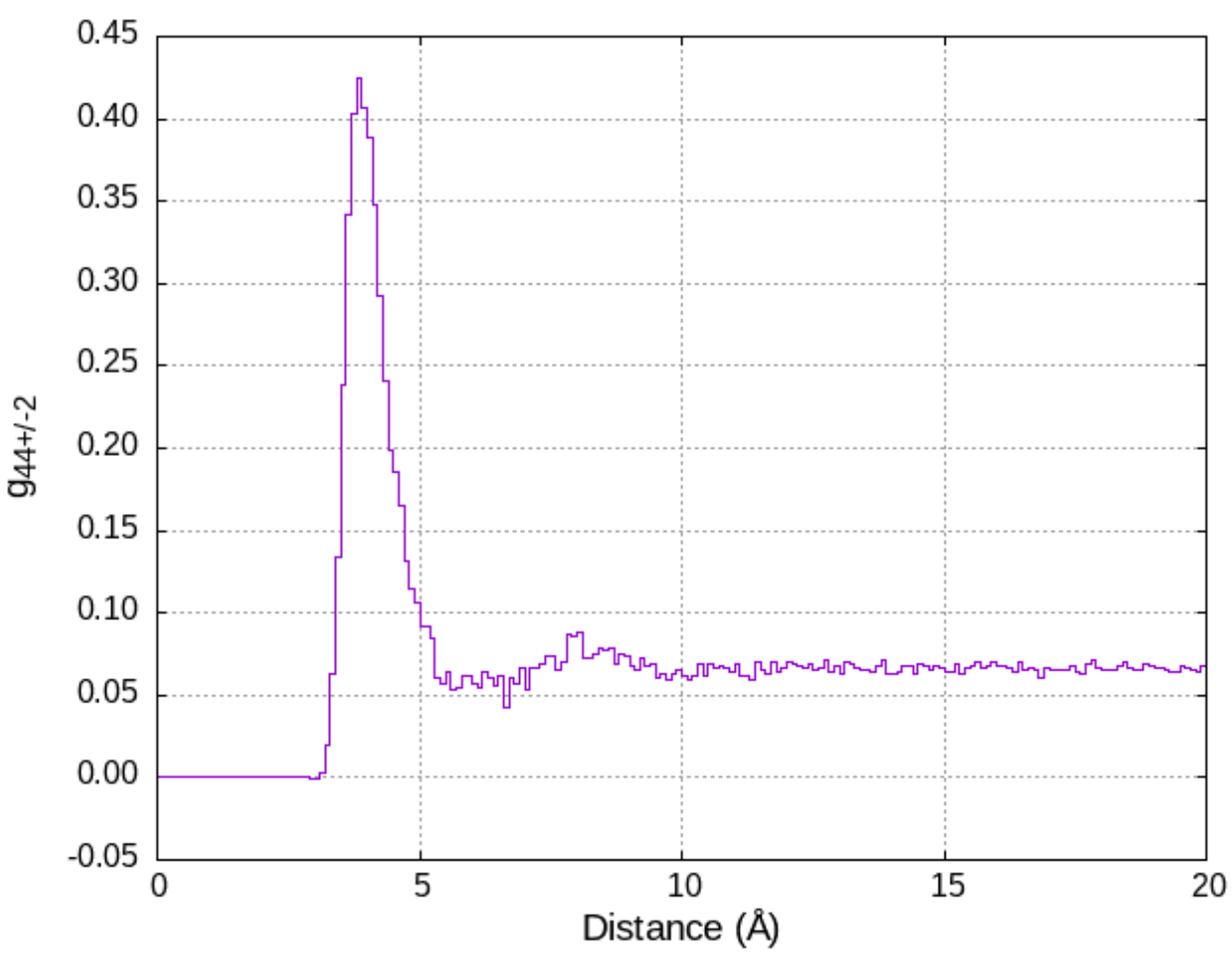

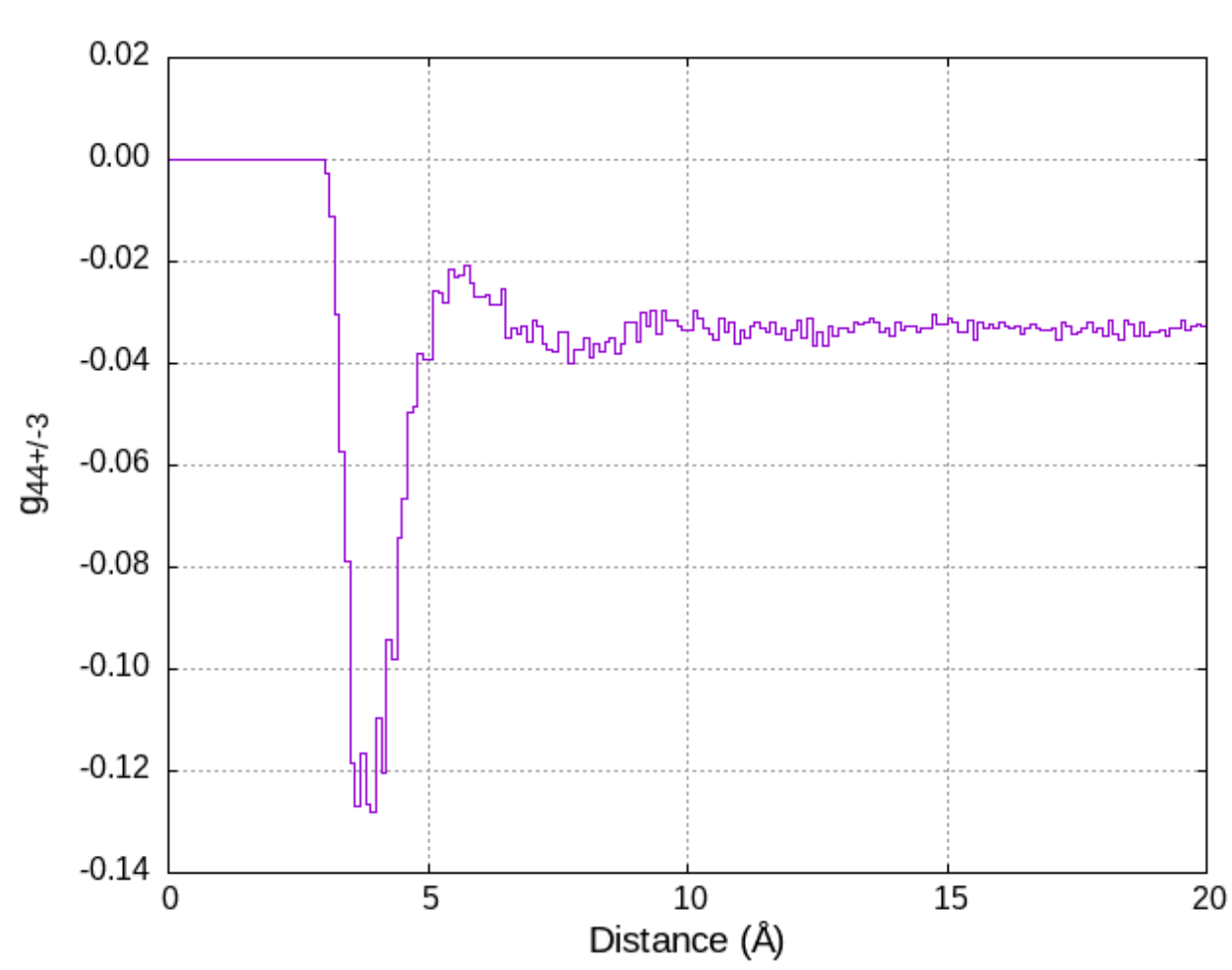

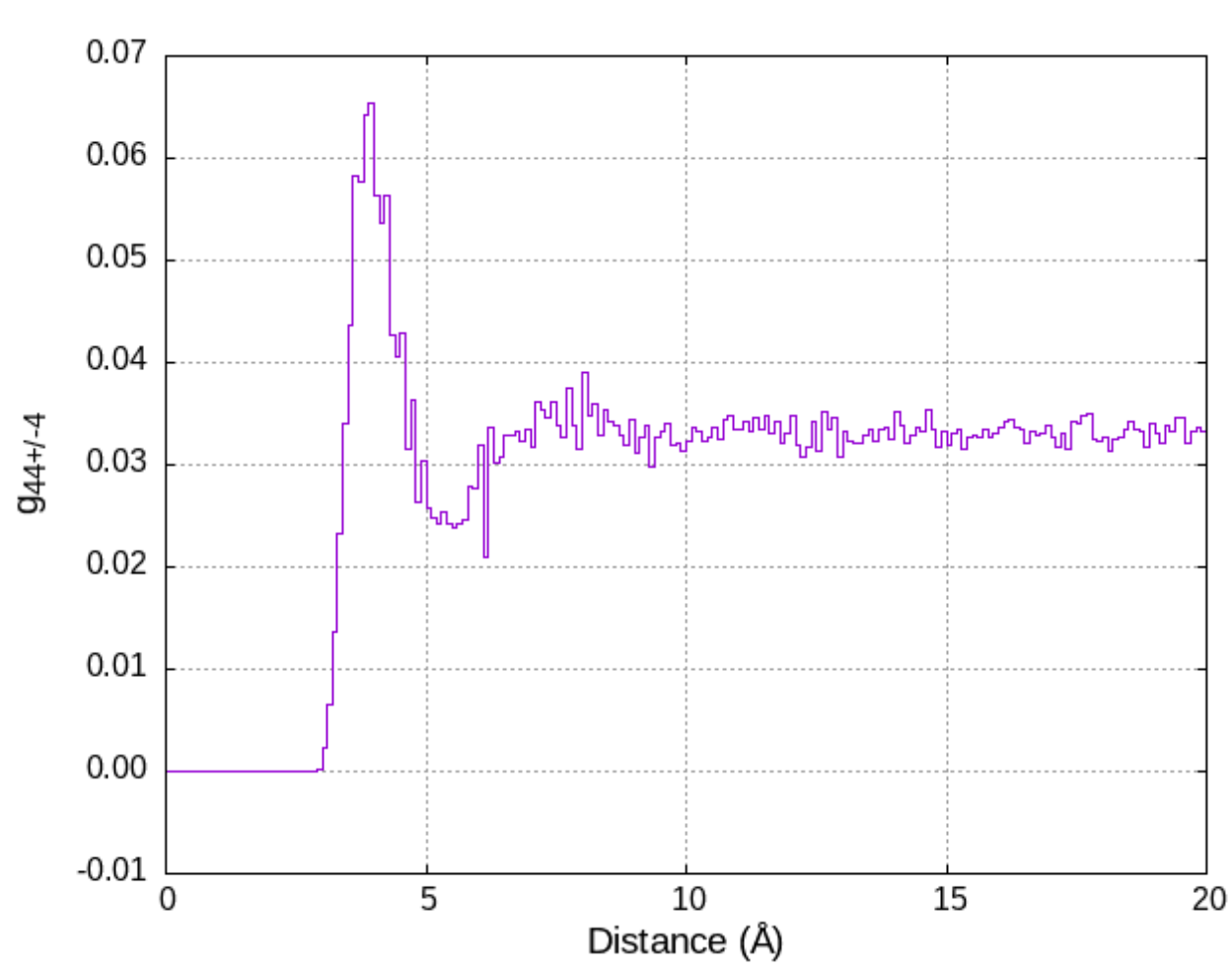

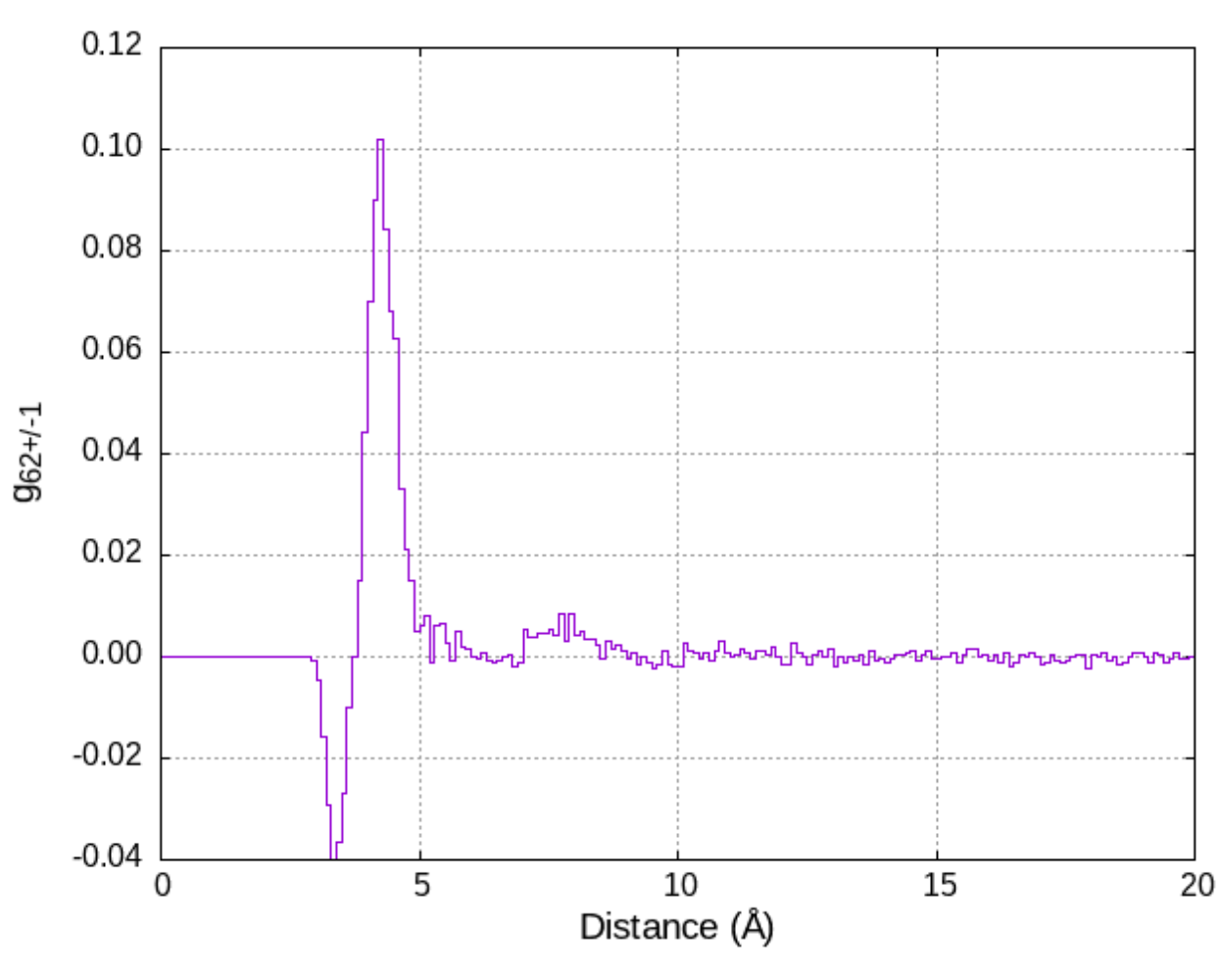

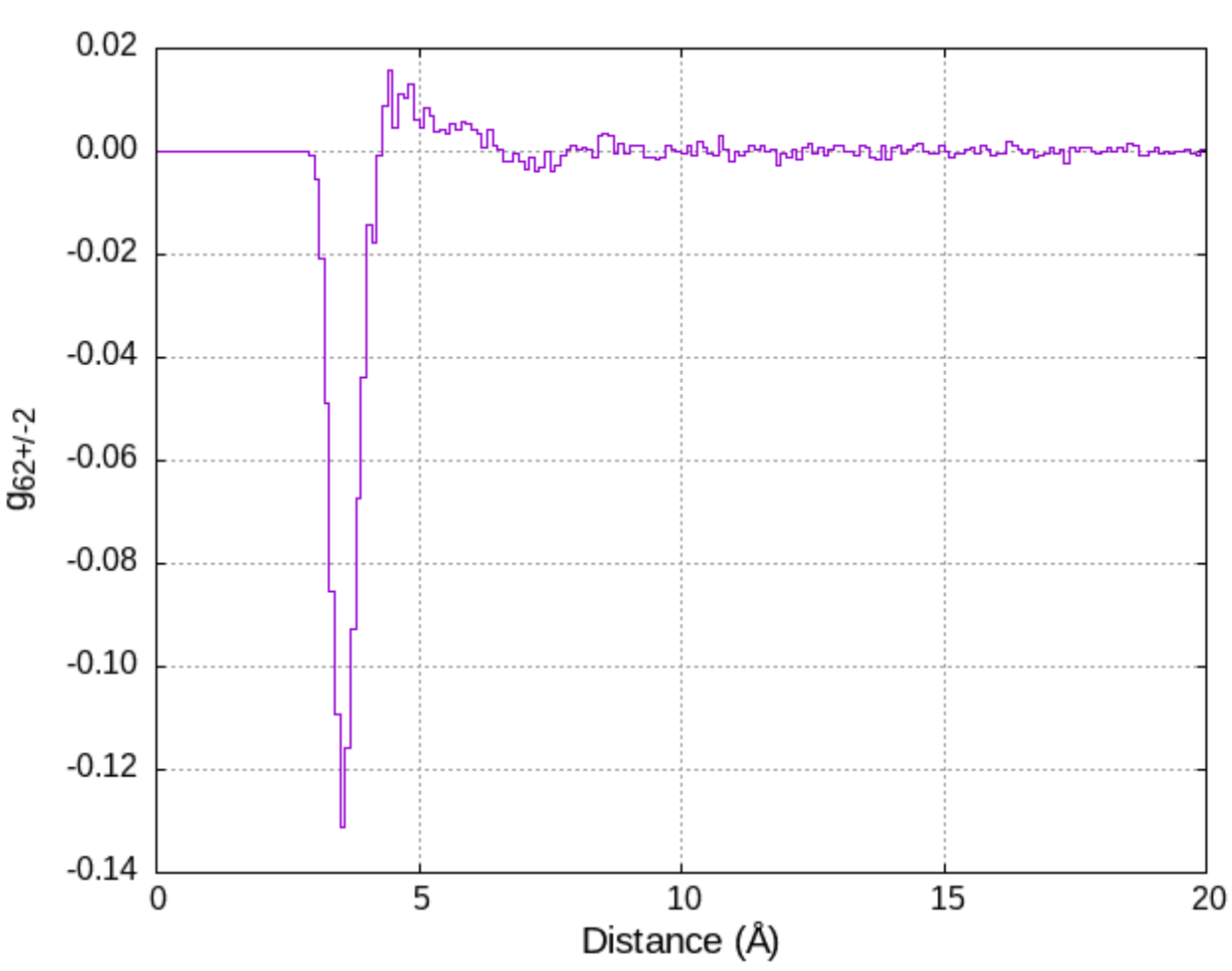

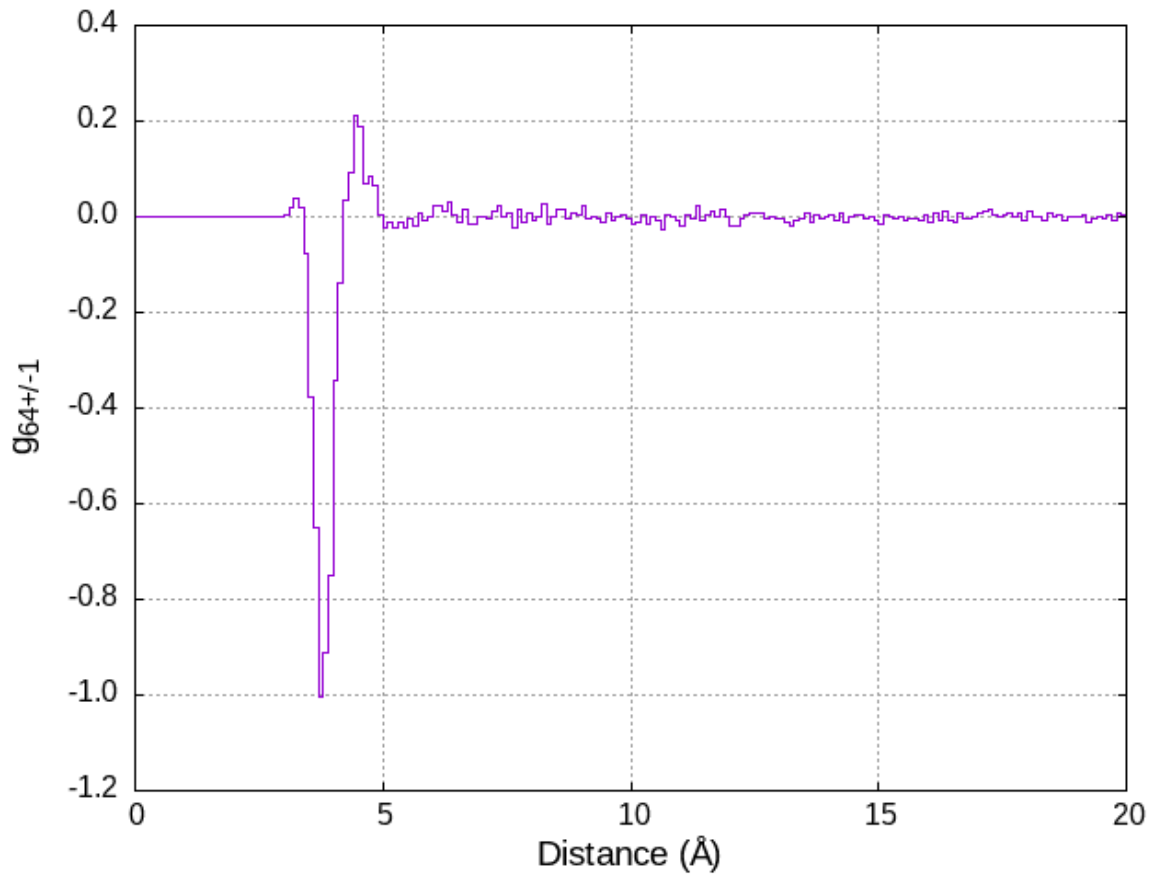

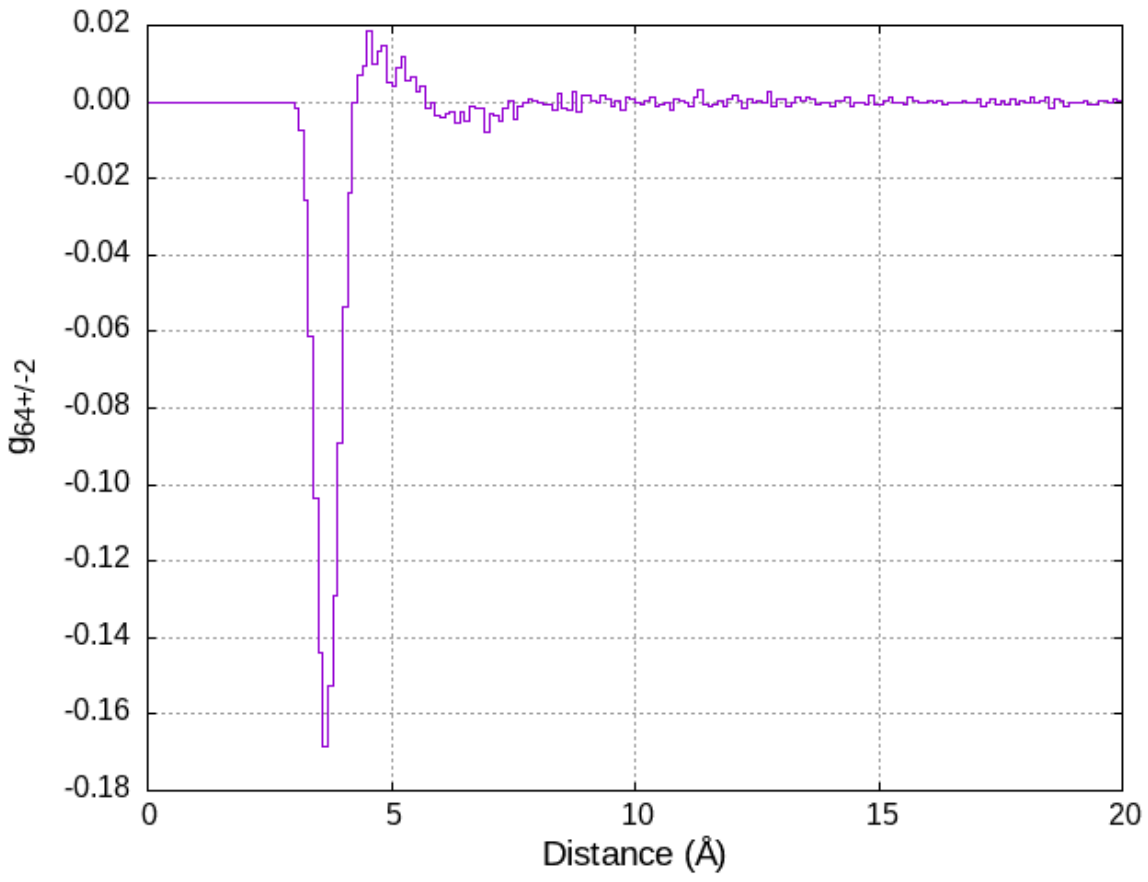

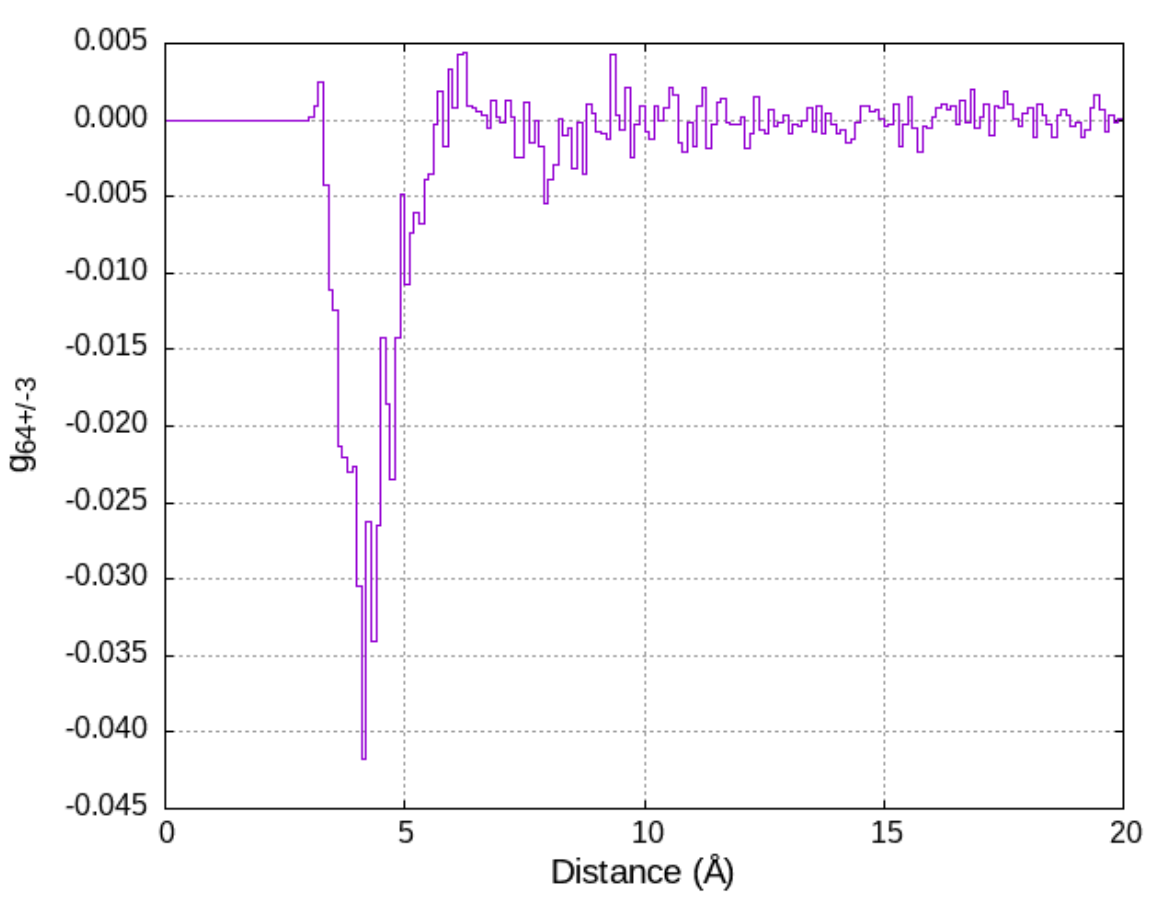

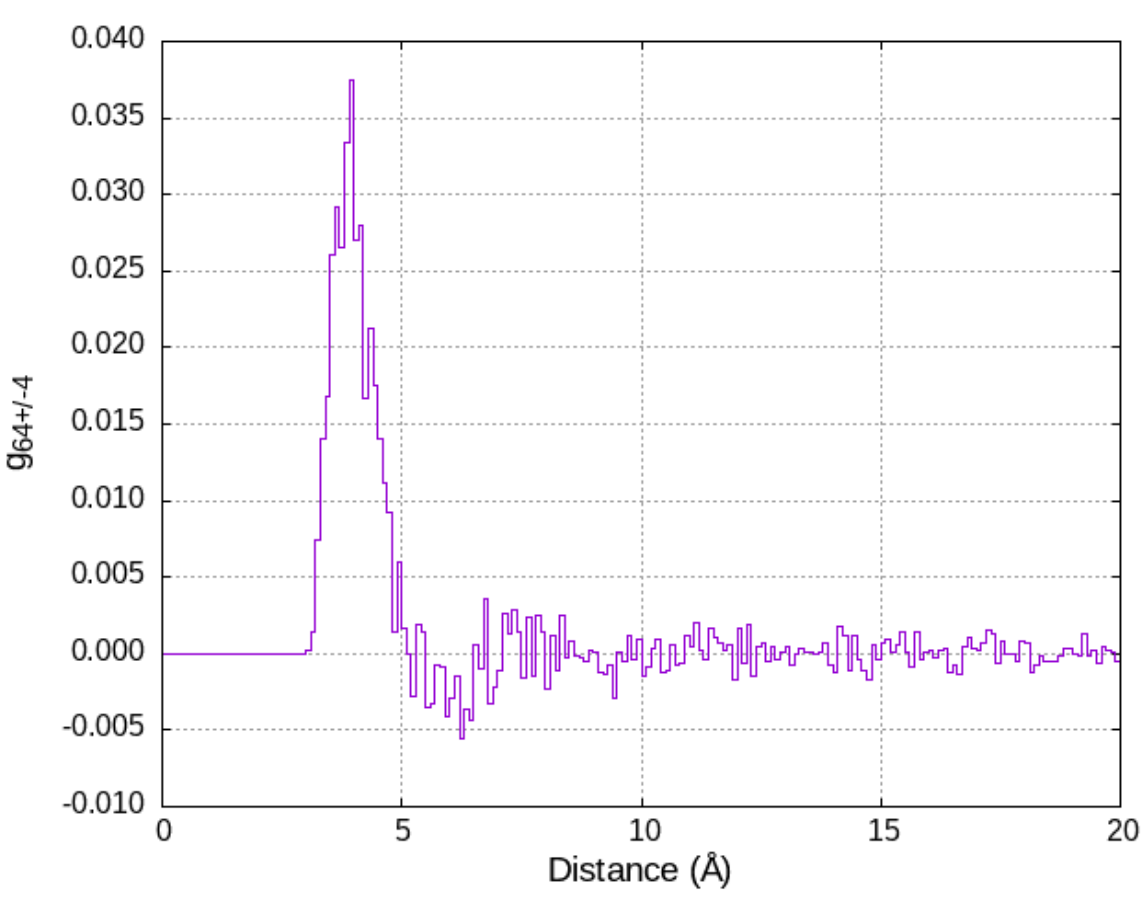

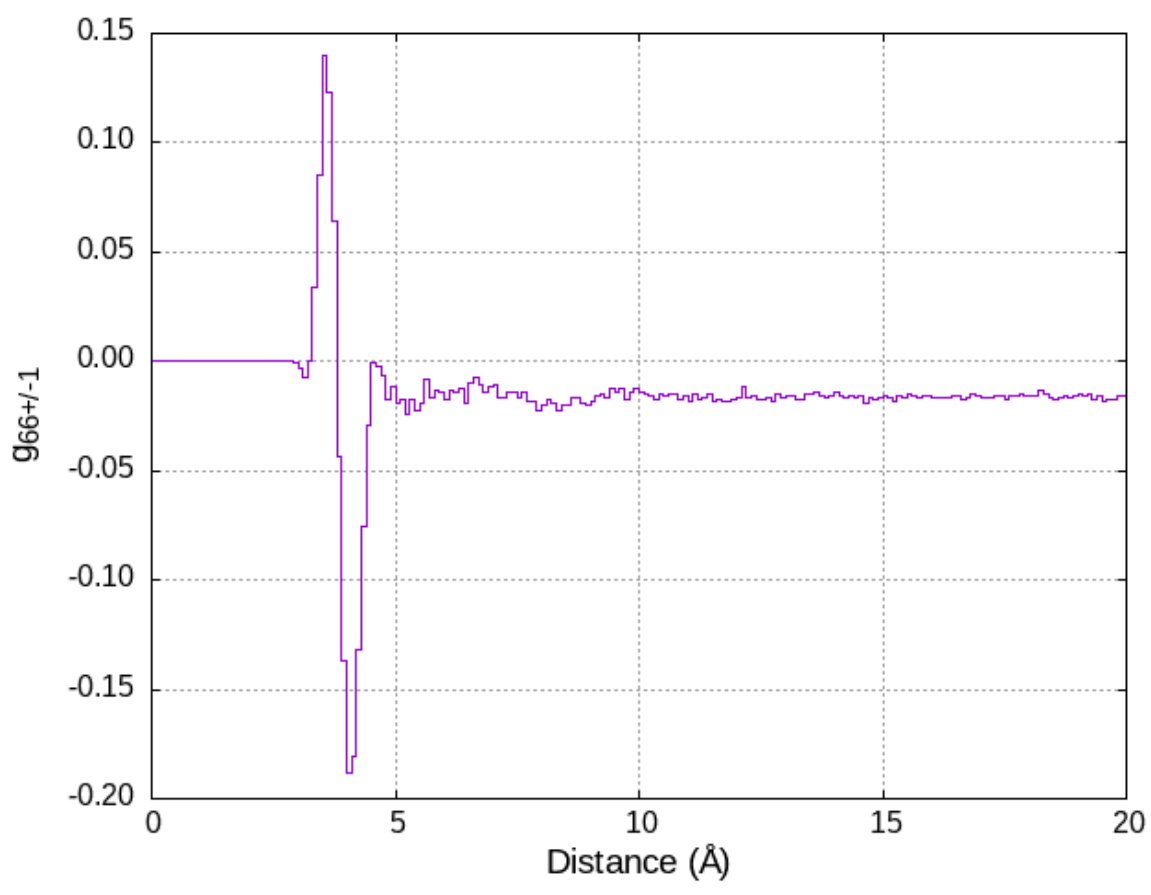

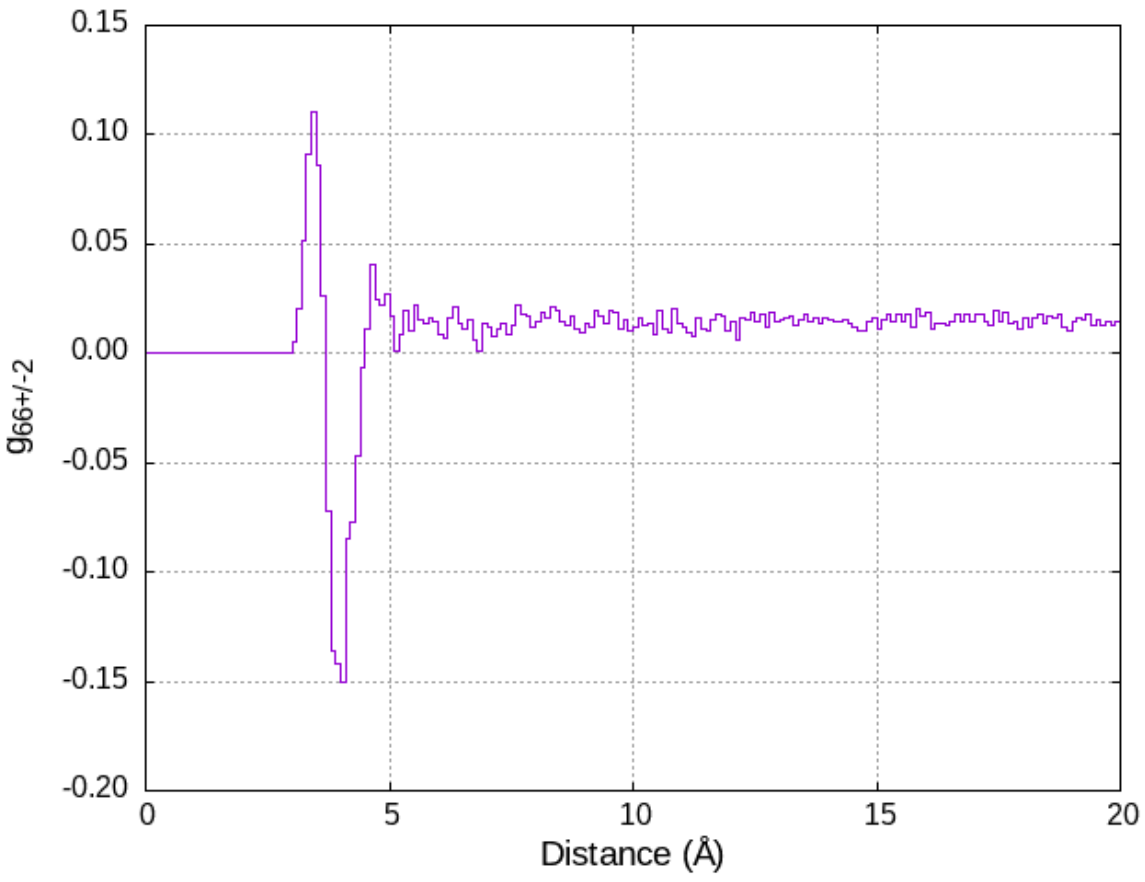

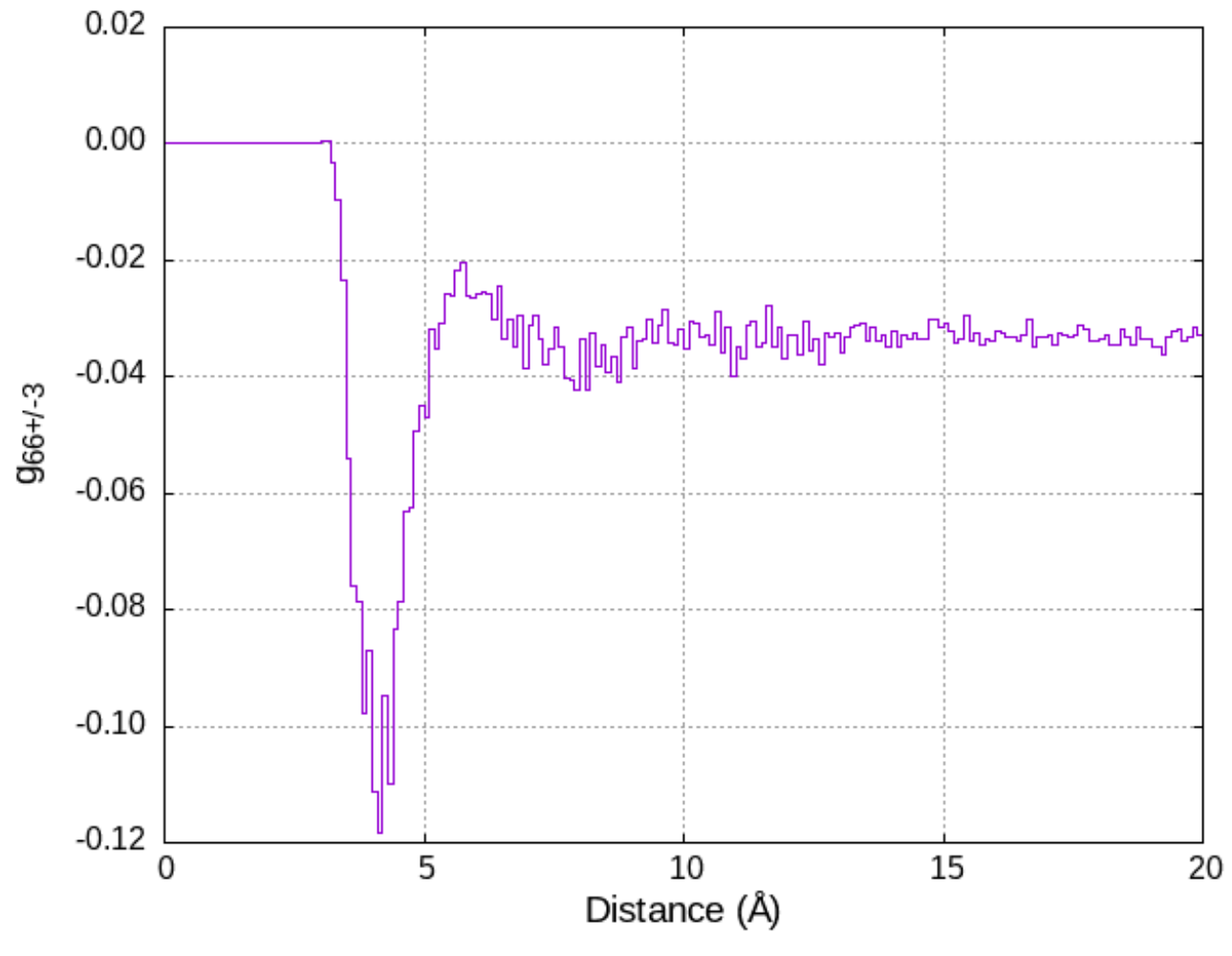

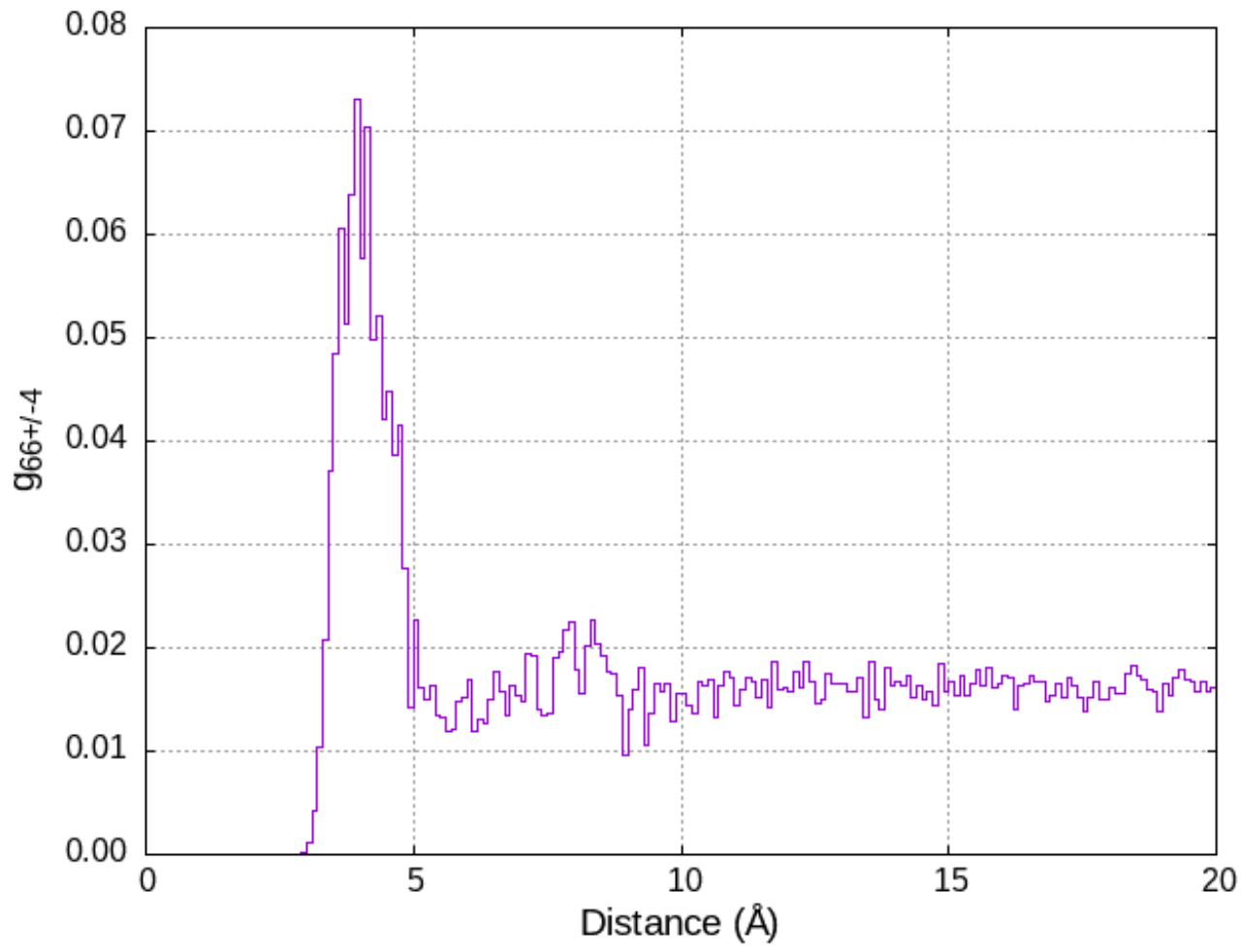

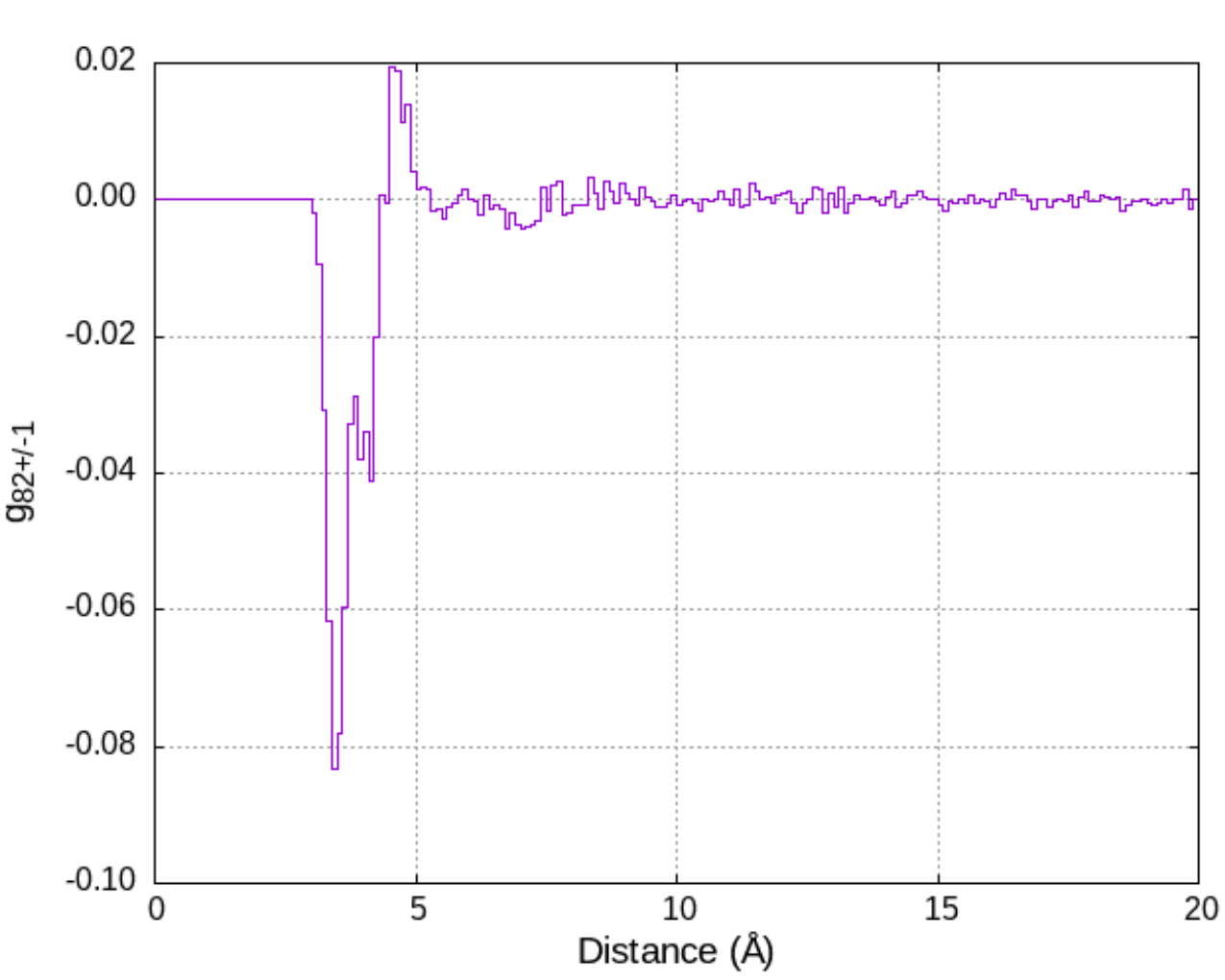

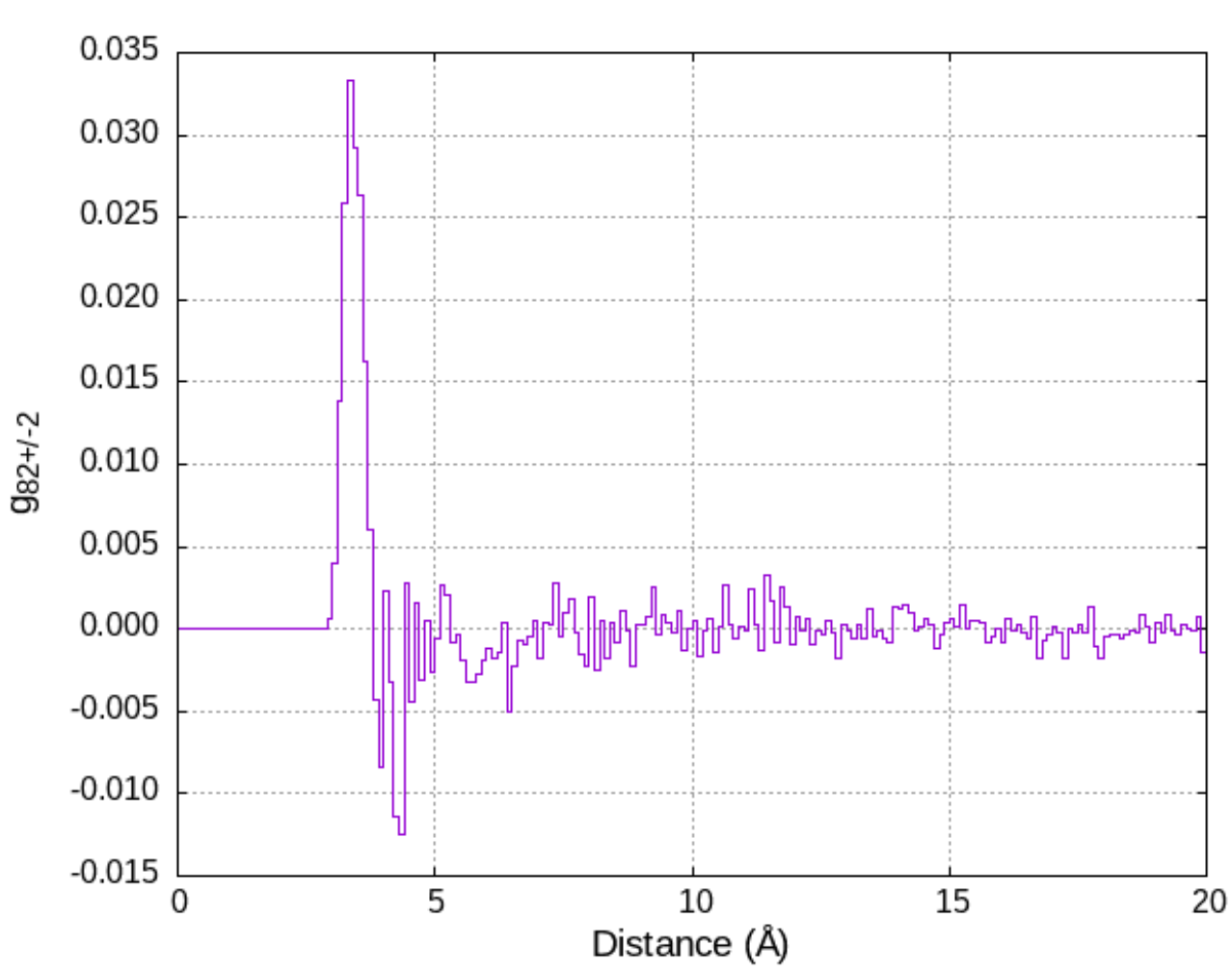

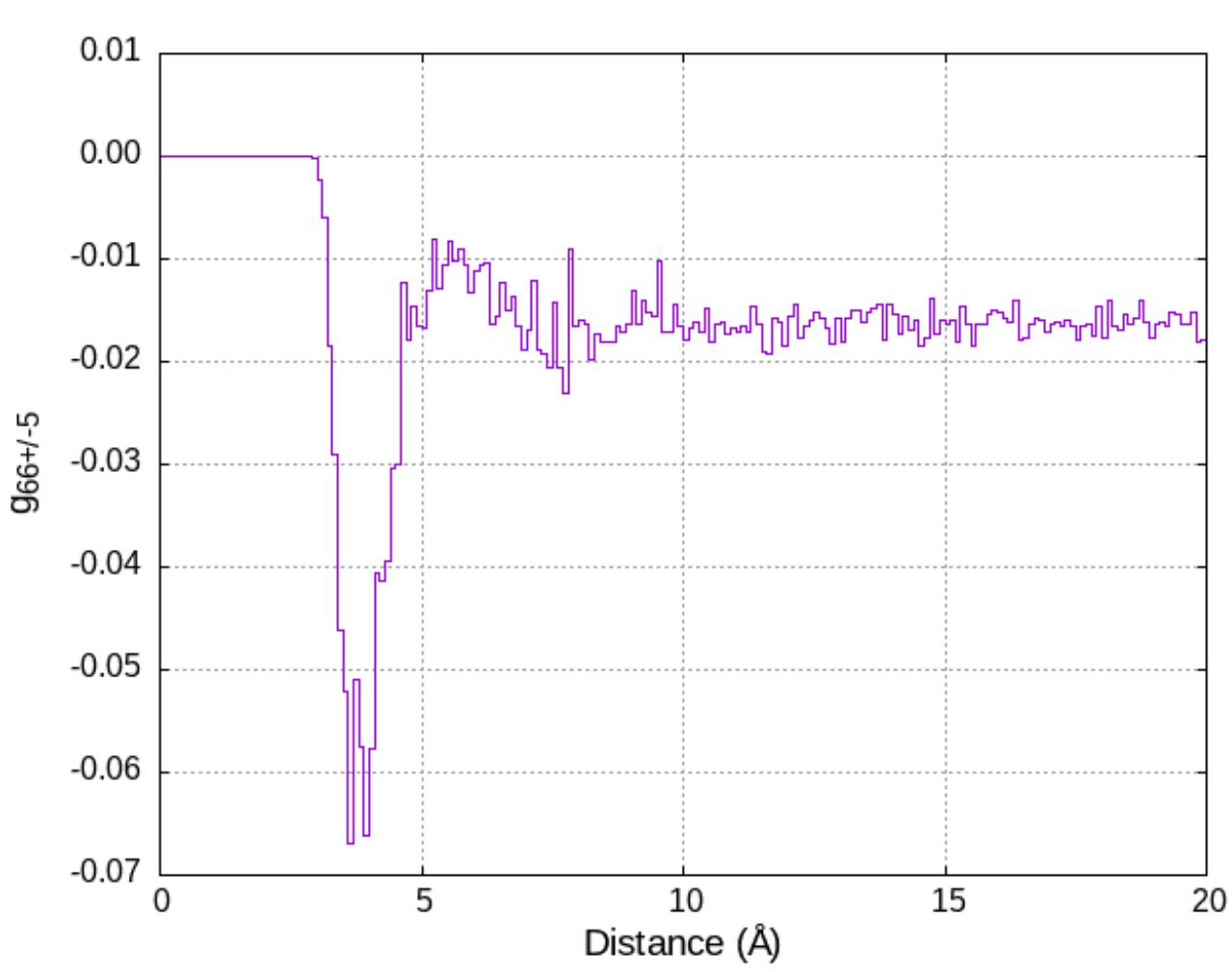

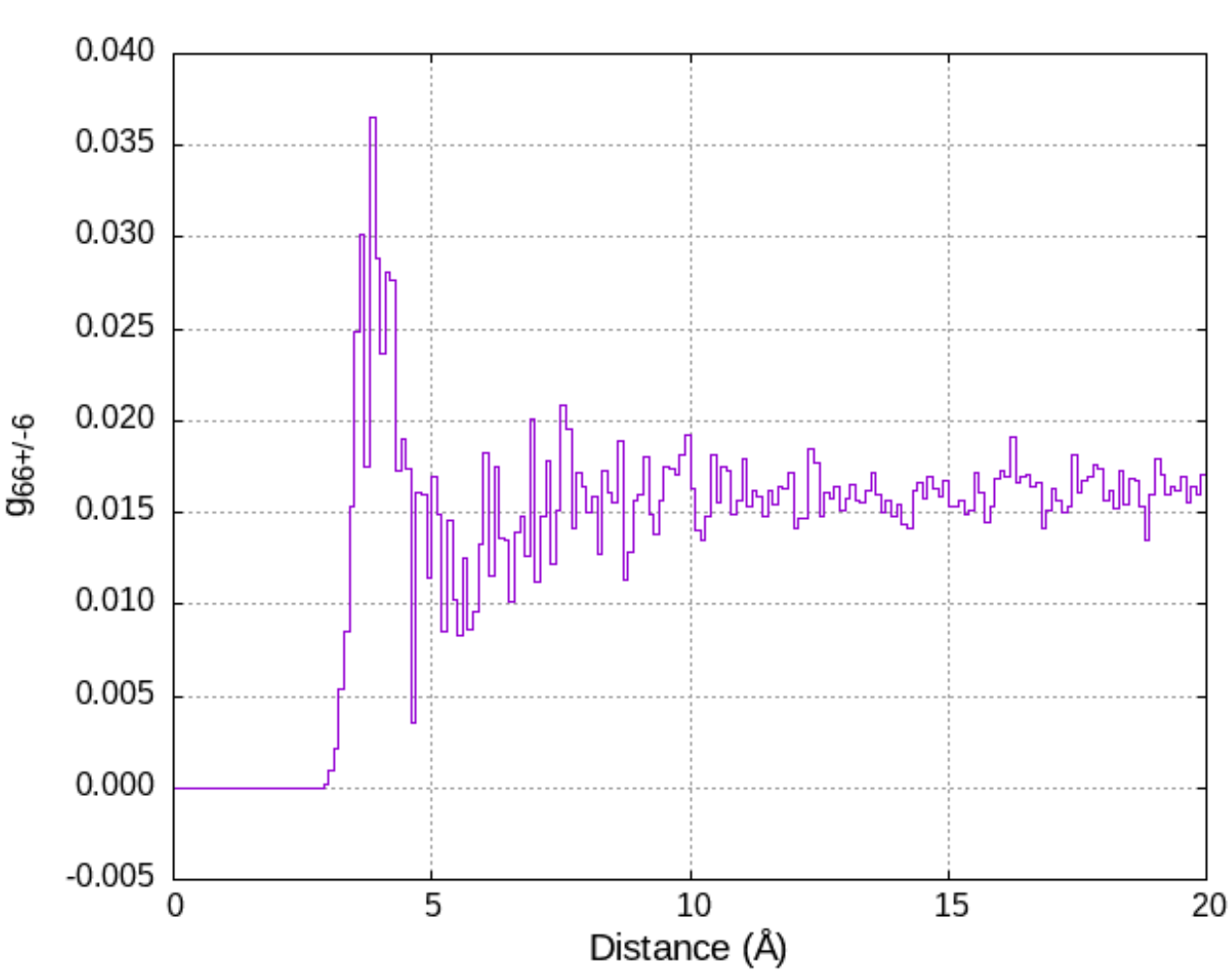

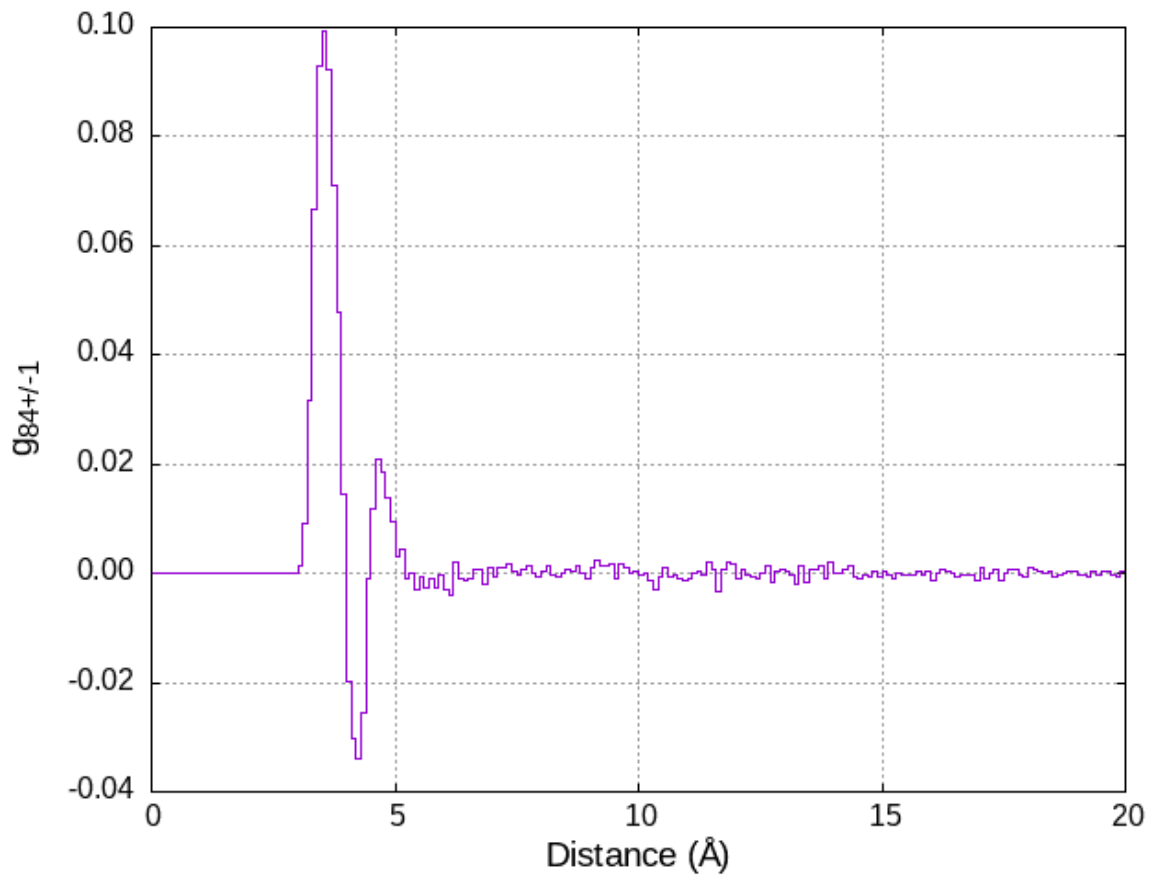

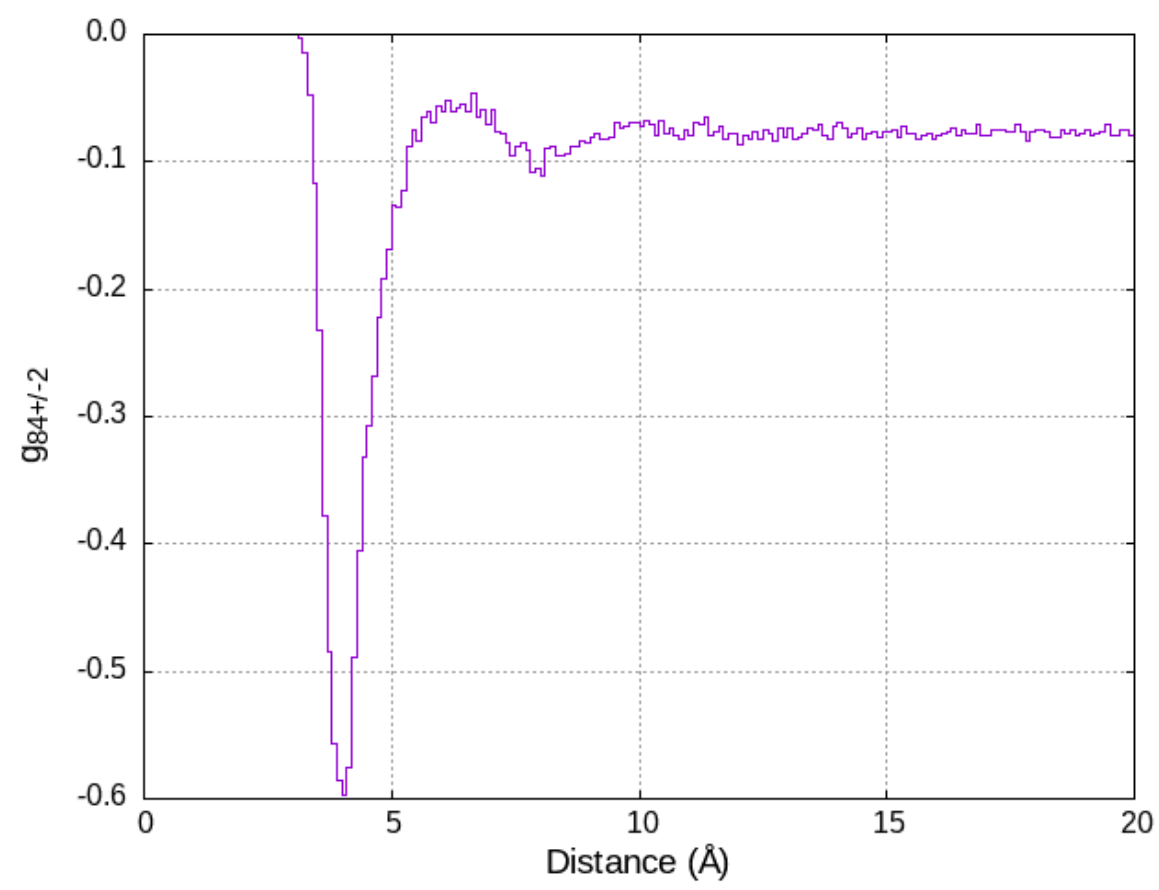

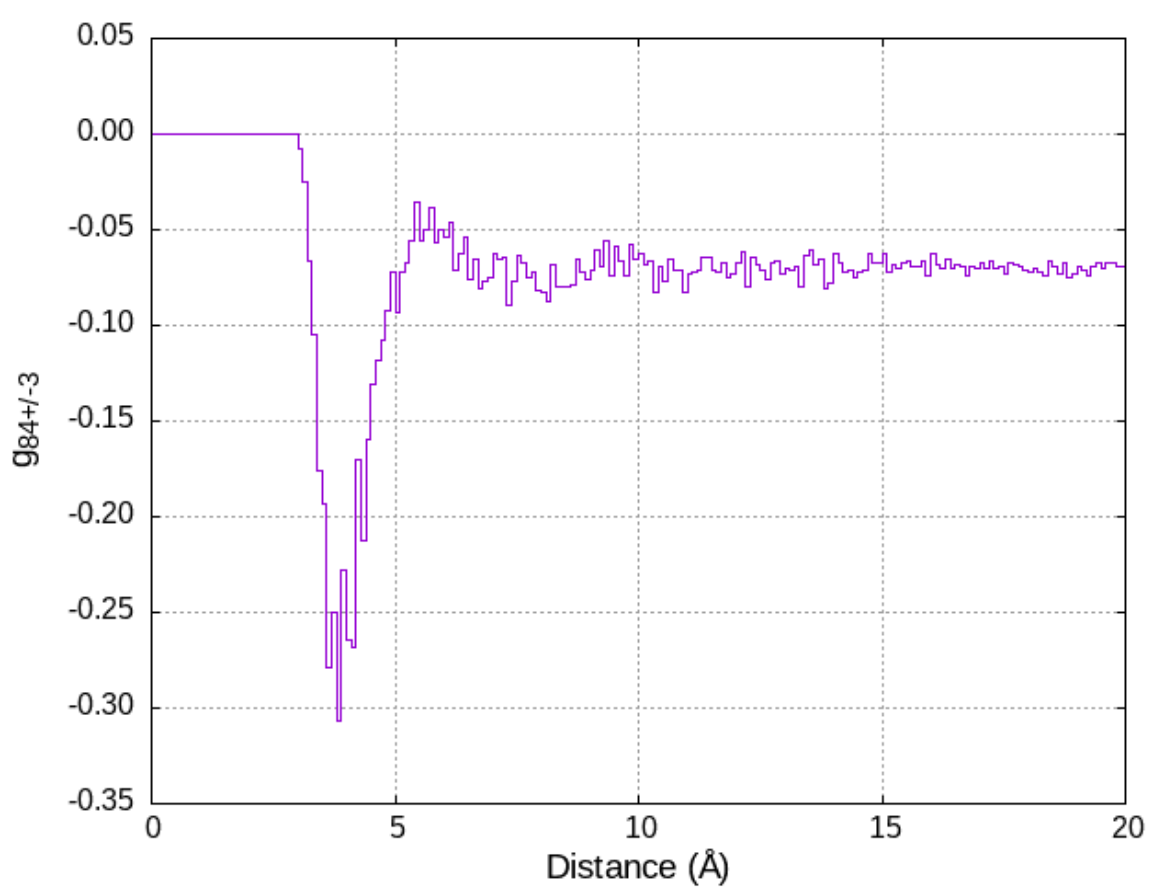

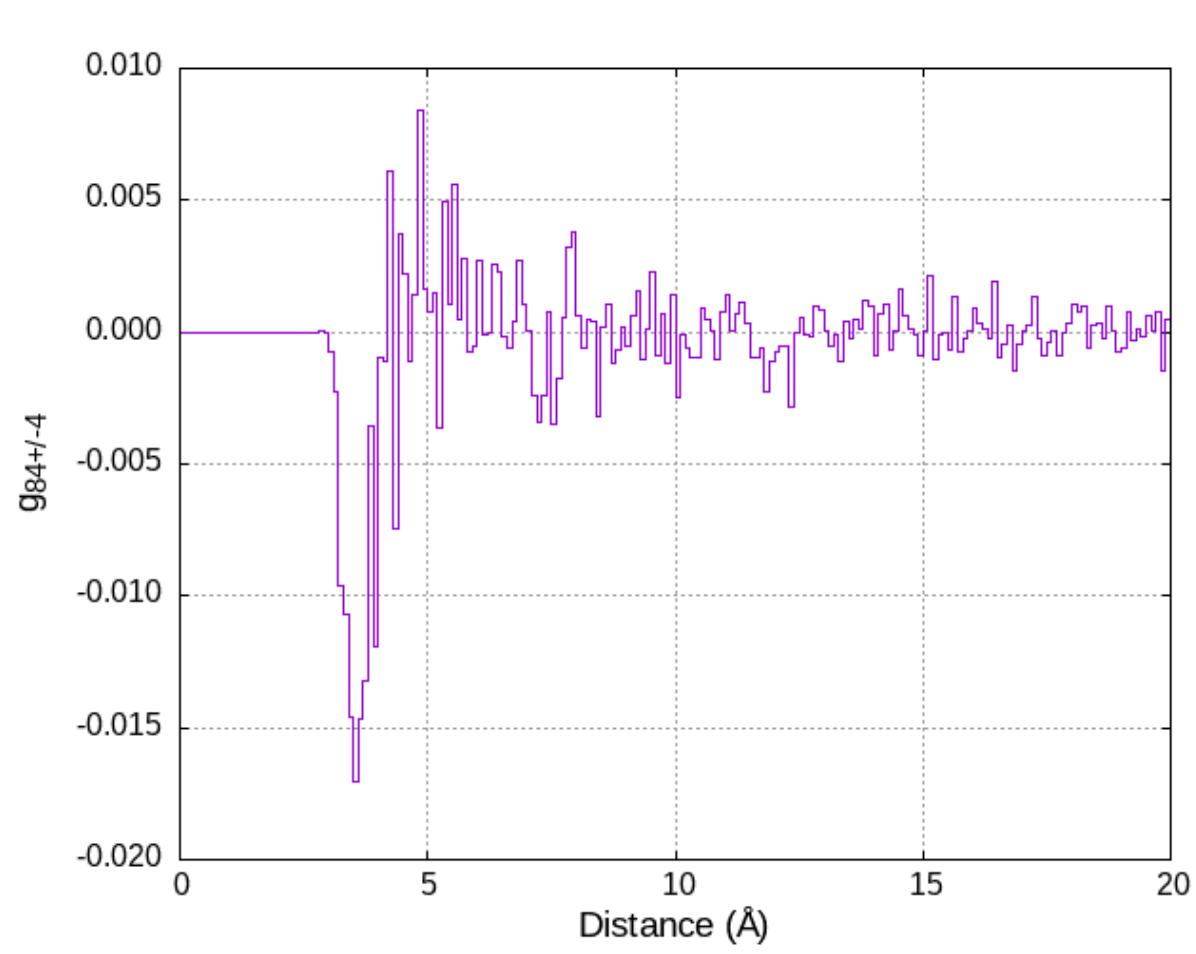

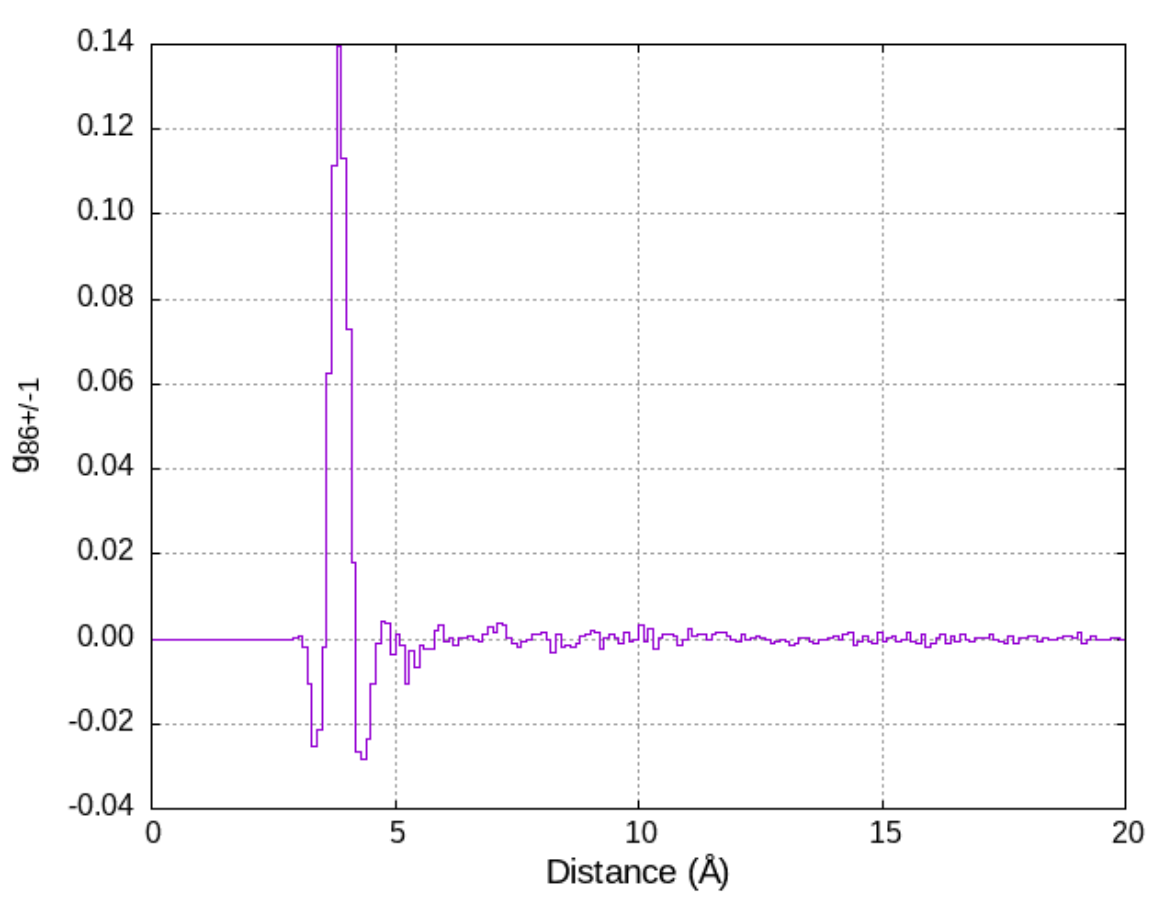

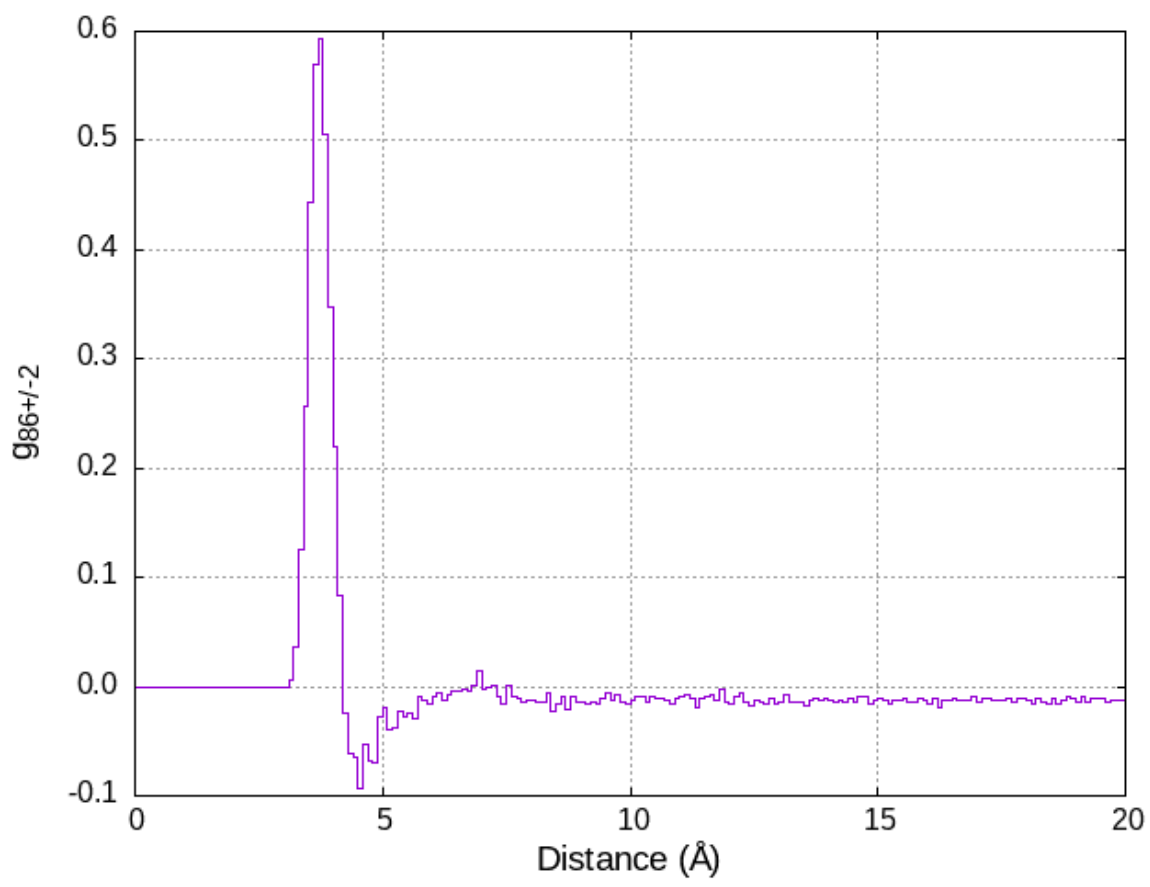

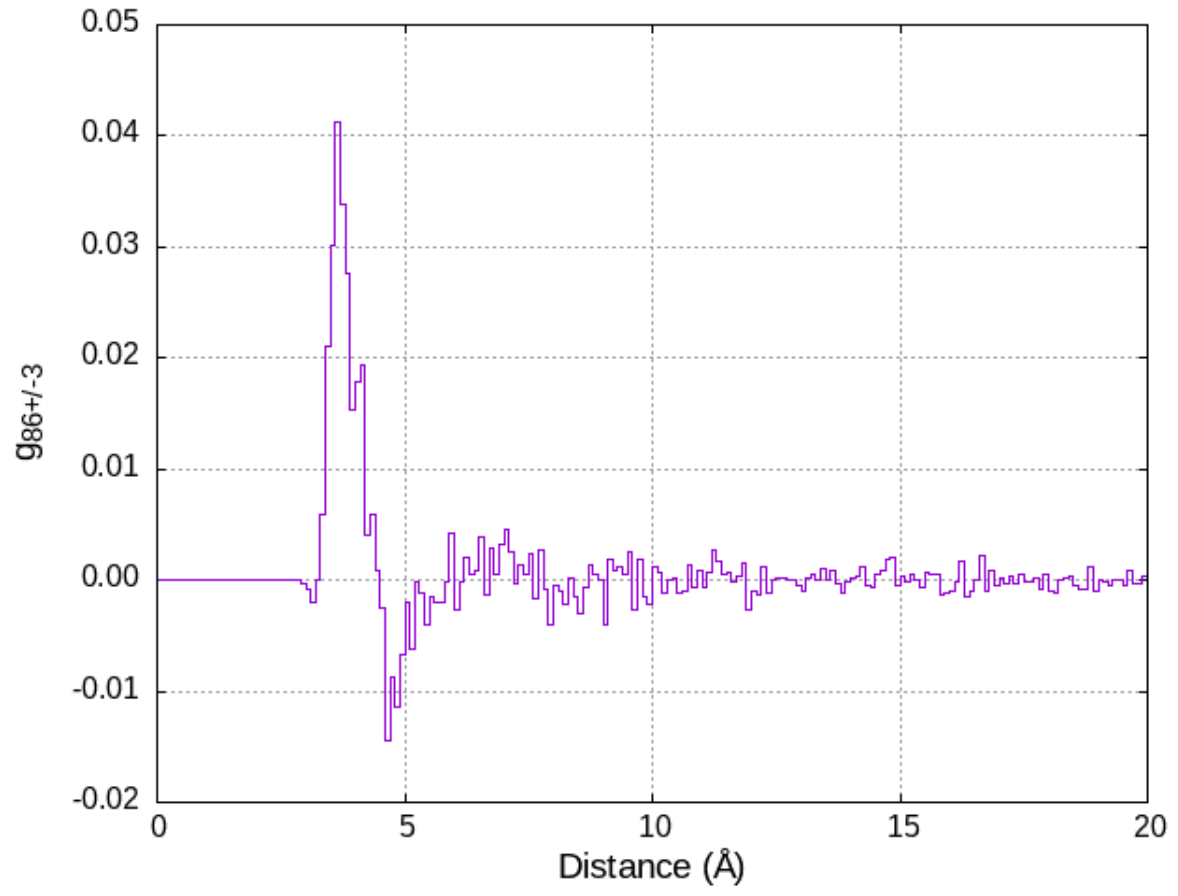

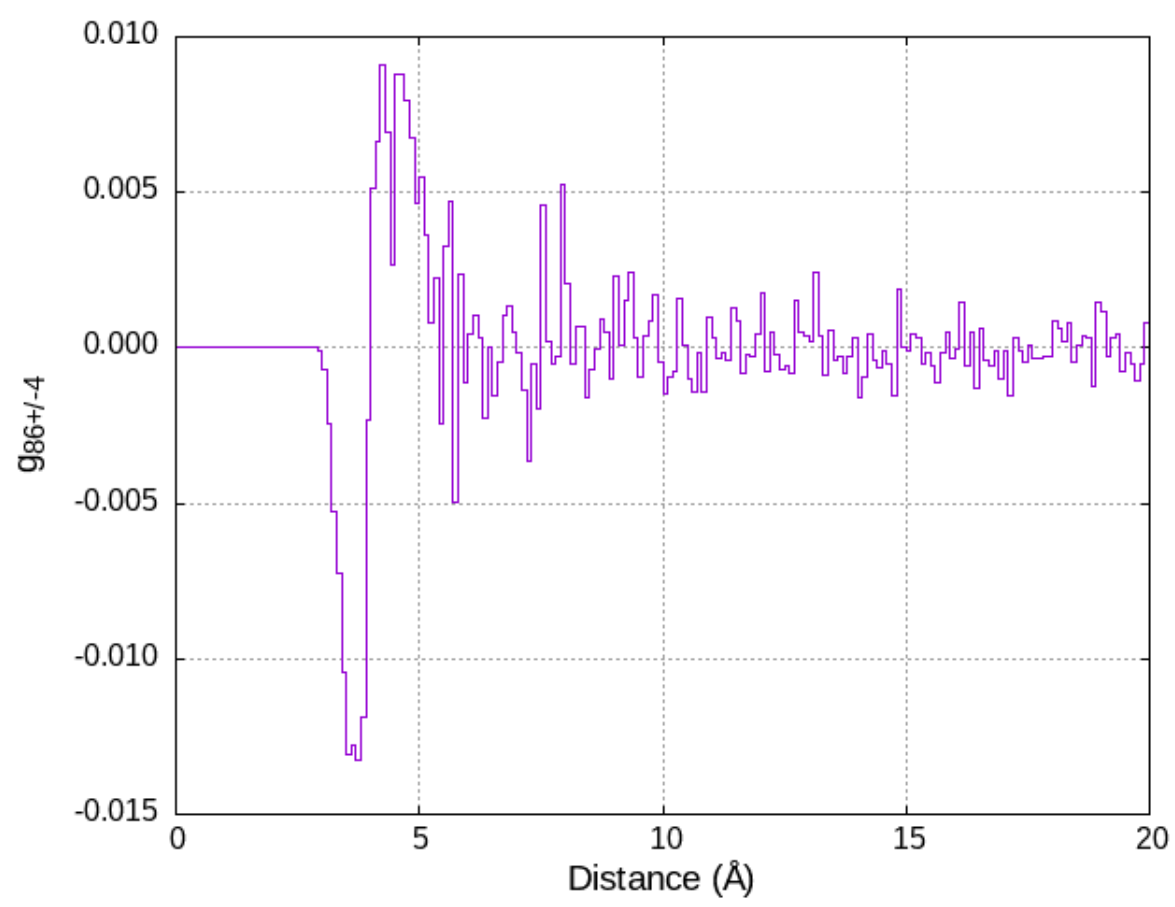

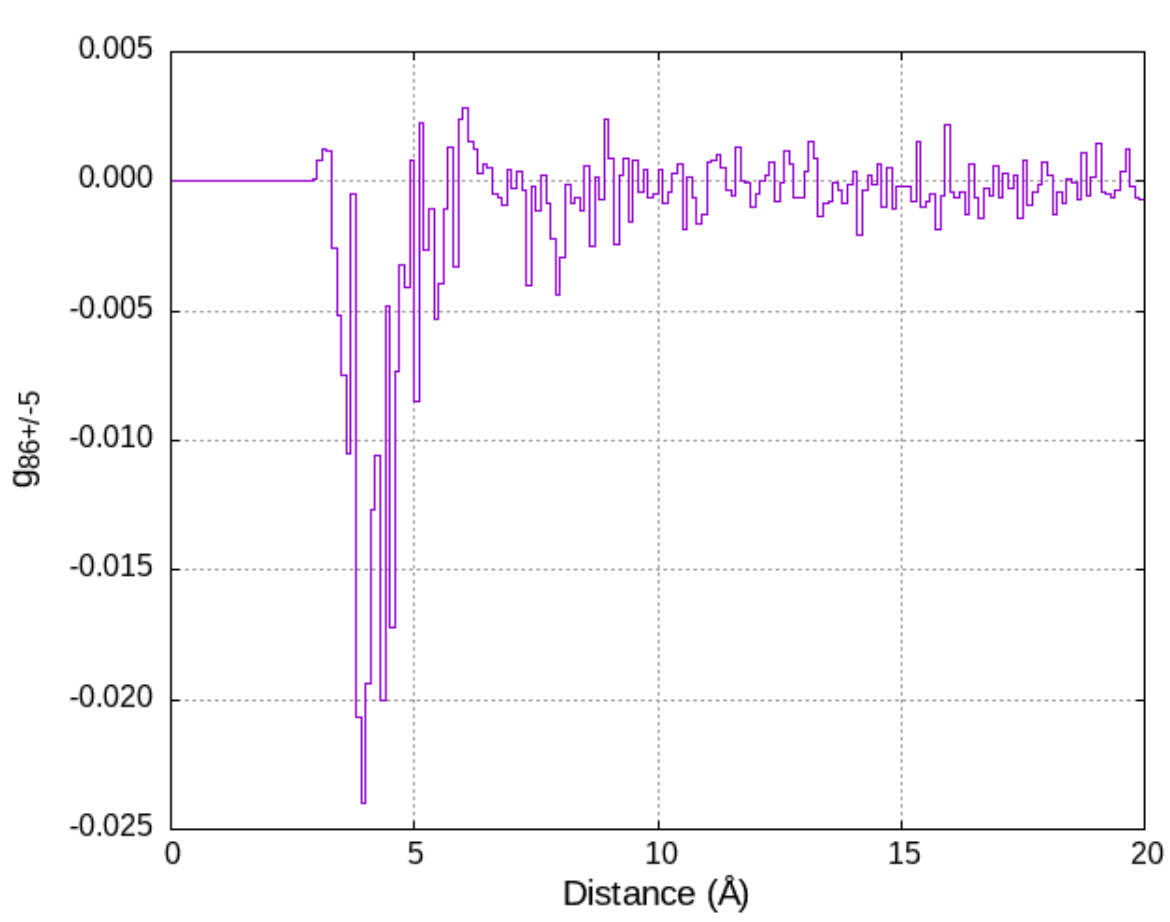

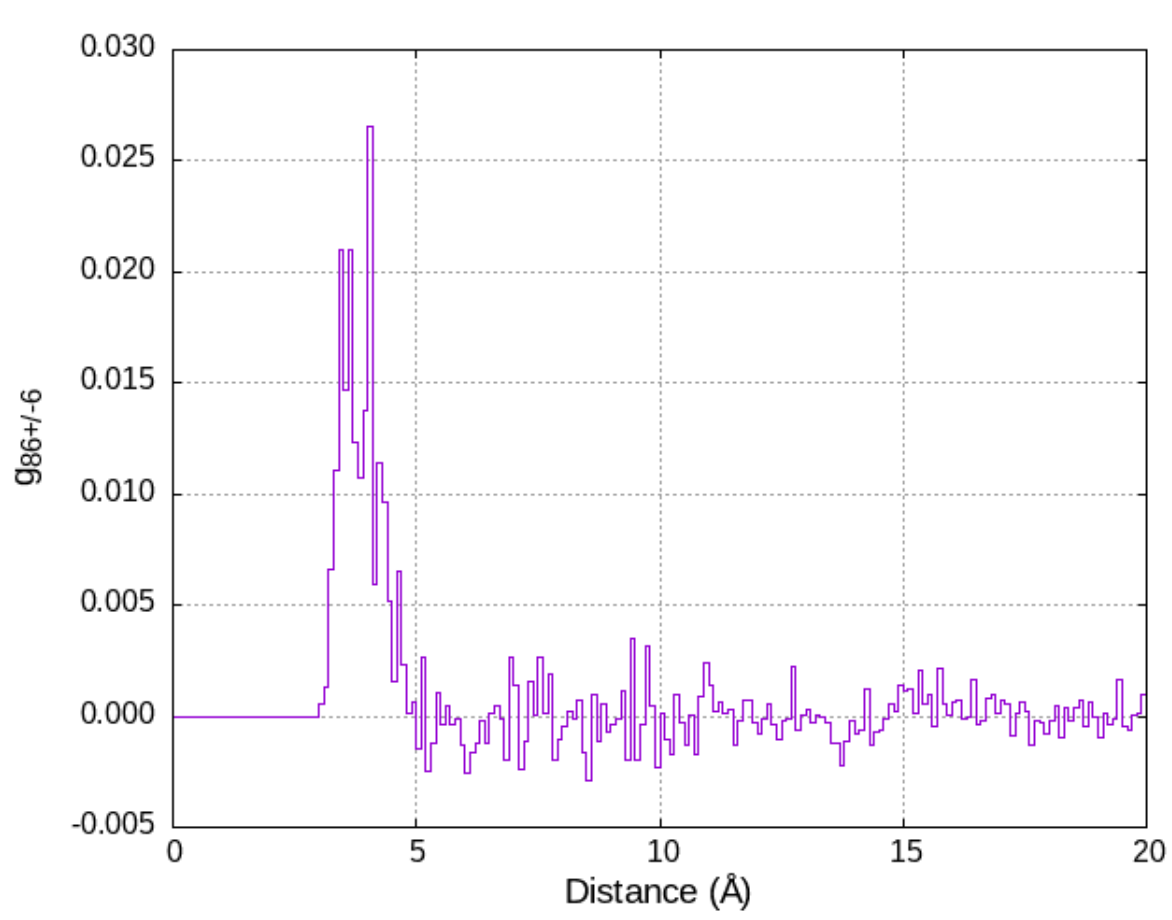

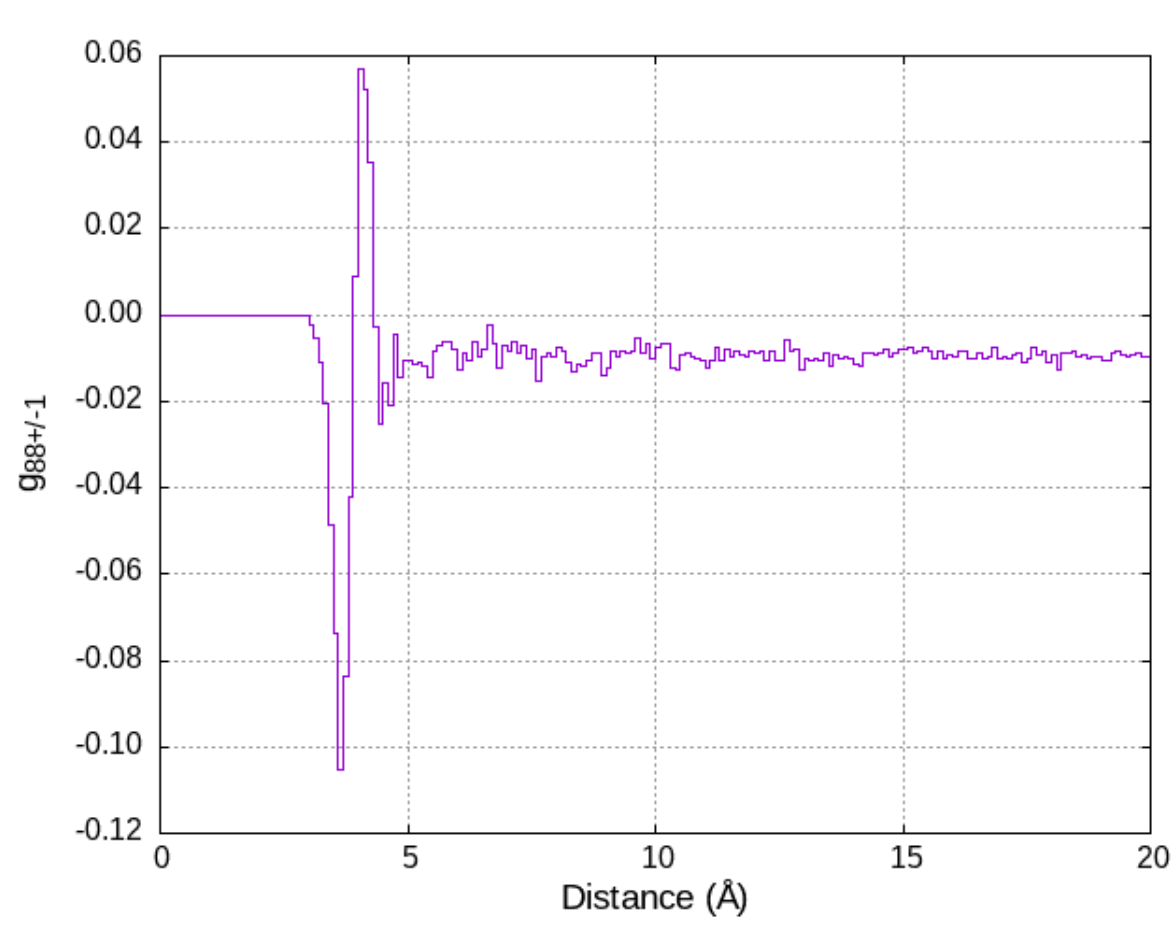

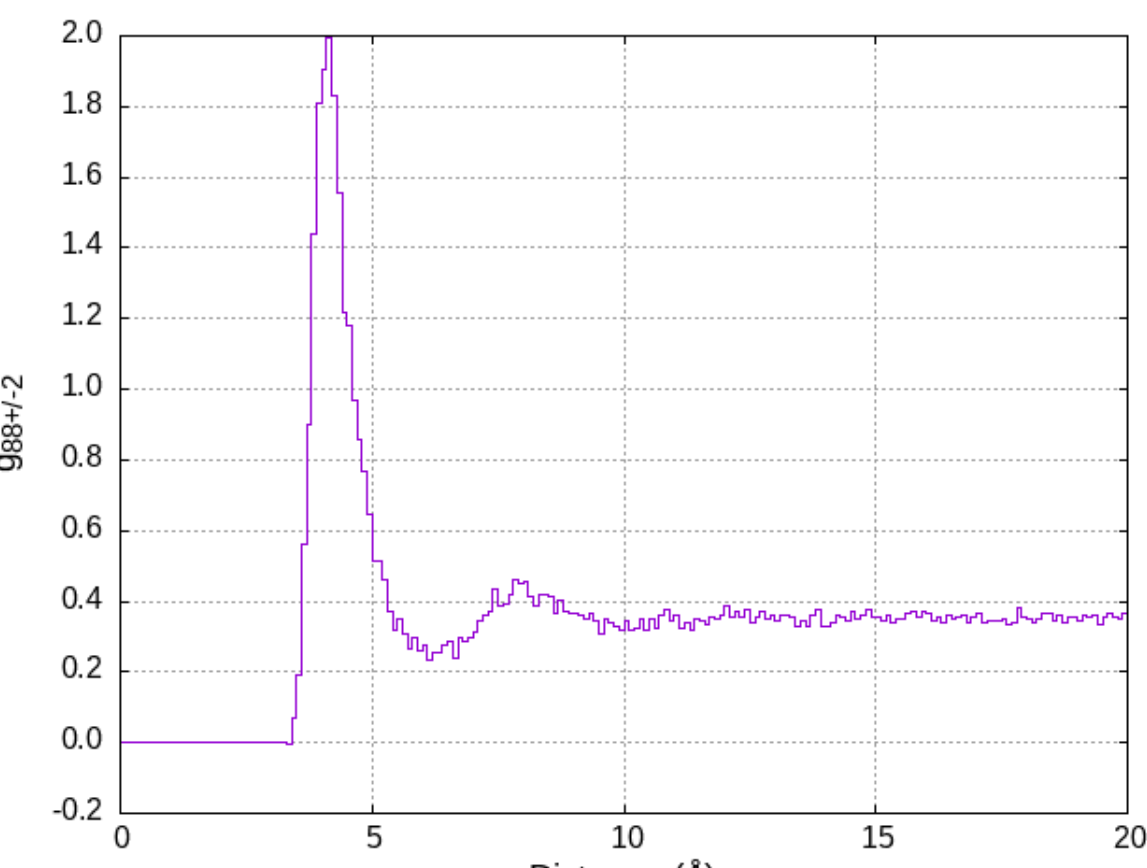

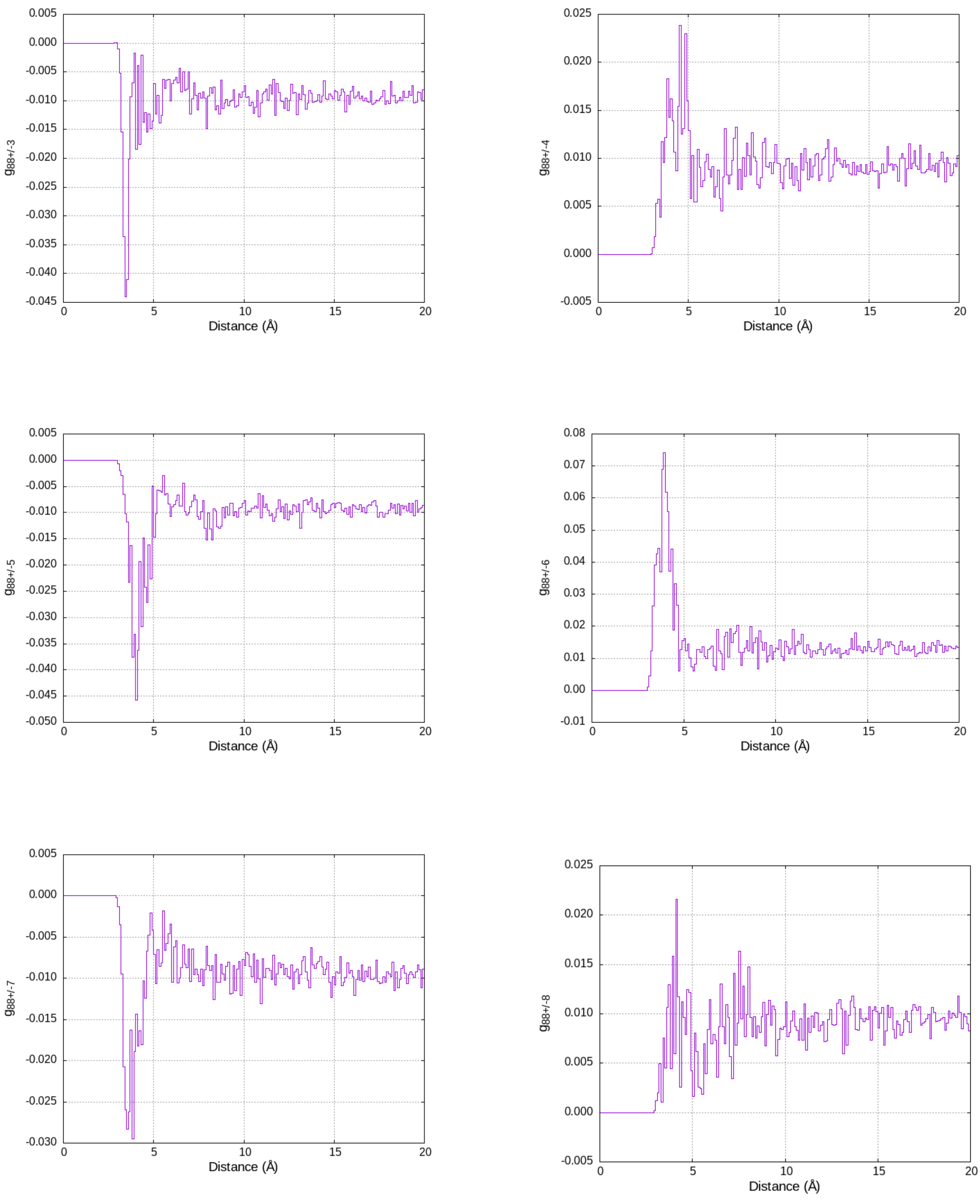

Fig. 5. $g_{ll'm}(r)$ computed according to Eq. 3. Each figure is identified by its title on the vertical axis or simply denoted as ($ll'm$).

Let us examine the above statements (i) to (iii). Regarding (i), it is seen in all figures of Fig. 5 that the first prominent positive and/or negative peak exists at $r \leq 5$ Å. This distance corresponds to the first peak of $g(r)$ of Fig. 3 at $r$ = 4.05 Å, indicating that the orientational correlation is most intense between the neighboring molecules. Regarding (ii), the peak values are either positive or negative. To examine the signs of the peaks, $\{Y_{l\text{-}m}(\omega_1)\ Y_{l'm}(\omega_2)\}$ were calculated for a pair of molecules with simple configurations, (1) T-shape, (2) Parallel, and (3) Cross, as shown in Fig. 6. The z-axis in blue passes through the C atoms in molecules 1 and 2: In (1), molecule 2 is vertical to molecule 1. In (2), molecule 2 is parallel to molecule 1 and positioned exactly above molecule 1. In (3), molecule 2 is rotated by 90˚ from (2). Note that $r$ is not included in $\{Y_{l\text{-}m}(\omega_1)\ Y_{l'm}(\omega_2)\}$. The parameters of $\omega_1$ and $\omega_2$ as well as computed results of $g_{220}(r)$, $g_{22+/-1}(r)$, and $g_{22+/-2}(r)$, denoted as (220), (221), and (222), respectively, are shown in Table 1.

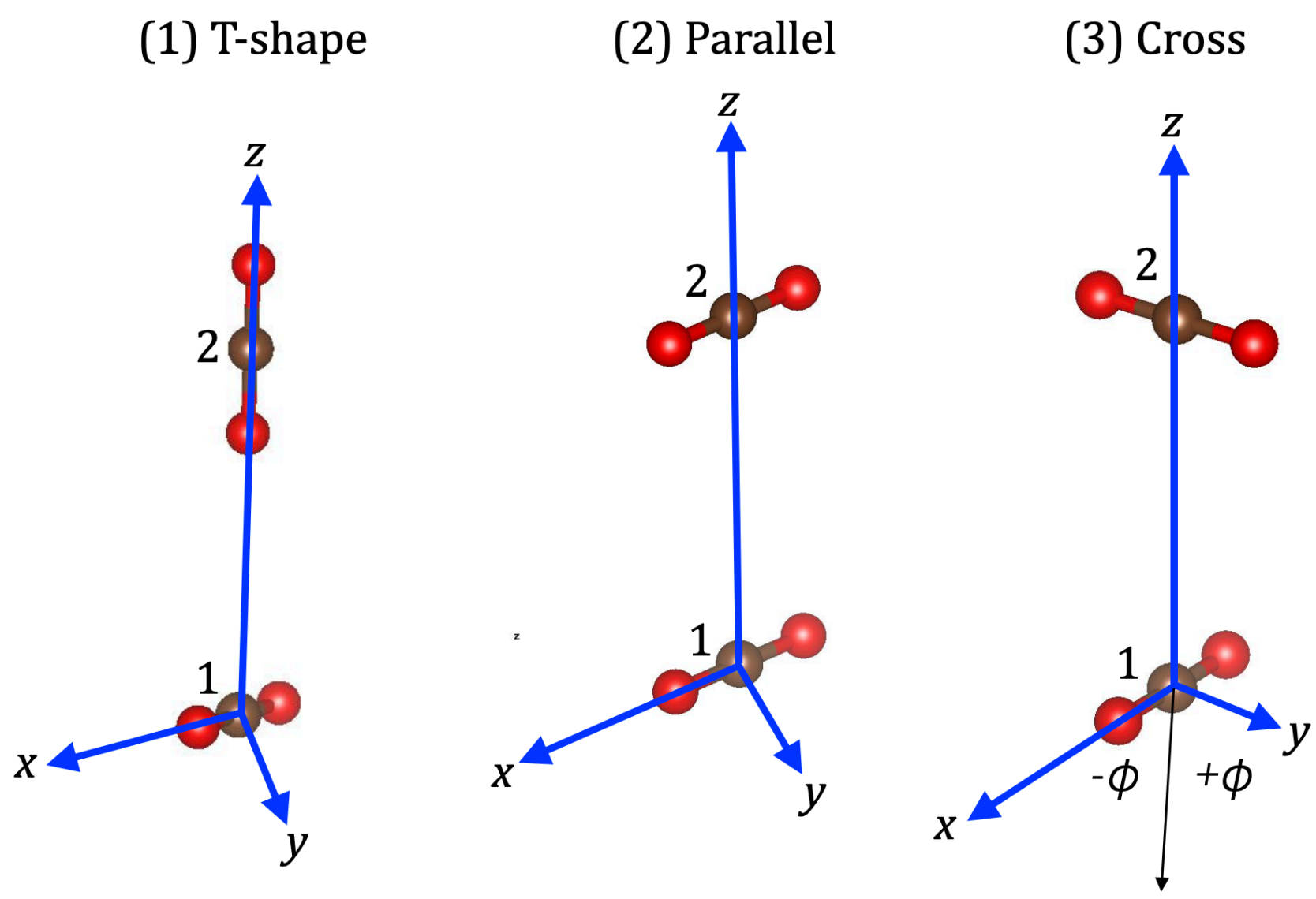

Fig. 6. Orientational relations of molecular pairs: (1) T-shape, (2) Parallel, and (3) Cross. Their parameters are summarized in Table 1. Note that $\varphi$ = 45˚: it is measured from a plane intersecting the angle $2\varphi$ between molecules 1 and 2. See Appendix III for details.

Table 1. Parameters for the molecular orientations in Fig. 6, and the calculated values of $\{Y_{l\text{-}m}(\omega_1)\ Y_{l'm}(\omega_2)\}$ denoted as $(ll'm)$.

| | (1) T-shape | (2) Parallel | (3) Cross |
|---|---|---|---|
| $\theta_1$ (deg.) | 90 | 90 | 90 |
| $\phi_1$ (deg.) | 0 | 0 | -45 |
| $\theta_2$ (deg.) | 0 | 90 | 90 |
| $\phi_2$ (deg.) | 0 | 0 | 45 |
| (220) | -0.20 | -0.20 | 0.10 |
| (221) | 0.0 | 0.0 | 0.0 |
| (222) | 0.0 | 0.0 | -0.15 |

From Table 1, it was not clear why the signs of $(ll'm)$ are different in different molecular configurations. It was only inferred that the signs depended on the properties of spherical harmonics involved in Eq. 7. Regarding (iii), $g_{ll'm}(r)$ is finite at long distances r > 15 Å (in practice, $r \geq 10$ Å) in 21 cases of Fig. 5, even though there is no orientation-dependent molecular interactions. This is only due to the properties of spherical harmonics, and has nothing to do with molecular orientational correlations. In addition to the above items (i) to (iii), it should be noted that $g_{88+/-2}(r)$, $g_{64+/-1}$(r), $g_{22+/-2}$(r), and $g_{220}$(r) had large peaks amongst others. It would be of interest to further investigate how $g_{ll'm}(r)$ change under different $P$ and $T$.

It should be emphasized that the numerical values of $(ll'm)$ do not make orientational correlations concretely visible, and their signs (plus or minus) just arise from the properties of spherical harmonics: this is unfortunate because Eq. 3 is a mathematically exact expression for $g_{ll'm}(r)$. At the present stage, therefore, it is not certain what the numbers of $(ll'm)$ and their signs mean. Datchi *et al.* [23] depicted cosines between adjacent molecules and graphically depicted the values. This technique is more informative than Eq. 3 at the present stage. Even so, it is necessary to further investigate the behavior of $g_{ll'm}(r)$ under different *P-T* conditions to make clear the relation between molecular orientational correlations and $g_{ll'm}(r)$ in liquid $CO_2$.

Shown in Fig. 7 below is a global view of molecular configuration in the cubic box used for *NVT*-MC simulation in the present study.

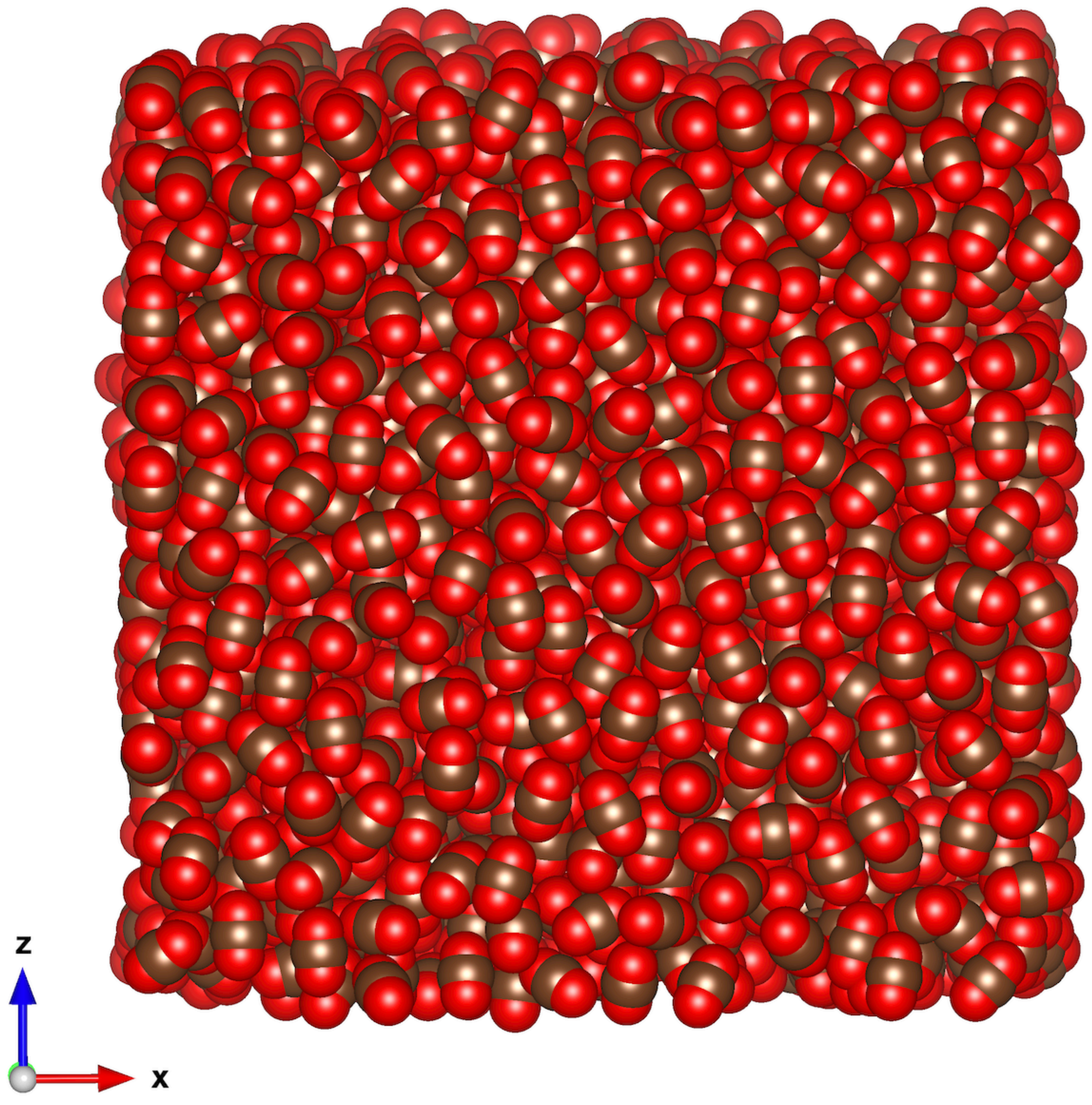

Fig. 7. A molecular configuration of liquid $CO_2$ at 250 K and 50 bar. The y-axis is in the opposite direction to the plane.

In Fig. 7, a significant numbers of molecules tend to be oriented nearly parallel to the z-axis. To examine the molecular orientations quantitatively, the unit vector of each molecule was brought to the same origin, and the end of the unit vector was plotted on a hemisphere with a unit radius. The result is shown in Fig. 8. It was found that the dots were more dense in the z-direction, indicating that the $CO_2$ molecules tend to be oriented in the z-direction. This is presumably due to the fact that the molecule has a rod-like structure. Further investigation will be done in the future works.

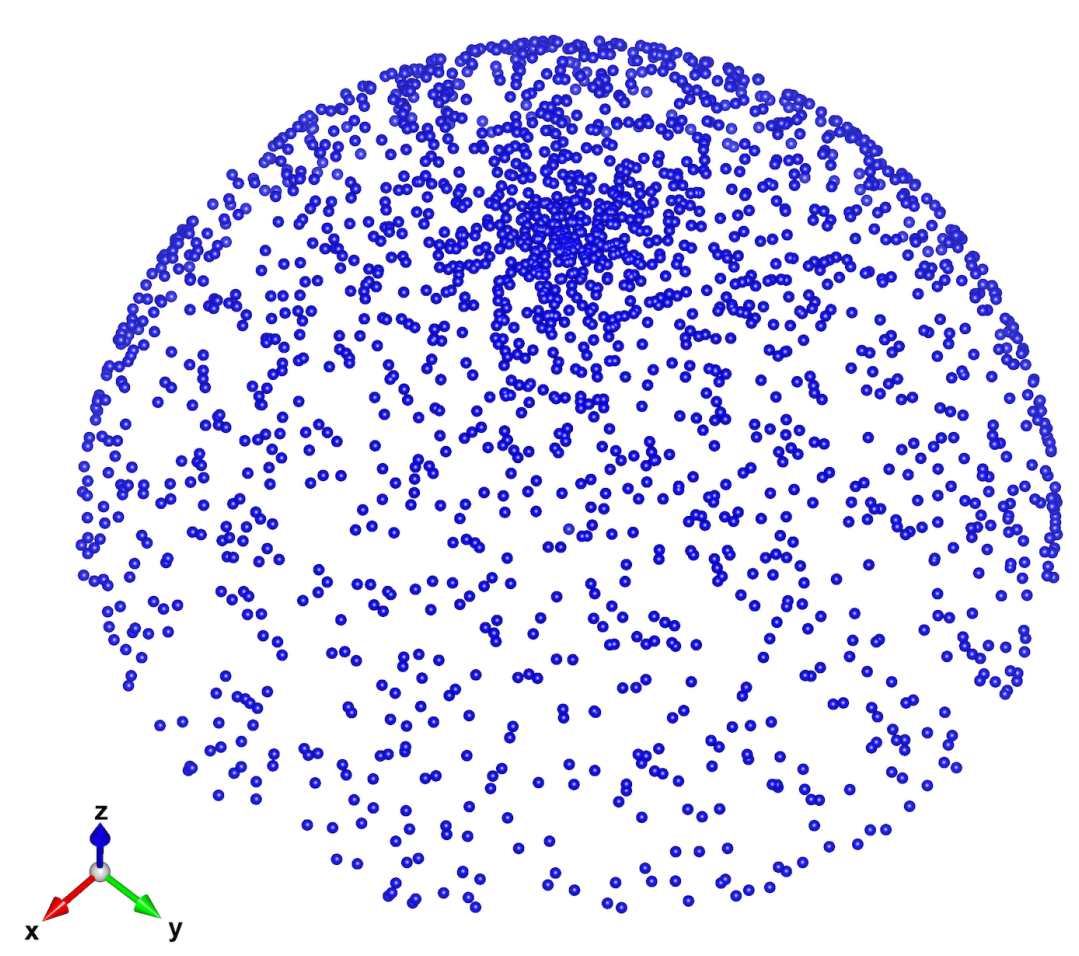

Fig. 8. Projection of molecular directions on the hemisphere with a unit radius.

## IV. Conclusion

Based on the pressure-constant MC simulation and the Kihara potential, the radial distribution function between molecular centers $g(r)$ and its expansion by spherical harmonics, in which the expansion coefficients $g_{ll'm}(r)$ express orientational correlation, were computed for liquid $CO_2$ up to $l$, $l'$, $|m| \leq 8$ and $r \leq 20$ Å at $T$ = 250 K and $P$ = 50 bar. It was shown that even though most spherical harmonics are complex, computations were able to be carried out with real values by defining $g_{ll'+/-m}(\mathrm{r})$, or equivalently using Eq. 9. It was found that $g_{ll'm}(r)$ had a prominent structure at around a distance where the radial distribution $g(r)$ had a major peak. It was also found that $g_{ll'+/-m}(r)$ had large peaks even for $l$ being as large as $l$ = 8. It is expected that further studies will be able to make clear the relation between $g_{ll'+/-m}(r)$ and molecular orientational correlations. The implication of the values of the expansion coefficients as well as their positive/negative peaks, relating to orientational correlation, are yet to be investigated further.

## Appendix I. Kihara core potential and Kihara potential

The Kihara potential is a unique empirical model of intermolecular potential for simple molecules such as $N_2$ [24] and $CO_2$ [25] as well as spherical molecules. Most often, intermolecular potential is assumed to be a sum of inter-atomic potentials defined as a function of atomic center distances. By contrast, the Kihara core potential assumes a solid core inside a molecule, and defines the intermolecular potential as a function of the nearest core-core distance, $\rho$, as shown in Fig. A1. For the case of $CO_2$, the core is a rod with a length $l$ and zero diameter. The Kihara core potential was initially used in the research of virial coefficients, and confirmed to be a good empirical potential for spherical and non-spherical molecules [10, 26-33]. This justified the Kihara potential to be used for the study of simple molecular fluids.

Fig. A1. Concept of Kihara core potential for $CO_2$. $l$: core length, $r$: molecular center-to-

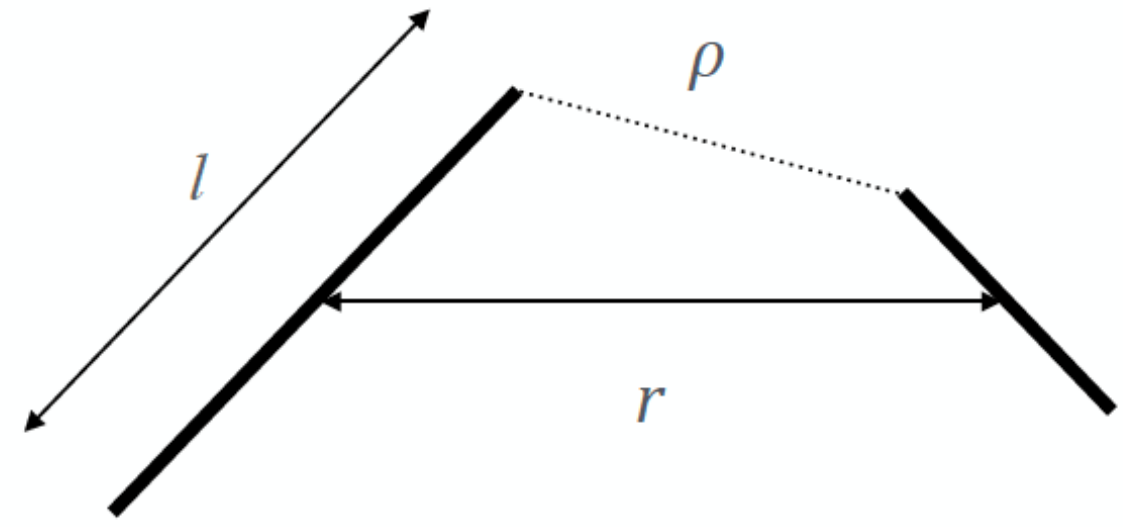

center distance, and $\rho$: the shortest core-core distance [25].

The Kihara potential for $CO_2$ molecule consists of the core-core potential and point EQQ potential. Thus, the former is referred to as the Kihara core potential in the present article. The Kihara core potential is expressed by a Lennard-Jones (LJ) potential $V_c(\rho)$,

$$V_c(\rho) = U_0 \left[ \frac{6}{n-6} \left( \frac{\rho_0}{\rho} \right)^n - \frac{n}{n-6} \left( \frac{\rho_0}{\rho} \right)^6 \right], \qquad \text{(A1)}$$

where $V_c(\rho)$ expresses the 12-6 LJ potential if $n$ = 12, and the 9-6 LJ potential if $n$ = 9. Five sets of the parameter values for $l$, $\rho_0$, $U_0$ and $n$ have been examined in Ref. 25 in reference to the Raman spectra of solid $CO_2$ and their pressure dependence using empirical lattice constants. In the present article, $n$ = 9, $U_0$ = 232 K (1 K = 1.38 x $10^{-23}$ J), $\rho_0$ = 3.27 Å, and $l$

= 2.21 Å (model V of Ref. 25) were used. It should be noted that the core length *l* used in the present simulation was smaller than the experimental O-O distance, 2.30 Å. In the actual computations, the maximum range of the Kihara core potential was limited to 10 Å, and the van der Waals potential that had no orientation dependence,

$$V_c(\rho) = -U_0\left[\frac{n}{n-6}\left(\frac{\rho_0}{r}\right)^6\right] \tag{A2}$$

was assumed [10] from $r$ = 10 to 15 Å. The EQQ potential between $CO_2$ molecules was expressed by a point electrostatic quadrupole moment, Q = - 4.3 x $10^{-26}$ esu.$cm^2$, that was placed in the center of molecule and assumed to work up to $r$ = 15 Å. For the *Pa3* structure of solid CO2 phase I at ambient pressure and $T$ = 5 K, where the lattice constant $a$ = 5.54 Å, the EQQ potential energy accounted for 43% of the total energy of the cubic box, a similar magnitude of the Kihara core potential energy. The EQQ potential is strongly orientation dependent, and determines the molecular orientations in the *Pa3* structure.

**Appendix II. Computational procedure**

The *NPT*-MC simulation was carried out in the following procedure:

(i) $CO_2$ molecules of 2048 were contained in a cubic box of volume $V$ with the edge length of $A_L$ ($V = A_L^3$). The periodic boundary condition was applied in the potential computations. In the present results of *NPT*-MC simulation at 250 K and 50 bar, $A_L$ = 53.2 Å, and hence the molar volume was 44.3 $cm^3$/mol that was 6.6% greater than the NIST datum, 41.56 $cm^3$/mol. In terms of $A_L$, the difference was 2.1%. In solid $CO_2$ phase I, the Kihara potential showed an interesting behavior: at $P$ = 1 bar, the computed lattice constants were greater than experimental lattice constants at higher temperatures ($T \leq$ 194 K), while they agreed well with experiment at high pressures ($P \leq$ 10 GPa). This is presumably because of the unique feature of the Kihara potential as described in Appendix I, and the Kihara potential used seemed to be slightly too repulsive.

(ii) The molecular coordinates include the molecular center $r_i = \{x_i, y_i, z_i\}$ and the molecular orientation $\omega_i = \{\theta_i, \varphi_i\}$ of the $i$-th molecule, where $i$ runs from 1 through 2048. The coordinates relating to molecule $i$, $\{r_i, \omega_i\}$, will be denoted altogether as $\Psi_i = \{r_i, \omega_i\}$ and all of $\Psi_i$ will be denoted as $\Psi = \{\Psi_i\}$. Thus, there are 2049 parameters, $V$ and $\Psi$, for the *NPT*-MC simulation.

(iii) In MC simulation, random numbers $q$ are sequentially used, where $0 \leq q \leq 1$. At the beginning of the computation, a variable for a MC trial was randomly chosen from $\{V, \Psi\}$, and replaced with a new value that was determined in the following way: When the volume $V$ of the cubic box was chosen as a variable for the MC trial, a new volume $V'$ can be determined by the equation:

$$ln(V') = ln(V) + (2q - 1)V_m. \qquad (A3)$$

Namely, the new volume $V'$ is determined from the natural logarithm of $V$, $ln(V)$, in which

$$ln(V) - V_m \leq ln(V') \leq ln(V) + V_m. \qquad (A4)$$

In the present study, $V_m = 0.02$ was the standard value, though other values had been tested. In the new volume $V'$, the molecular centers of all molecules were uniformly displaced according to the volume change, and $\Delta\varepsilon$, defined by the next equation, were computed [1].

$$\Delta\varepsilon = -\beta\,[U(\Psi_i, V') - U(\Psi_i, V) + P\,(V' - V) - (N + 1)\,\beta^{-1}\,ln(V'/V)]. \qquad (A5)$$

In Eq. A5, $U(\Psi_i, V')$ and $U(\Psi_i, V)$ are the total molecular interaction energy in the cubic box with the new and a current volumes, $V'$ and $V$, respectively, and $\beta = 1/k_BT$, $k_B$ being the Boltzmann constant. The new volume was accepted only if $exp(\Delta\varepsilon) \geq q$, and otherwise the current values of $\{V, \Psi\}$ were restored.

(iv) On the other hand, if the $i$-th molecule was chosen for the MC trial, the new values $\Psi_i'$ were determined using a set of predetermined maximum values for the parameters, $\Psi_m = \{X_m, \Omega_m\} = \{x_m, y_m, z_m, \theta_m, \varphi_m\}$ in such a way as:

$x_i' = x_i + (2q - 1)\, x_m$ (A6)

for $x_i$, and other parameters were similarly determined. In the present article, the values used were $\{x_m, y_m, z_m\}$ = 3 Å, $\theta_m$ = 90°, $\varphi_m$ = 180° down to $\{x_m, y_m, z_m\}$ = 2 Å, $\theta_m$ = 45°, $\varphi_m$ = 90°. The former set of values was used to include rare events, while the latter was the standard set of values. The total number of MC trial was 1.8 x $10^9$. Note that in the previous articles of solid $CO_2$ by the present author [13-17], the first molecule was always pinned at the origin. This restriction was lifted in the present article, and instead, molecules moved out from the cubic box were returned to the equivalent positions in the cubic box at the beginning/end of the computation: some molecules moved out of the cubic box during the *NPT*-MC simulations. For MC simulation, the acceptance ratio is usually set so that it becomes 35 - 50 % [4]. In the present study, the acceptance ratio depended on $\{x_m, y_m, z_m\}$ and $\{\theta_m, \varphi_m\}$, and it was less than a few % for orientational changes: however, for a set of $10^8$ MC trials, the absolute number of acceptance was 2 to 3 x $10^6$ that seemed to be sufficient to sample different configurations. On the other hand, the acceptance ratio for volume was 20 to 60%: it depended on the value of $V_m$. In both cases, the total energy was stable. The number of acceptance under a stable total energy, indicative of the system being in equilibrium, is more important than the acceptance ratio because the latter can be increased only by using small values of $\{x_m, y_m, z_m\}$ and $\{\theta_m, \varphi_m\}$ that may however fail samplings of rare events.

(v) In the last computation with $10^8$ MC trials, $A_L$ and the total energy were stored every time the new volume was accepted. Successively, the averages of $A_L$ and the total energy were evaluated over the last 100 data. The average value of $A_L$ was used for the next computations as the edge length of the cubic box, and the molecular center positions in the last computation were uniformly moved according the difference between the $A_L$ determined in the very last computation of *NPT*-MC simulation and the averaged value of $A_L$. For molecular orientations, the last results of *NPT*-MC simulation were used as the initial values for the next *NVT*-MC simulation. The averages of the molar volume and the total energy were also calculated.

(vi) Successively, 100 sets of *NVT*-MC simulation were undertaken using the average $A_L$ (equivalently the volume of the cubic box) determined in (v). Each set of *NVT*-MC simulation consisted of $10^7$ MC trials, and typically 2 x $10^6$ new orientations were accepted. The last data $\Psi$ of each set were stored. After this step was over, each of the data $\Psi$ was used for the computation of the radial distribution function $g(r)$ and its expansions with spherical harmonics $g_{ll'm}(r)$, and then the 100 results were averaged to obtain the final results.

There was a problem in this step: the total energy in the first *NVT*-MC simulation was approximately 3% higher than the averaged total energy obtained by *NPT*-MC simulation in (v). This had occurred also in solid $CO_2$ when *NPT*-MC simulation was switched to *NVT*-MC simulation. The cause has not yet been known and is left for future investigation.

**Appendix III. A method to evaluate sin φ and cos φ**

In Sec. 2, it was mentioned that in Eqs. 2 and 3 and all other related equations, the molecular orientation $\omega$ in spherical harmonics $Y_{ll'm}(\omega)$ refers not to the original coordinate system that was used for *NPT*-MC simulation but to the blue axis shown in Fig. 2, passing through molecules 1 and 2. Let the unit vector representing the direction of the blue axis be $\boldsymbol{e}_0$, and the unit vectors representing the directions of molecules 1 and 2 be $\boldsymbol{e}_1$ and $\boldsymbol{e}_2$, respectively. In the present case of $CO_2$ molecules, cos $\theta_i$ and hence sin $\theta_i$ ($i$ = 1 and 2) can be evaluated from the inner product between the unit vectors of the axis $\boldsymbol{e}_0$ and molecules $\boldsymbol{e}_i$. The issue here is that the coordinate axes that is vertical to $\boldsymbol{e}_0$ cannot be uniquely determined, and hence the angles $\varphi_1$ and $\varphi_2$, associated with molecules 1 and 2, respectively, cannot be uniquely determined, either. Unless the coordinate axes vertical to $\boldsymbol{e}_0$ is properly determined so that $|\varphi_1| = |\varphi_2|$, the value of

$\langle Y_{lm}^*(\omega_1)\ Y_{l'\text{-}m}^*(\omega_2)\rangle_{\text{shell}}$ becomes a complex number in general. It is thus necessary to find a way that $|\varphi_1| = |\varphi_2|$. In the present article, this was achieved by using external products of those vectors. In Fig. A3, all vectors are depicted in the same coordinate system, and the vector $\boldsymbol{e}_0$ is set vertically, for simplicity. Now, define vectors $\boldsymbol{e}_3$ and $\boldsymbol{e}_4$ by $\boldsymbol{e}_1 \times \boldsymbol{e}_0 = \boldsymbol{e}_3$ and $\boldsymbol{e}_2 \times \boldsymbol{e}_0 = \boldsymbol{e}_4$, respectively. It is seen that both vectors $\boldsymbol{e}_3$ and $\boldsymbol{e}_4$ are located on the same plane vertical to $\boldsymbol{e}_0$. The angle between $\boldsymbol{e}_3$ and $\boldsymbol{e}_4$ is $2\varphi$. From a trigonometrical relation of cos($2\varphi$), cos($\varphi$) and then sin($\varphi$) are obtained. The vector $\boldsymbol{e}_5$ is depicted only for explanation that a plane defined by $\boldsymbol{e}_0$. and $\boldsymbol{e}_5$ bisects the $2\varphi$ angle between molecules 1 and 2. Note that the angle $\varphi$ between $\boldsymbol{e}_3$ and $\boldsymbol{e}_5$ is negative while the angle $\varphi$ between $\boldsymbol{e}_4$ and $\boldsymbol{e}_5$ is positive as they are measured from $\boldsymbol{e}_5$.

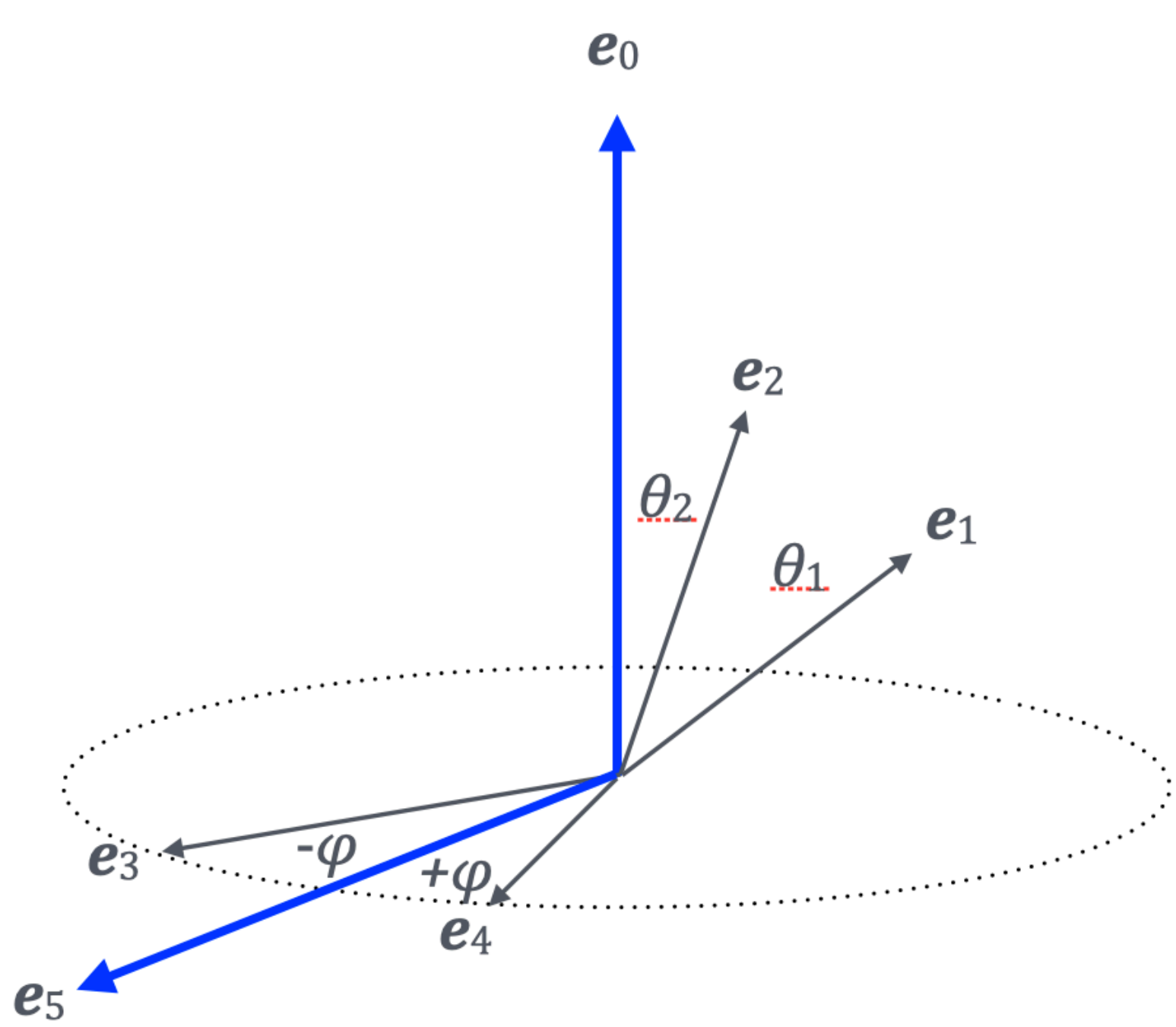

Fig. A3. The coordinate axes and molecular orientation.